\documentclass[a4paper,11pt]{article}

\pdfoutput=1
\usepackage{jheppub}
\usepackage[T1]{fontenc}
\usepackage{diagbox}
\usepackage{adjustbox}
\usepackage{multirow}
\usepackage{pifont}
\usepackage{array}
\usepackage{mathtools}
\usepackage{booktabs}
\usepackage[dvipsnames]{xcolor}
\usepackage{tcolorbox}
\usepackage{float}
\usepackage{placeins}

\usepackage{verbatim}

\usepackage{graphics}
\usepackage{graphicx}
\usepackage{epsf}
\usepackage{epsfig}
\usepackage{float}
\usepackage{makecell}
\usepackage{multirow}
\usepackage{color}
\usepackage{xcolor}
\usepackage{simplewick}
\usepackage{longtable}

\usepackage[utf8]{inputenc}
\usepackage{amsmath}
\usepackage{amssymb}
\usepackage{subfig}
\usepackage[normalem]{ulem}

\usepackage{mathtools}
\usepackage{amsmath}
\usepackage{diagbox}

\usepackage{bm}       
\usepackage{bbm} 

\usepackage{relsize}  

\usepackage{autobreak}

\usepackage{makeidx} 
\usepackage{bbding} 
\usepackage{listings} 
\usepackage{ytableau}
\usepackage[utf8]{inputenc}
\usepackage[T1]{fontenc}
\usepackage{lmodern}

\usepackage{tikz}
\usepackage[vcentermath]{youngtab}
\usepackage{slashed} 
\usepackage[font=small]{caption} 
\usepackage{times}

\usepackage{lscape}

\newcounter{magicrownumbers}
\renewcommand{\themagicrownumbers}{\arabic{magicrownumbers}}
\newcommand{\myrownumber}{\refstepcounter{magicrownumbers}\themagicrownumbers}

\newcommand\undermat[2]{
	\makebox[0.5pt][l]{$\smash{\underbrace{\phantom{%
					\begin{matrix}#2\end{matrix}}}_{ \let\scriptstyle\textstyle\text{\large $#1$}}}$}#2}
\newcommand\overmat[2]{
	\makebox[-1pt][l]{$\smash{\overbrace{\phantom{%
					\begin{matrix}#2\end{matrix}}}^{ \let\scriptstyle\textstyle\text{\large $#1$}}}$}#2}    

\usepackage{hyperref} 
\usepackage{cleveref} 

\hypersetup{colorlinks = true, 
	    linkcolor = black, 
	    urlcolor = blue,
            citecolor = blue} 

\title{\boldmath{Complete CP Eigen-bases of Mesonic Chiral Lagrangian up to $p^8$-order} }

\author[a,b,c]{Xuan-He Li,}
\author[b,c]{Hao Sun,}
\author[a,b,c]{Feng-Jie Tang,}
\author[a,b,c,d]{Jiang-Hao Yu,}
\affiliation[a]{School of Fundamental Physics and Mathematical Sciences, Hangzhou Institute for Advanced Study, University of Chinese Academy of Sciences, Hangzhou 310024, China}
\affiliation[b]{Institute of Theoretical Physics, Chinese Academy of Sciences,\\Beijing 100190, P. R. China}
\affiliation[c]{School of Physical Sciences, University of Chinese Academy of Sciences,\\Beijing 100049, P.R. China}
\affiliation[d]{International Centre for Theoretical Physics Asia-Pacific,\\Beijing/Hangzhou, China}

\emailAdd{jhyu@itp.ac.cn}

\abstract{
Chiral perturbation theory systematically describes the low energy dynamics of meson and baryons using nonlinear Nambu-Goldstone fields. Using the Young tensor technique, we construct the pure mesonic effective operators up to $p^8$-order, one-to-one corresponding to contact amplitudes with the on-shell Adler zero condition. The off-shell external sources, non-vanishing under equation-of-motion conditions, are also added to the operator bases. We also show the invariant tensor bases using the Young tableau is equivalent to the trace bases with Cayley-Hamilton relations. Separated into different $CP$ eigenstates, at $\mathcal{O}(p^8)$ we obtain the operator lists of the 567 $C$+$P$+ operators, 483 $C$+$P$- operators, 376 $C$-$P$+ operators, and 408 $C$-$P$- operators for $SU(2)$ case, while there are 1959 $C$+$P$+ operators, 1809 $C$+$P$- operators, 1520 $C$-$P$+ operators, and 1594 $C$-$P$- operators for $SU(3)$ case, consistent with results using the Hilbert series.

}

\begin{document}
\maketitle
\flushbottom

\section{Introduction}
\label{sec:intro}
At the low energy, non-perturbative effects of the quantum chromodynamics (QCD) make it challenging to deal with the strong interactions among hadrons.
Fortunately, the chiral perturbation theory (ChPT)~\cite{Weinberg:1968de,Weinberg:1978kz,Gasser:1983yg,Gasser:1984gg,Gasser:1987rb} provides a systematic approach to describe the low-energy QCD phenomena.
In the three-flavor case, the ChPT is based on the fact that the chiral symmetry $ SU(3)_L \times SU(3)_R $  is spontaneously broken down to $SU(3)_V$, resulting in the Goldstone bosons that form a pseudoscalar meson octet.
Identify these mesons, along with the external scalar and vector sources as fundamental building blocks, one can construct the effective Lagrangian of pure mesons using the $G/H$ coset construction framework~\cite{Weinberg:1968de, Coleman:1969sm, Callan:1969sn}.

The effective Lagrangian of the ChPT can be systematically constructed order by order, using a chiral expansion in terms of a generic low-energy momentum transfer $p$, and thus the $n$-th order is $p^n$. The leading order (LO) is counted as $\mathcal{O}(p^2)$, including momentum-squared and the light quark masses. For pure meson system, due to the Lorentz derivative indices always being contracted, all purely mesonic operators are of the even power~\cite{scherer2012primer}:
\begin{equation}
    \mathcal{L}_{2n} = \sum_{i} c_{i}^{2n} \mathcal{O}_{i}^{2n}\,, 
\end{equation}
where $\mathcal{O}_{i}^{2n}$ belongs to the $ \mathcal{O}(p^{2n})$, and \( c_{i}^{2n} \) represents the low-energy constants (LECs).

The effective operators at the order $\mathcal{O}(p^2)$ and $\mathcal{O}(p^4)$, or called LO and next-to-leading order (NLO), were first introduced in Refs.~\cite{Wess:1971yu,Witten:1983tw,Gasser:1983yg,Gasser:1984gg}. 
Subsequently construction of these operators has been extended to order $\mathcal{O}(p^6)$, or NNLO, as detailed in Refs.~\cite{Fearing:1994ga,Bijnens:1999hw,Bijnens:1999sh,Ebertshauser:2001nj,Bijnens:2001bb}.
Recently, operators of order $\mathcal{O}(p^8)$ have received a lot of interest since the advancement of calculations for the $\mathcal{O}(p^2)$ LECs have been developed to two loops~\cite{Bijnens:2017wba}, which need $\mathcal{O}(p^8)$ operators as the counter-terms.
The $CP$-even operator constructions at the $\mathcal{O}(p^8)$ have been presented in Refs.~\cite{Bijnens:2018lez,Hermansson-Truedsson:2020rtj,Bijnens:2023hyv}.
However, the mesonic operators with other $C$ or $P$ eigenvalues are also important. For example, the decay of $\eta$ and $\eta'$ mesons is related to the $CP$ properties of the operators apart from the baryon interactions, for example, see Ref.~\cite{Akdag:2022sbn}. Different $CP$ meson operators are also indispensable in the calculation of the non-vanishing electric dipole moments (EDM), which in the foreseeable future would be expected as evidence of the
P and T violation beyond the CP phase of the standard model (SM)~\cite{Engel:2013lsa,Bsaisou:2014oka,deVries:2011an}.

Constructing higher order mesonic operators is quite a challenging task since many redundant relations make operators in one type not independent and it is prone-to-error to eliminate these redundancies. There are three major difficulties in finding complete and independent operators. 
Firstly, when the number of derivatives in the operators increases, although the internal group remains unchanged, there are correspondingly more ways of Lorentz indices contraction available, which are related by the relations such as the equations of motion (EOMs) of the fields, the integration-by-part (IBP), the Fierz identities, and so on. Besides, the operators containing external sources should not take EOM condition, which makes the situation even worse: some fields needs EOM but others are not. 
Secondly, for the internal flavor structure, considering the trace basis when the number of fields increases, there are multiple Cayley-Hamilton relations to consider. 
Thirdly, the presence of repeated fields introduces further redundancy, which in turn increases the difficulty in constructing the operators in the traditional way.

To overcome all the above problems, we utilize the on-shell operator amplitude correspondence and the Young tensor technique~\cite{Li:2020gnx,Li:2020xlh,Li:2022tec}. This method can conveniently eliminate the redundancies caused by multiple derivatives, multiple fields, and repeated fields, and has already been used in the standard model effective field theory  (SMEFT)~\cite{Li:2020gnx,Li:2020xlh,Li:2022tec,Ren:2022tvi}, the low energy effective field theory (LEFT)~\cite{Li:2020tsi}, the Higgs effective field Theory (HEFT)~\cite{Sun:2022aag, Sun:2022snw,Sun:2022ssa}, the gravity effective field theory~\cite{Li:2023wdz}, and other effective theorys (EFTs) involving new light fields~\cite{Li:2021tsq, Song:2023jqm,Song:2023lxf}.
This technique has also been applied to theory with spontaneous symmetry breaking: the electroweak chiral Lagrangian in Ref.~\cite{Sun:2022snw,Sun:2022ssa} and the ChPT in Refs.~\cite{Low:2022iim}. Without including the external sources, Refs.~\cite{Low:2022iim} writes down the purely mesonic operators at the $\mathcal{O}(p^6)$ and $\mathcal{O}(p^8)$, agreed with results presented in Refs.~\cite{Bijnens:1999sh,Bijnens:2001bb,Bijnens:2018lez,Bijnens:2023hyv}. At the same time, when there is no external sources, the on-shell operator basis of the pure mesonic system, in terms of
the kinematic invariants, was also constructed based on the on-shell soft theorem~\cite{Cheung:2014dqa,Cheung:2015ota} and the Adler zero condition~\cite{Low:2019ynd,Dai:2020cpk,Kampf:2021jvf}.
However, when the external sources are included in the building blocks, things become complicated because the external sources are off-shell object, and the on-shell amplitudes are not suitable.

In this paper, following the treatment in Ref.~\cite{Ren:2022tvi}, we extend the Young tensor technique to take the off-shell external sources into account, and furthermore, identify different $CP$ eigen operators using the trace basis with the Caylay-Hamilton relations. Thus we can obtain the complete and independent operators with external sources up to $\mathcal{O}(p^8)$ order, which are divided into four parts, $C$+$P$+, $C$+$P$-, $C$-$P$+, and $C$-$P$- according to their $C,P$ properties, and thus surpass the results presented in Refs.~\cite{Bijnens:2018lez,Bijnens:2023hyv}. 
The procedures utilized in this paper are implemented in the Mathematica package ABC4EFT \footnote{https://abc4eft.hepforge.org/}~\cite{Li:2020gnx,Li:2020xlh,Li:2022tec}.
Besides, the Hilbert series~\cite{Lehman:2015via,Henning:2015alf,Henning:2015daa,Marinissen:2020jmb,Graf:2020yxt,Bijnens:2022zqo} provides a complementary check for the construction of effective operators since it counts the operator numbers of each type. The counting result of $SU(3)$ case has been presented in Ref.~\cite{Graf:2020yxt}, and in this paper, we present the counting results of the $SU(2)$ case, which has been used as a crosscheck.
Including $p^2$, $p^4$, $p^6$, $p^8$, there are 2008 $SU(2)$-operators and 7223 $SU(3)$-operators, of which 
640 are $C$+$P$+, 532 are $C$+$P$-, 399 are $C$-$P$+, and 437 are $C$-$P$- in the $SU(2)$ case, 
while 2090 are $C$+$P$+, 1906 are $C$+$P$-, 1570 are $C$-$P$+, and 1657 are $C$-$P$- in the $SU(3)$ case. More details can be found in Sec.~\ref{sec:3_6_HS_OC}.



The paper is organized as follows.
Sec.~\ref{sec:2_YTO} provides an introduction to the Young tensor technique, beginning with a review of  the basics of the ChPT in Sec.~\ref{sec:2_1_ChPT}. 
In Sec.~\ref{sec:2_2_Ytt}, we discuss the amplitude-operator correspondence, employing the Young tensor technique, alongside additional considerations including the Adler zero condition and 
the off-shell structure of the external sources. Using examples of operators including the $SU(3) \, D f_{-} u \Sigma{+}^2$ with 4 fields and the $SU(2) \, D u^3 \Sigma_{+}^2 $ with 5 fields, we demonstrate the application of the Young tensor technique to construct corresponding bases.
In Sec.~\ref{sec:3_OB} based on the bases introduced in Sec.~\ref{sec:2_YTO}, we present technical details for constructing operators with different CP eigenvalues and provide the number of operators as determined by the Hilbert Series. 
We convert the invariant bases of the SU(3) 4-field and 5-field operators into the trace bases to construct operators with different CP eigenvalues.
Sec.~\ref{sec:3_6_HS_OC} presents the number of operators determined by Young tensor with cross-check from the Hilbert Series.
Conclusions are presented in Sec.~\ref{sec:4_Conclusions}.
In Appendix \ref{app:A1_Cayley}, we present the Cayley-Hamilton Theorem. In Appendix \ref{app:p4}, \ref{app:p6}, \ref{app:p8}, we detail the chiral dimension 4, 6, 8 purely mesonic operators for the $SU(2)$ and the $SU(3)$ cases. 

%

\section{Young Tensor Operator} 
\label{sec:2_YTO}
\subsection{Brief review of chiral perturbation theory}
\label{sec:2_1_ChPT}
In the low-energy QCD theory, the Lagrangian can be written as:
\begin{equation}
    \mathcal{L}_0 = \bar{q_L} i \slashed{D} q_L + \bar{q_R} i \slashed{D} q_R -\frac{1}{4} G_{\mu\nu}^{a} G^{a\mu\nu},
\end{equation}
where $ D_{\mu} =\partial_{\mu} + i g G^{a}_{\mu} t^{a} $, $q_{L/R}$ is the light quark field, $G$ is the gluon field.
It has chiral symmetry $SU(N_f)_L \times SU(N_f)_R$, where $N_f$ is the number of light quarks. 

At low energies, the spontaneous symmetry breaking occurs from the $SU(N_f)_L \times SU(N_f)_R$ to the $SU(N_f)_V$ subgroup due to the quark condensate.
Then the strong interaction could be described by the ChPT~\cite{Weinberg:1968de,Weinberg:1978kz,Gasser:1983yg,Gasser:1984gg,Gasser:1987rb}. In the ChPT, the light boson fields can be treated as the Goldstones $\phi$, and $\phi$ can be written as a unitary matrix $u$:
\begin{equation}
    u(x) = \exp\left( \sum_a \frac{\phi(x)^a \lambda^a}{2F_{0}} \right) \, ,
\end{equation}
which contains 8 massless Goldstone bosons when $N_f=3$, defined as the pseudoscalar meson octets
\begin{equation}
\begin{aligned}
    & \sum_{a=1}^8 \phi_a (x) \lambda_a 
    &= \begin{pmatrix}
    \pi^0+\frac{1}{\sqrt{3}}\eta &\sqrt{2}\pi^+&\sqrt{2}K^+\\
    \sqrt{2}\pi^-&-\pi^0+\frac{1}{\sqrt{3}}\eta&\sqrt{2}K^0\\
    \sqrt{2}K^- &\sqrt{2}\bar{K}^0&-\frac{2}{\sqrt{3}}\eta
    \end{pmatrix}\, ,
\end{aligned}
\end{equation}
where $\lambda^a$ are the Gell-Mann matrices. For the SU(2) case the matrix is
\begin{equation}
\begin{aligned}
    & \sum_{I=1}^3 \phi_I (x) \lambda_I 
    &= \begin{pmatrix}
    \pi^0 &\sqrt{2}\pi^+\\
    \sqrt{2}\pi^-&-\pi^0
    \end{pmatrix}\, .
\end{aligned}
\end{equation}

The transforming property of the Goldstone field $u(x)$ can be written as:
\begin{equation}
    u(x) \rightarrow u(x') = g_R u(x) h(g,x)^{-1} = h(g,x) u(x) g_{L}^{-1},
\end{equation}
where $g = (g_L,g_R) \in SU(3)_L \times SU(3)_R$ and $h(g,u)\in SU(3)_V$.

Upon introducing the external sources \cite{Gasser:1983yg,Gasser:1984gg} with traceless vector sources $v_{\mu}$, $a_{\mu}$ and scalar sources $s$, $p$, the Lagrangian becomes 
\begin{equation}
    \mathcal{L} = \mathcal{L}^0_{QCD}+
    \overline q\gamma^\mu\left(v_\mu+a_\mu\gamma_5\right)-\overline q \left(s-ip\gamma_5\right)q \, ,
\end{equation}
where $q = (q_L \,, q_R)^T $ and the sources undergo redefinitions:
\begin{equation}
\label{eq: sources}
\begin{aligned}
    \Sigma &= 2B(s + ip), 
    \\
    F_{L}^{\mu \nu} &= \partial ^{\mu}\ell ^{\nu}-\partial ^{\nu}\ell ^{\mu} -i\left[ \ell ^{\mu},\ell ^{\nu} \right] \, ,
    \\
    F_{R}^{\mu \nu} &= \partial ^{\mu}r ^{\nu}-\partial ^{\nu}r ^{\mu} -i\left[ r ^{\mu},r ^{\nu} \right] \, ,
\end{aligned}
\end{equation}
where $r_\mu = v_\mu + a_\mu$, $\ell_\mu = v_\mu - a_\mu$. And since $\ell_\mu$ and $r_\mu$ are traceless, the field strength tensors $F_{L}^{\mu \nu}$ and $F_{R}^{\mu \nu}$ are also traceless.


Integrating out the quark degrees of freedom, one can choose the basic building blocks as
\begin{equation} \label{eq: u-Sigma-f}
\begin{aligned}
    u_{\mu}&\, = i\Big[ u^{\dagger} (\partial _{\mu}-ir_{\mu})u-u(\partial _{\mu}-i\ell _{\mu} )u^{\dagger}\Big]
    \, ,
    \\
    \Sigma _{\pm}&\, = u^{\dagger}\Sigma u^{\dagger}\pm u\Sigma ^{\dagger} u
    \, ,
    \\
    f_{\pm}^{\mu \nu}&\, = u F_{L}^{\mu \nu }u^{\dagger} \pm u^{\dagger}F_{R}^{\mu \nu}u
    \, ,
    \\
    \tilde{f}_{\pm}^{\mu \nu} &\, = \epsilon^{\mu\nu\rho\lambda} f_{\pm \rho\lambda} \, .
\end{aligned}
\end{equation}
Under chiral symmetry, the building blocks are transformed as
\begin{equation} 
    X \to h(g,x) X h(g,x)^{-1} \, , h(g,x) \in SU(3)_V \,,
\end{equation} 
Additionally, we define the covariant derivative as:
\begin{equation}
    D_{\mu}  =  \partial_{\mu}+\Gamma_{\mu}~, \qquad
    \Gamma_{\mu} =  \frac{1}{2}\{u^{\dagger}(\partial_{\mu}-ir_{\mu
    })u+u(\partial_{\mu}-i\ell_{\mu})u^{\dagger}\}~.
\end{equation}

To construct chiral effective operators we also need to know the properties of building blocks under the P and C transformation, 
which is presented in Table.~\ref{Table:BuildingBlock-Transformations}.

\begin{table}[h]
\begin{center}
\begin{tabular}{c|cccc}
\hline
Fields                        & $SO(1,3)$                  & C           & P & Chiral Dim\\\hline
$u_\mu$                     & (1/2,1/2)  & $+u_\mu^T$                    & $-u_\mu$              & 1  \\
$f_{\pm\mu\nu}$            & (1,0)$\oplus$(0,1)       & $\mp f_{\pm\mu\nu}^T$         & $\pm f_{\pm\mu\nu}$            & 2 \\
$\tilde{f}_{\pm \rho\lambda}$& (1,0)$\oplus$(0,1) & $\mp \tilde{f}_{\pm \rho\lambda}^T$ &$\mp \tilde{f}_{\pm \rho\lambda}$ & 2\\
$\Sigma_\pm$               & (0,0)                      & $+\Sigma_\pm^T$               & $\pm \Sigma_\pm$            & 2 \\
$\langle\Sigma_\pm\rangle$ & (0,0)                      & $+\langle\Sigma_\pm\rangle$   & $\pm \langle\Sigma_\pm\rangle$            & 2 \\
\hline
\end{tabular} 
\caption{Here we present the Lorentz group representation, $CP$ and chiral dimension of building blocks in chiral Largrangian. We define $\tilde{f}_{\pm \mu \nu}=\epsilon_{\mu\nu\rho\lambda} f_{\pm}^{\rho\lambda}$. }
\label{Table:BuildingBlock-Transformations}
\end{center}
\end{table}
Actually, the effective Lagrangian of the ChPT is organized in terms of power of the mesonic momentum transfer $p$, which defines the power-counting scheme. Thus the field $u_\mu$ and the covariant derivative $D_\mu$ are of $\mathcal{O}(p)$, and the external sources $f_\pm$ are of $\mathcal{O}(p^2)$ because of their definition in Eq.~\eqref{eq: sources}. Besides the scalar sources $\Sigma_\pm$ are also of $\mathcal{O}(p^2)$ since they contribute to the meson masses, which are of $\mathcal{O}(p^2)$ due to:
\begin{equation}
    p^2 \sim m^2 
\end{equation}
The power of momentum $p$ is called chiral dimension, which is presented in Table.~\ref{Table:BuildingBlock-Transformations} as well. 

For the simplest case, consider a dimension-2 pure meson effective operator. To construct an operator that is invariant under chiral symmetry and Lorentz transformation, the indices must be fully contracted. According to chiral dimensions, a dimension-2 operator can only be composed of two $u_{\mu}$ or a single $\langle \Sigma_{\pm} \rangle$. Consequently, we can express the LO effective operators as follows:
\begin{equation}
     \langle u_{\mu} u^{\mu} \rangle \,,\quad  \langle \Sigma_{\pm} \rangle .
\end{equation}
Correspondingly, we can classify it under the $CP$ transformation \footnote{for detailed properties of the $CP$ transformation, see Sec.~\ref{sec:3_6_HS_OC}.}. Among these, the $C$+$P$+ operators are:
\begin{equation}
    \langle u_{\mu} u^{\mu} \rangle\,, \quad  \langle \Sigma_{+} \rangle.
\end{equation}
while the $C$+$P$- operator is:
\begin{equation}
    \langle \Sigma_{-} \rangle .
\end{equation}
Based on the chiral dimension and the $CP$ properties, we can assign different dimensions and the $CP$ eigenvalues to the corresponding operators.
However, if we want to construct complete and linearly independent effective operators, we need to consider the EOMs, the Fierz identities, and the IBP, which will introduce redundancies. Indeed, these redundancies can be conveniently handled in the spinor helicity formalism. Below, we will discuss the relationship between fields and the spinor helicity formalism, as well as how to use the Young tensor technique to eliminate redundancies.

\subsection{Young tensor technique}
\label{sec:2_2_Ytt}
Below, we will introduce the representation of operators using the spinor helicity formalism and express the operators in the irreducible singlet representations using the Young tensor technique, which eliminates redundancies in the operators~\cite{Li:2020gnx, Li:2020xlh, Li:2022tec}.

\subsubsection{On-shell amplitude and Young tensor technique}
\label{sec:2.2.1_Ytt }

%
To express operators using the spinor helicity formalism, we need to find their correspondence to on-shell amplitudes. We know that amplitudes consist of momenta and spinor wavefunctions, where the momenta are represented using the spinor helicity formalism
\begin{equation}
    p_i^\mu = \lambda_i^\alpha \sigma^\mu_{\alpha\dot\alpha} \tilde\lambda_i^{\dot\alpha} \,.
\end{equation}
Therefore amplitudes can always be expressed as functions of spinors, 
and the spinor indices must be contracted under the Lorentz invariance requirement:
\begin{equation} \label{eq SpinorHelicity-Contraction}
    \lambda_i^{\alpha} \lambda_{j\alpha} := \langle ij \rangle\, , \qquad 
    \tilde\lambda_{i\dot\alpha} \tilde\lambda_j^{\dot\alpha} :=[ij]\,.
\end{equation}
In general, the amplitude basis $\mathcal{O}$ can be presented by the little group of internal particle and decomposed into parts belonging to the internal group $T$ and the Lorentz group $M$, which also serves as the kinematic factor. The kinematic factor $M$ depends only on helicities $h$ of external praticles, which allows the amplitude to be expressed in terms of the Weyl spinors:
\begin{equation}
\begin{aligned} \label{eq:Amplitude_TM}
    \mathcal{O}\left( \phi_1^{a_1}(p_1),\dots,\phi_N^{a_N}(p_N) \right) &= T^{a_1,\dots,a_N} M(h_1,\dots,h_N) \, ,\\ M(h_1,\dots,h_N) &\sim \prod_{i=1}^{N}\lambda_i^{r_i-h_i}\tilde\lambda_i^{r_i+h_i}  \, , 
\end{aligned}
\end{equation}
for particle $\phi_i$ with momenta $p_i$, group indices $a_i$, mass dimension $r_i$, helicity $h_i$ and $i=1,\dots,N$. The total dimension of amplitude $ [\mathcal{O}] \equiv r = \sum_i r_i$, with total helicity $h=\sum_{i} h_i$. The kinematic factor $M$ should be 
consist of $n=\frac{r-h}{2}$ $\langle\cdot\rangle$ and $\tilde{n}=\frac{r+h}{2}$$[\cdot]$ from Eq.~\eqref{eq SpinorHelicity-Contraction}.

Hence, the gauge boson fields, the fermion fields and the scalar fields can also be represented using the spinor helicities formalism
\begin{eqnarray} \label{eq:dictionary-WeylSpinor}
    \lambda_i^{r_i\pm1}\tilde{\lambda}_i^{r_i\mp1} & \Leftrightarrow & D^{r_i-1}F_{{\rm L/R}\,i}\,, \nonumber \\
    \lambda_i^{r_i\pm1/2} \tilde{\lambda}_i^{r_i\mp1/2} & \Leftrightarrow & D^{r_i-1/2}\psi^{(\dagger)}_i \,,\nonumber \\
    \lambda_i^{r_i}\tilde{\lambda}_i^{r_i} & \Leftrightarrow & D^{r_i}\phi_i \,.
\end{eqnarray}

Since this paper focuses on constructing pure mesonic operators, mainly considering scalar fields and gauge boson fields, the Lorentz group representations of the corresponding fields can be referred to in Table.~\ref{Table:BuildingBlock-Transformations}.

Due to the relationships between operators such as the EOMs\footnote{The external sources possess no EOM and we will discuss this later.}, the Fierz transformations, and the IBP, complicated calculations are required to eliminate redundancies when writing operators. 
However, enabled by the operator amplitude correspondence, we can efficiently eliminate redundancies using the spinor helicity formalism:
\begin{equation}
\label{eq: relations}
\begin{aligned}
    &\text{Equations of Motion} : &\langle ii \rangle = [jj] = 0 \,, \\
    &\text{Schouten Identity} : &|i\rangle \langle jk \rangle + |j\rangle \langle ki \rangle + |k\rangle \langle ij \rangle = 0 \,, \\
    &&|i\rangle [jk] + |j\rangle [ki] + |k\rangle [ij] = 0 \,, \\
    &\text{Integration by Parts}:&\sum_i \langle ji \rangle [ik] = 0 \, .
\end{aligned}
\end{equation}

Referring to the previously mentioned Eq.~\eqref{eq:Amplitude_TM}, the basis of the amplitude can be expressed as the kinematic factor $M$ and the part related to the internal group $T$. We can reduce these two parts separately using the Young tensor technique.

For the kinematic factor $M$, representing the Lorentz structure, momentum on-shell conditions automatically eliminate redundancies due to the EOM. The key concerns are the momentum conservation and the Schouten identity. We will use the $SU(N)$ semi-standard Young tableau (SSYT)~\cite{Henning:2019enq,Li:2020gnx,Li:2020xlh,Li:2022tec}, to present all independent Lorentz structures with redundancies removed. For $N$ fields composed of $n$ $\langle\cdot\rangle$ and $\tilde{n}$ $[\cdot]$ in the spinor helicity formalism, the SSYT has the following form:
\begin{eqnarray}\label{eq:YD_shape}
\arraycolsep=0pt\def\arraystretch{1}
\rotatebox[]{90}{\text{$N-2$}}
	\left\{
	\begin{array}{cccccc}
		\yng(1,1) &\ \ldots{}&\ \yng(1,1)& \overmat{n}{\yng(1,1)&\ \ldots{}\  &\yng(1,1)} \\
		\vdotswithin{}& & \vdotswithin{}&&&\\
		\undermat{\tilde{n}}{\yng(1,1)\ &\ldots{}&\ \yng(1,1)} &&&
	\end{array}
	\right.\,,
	\\
	\nonumber 
\end{eqnarray}

The corresponding numbers filled in the Young tableau are given by
\begin{equation} \label{eq Numbers-i}
    \#i = \tilde{n} -2h_i \quad , \qquad i = 1,2,\dots,N \,.
\end{equation}
Thus, we can obtain the corresponding Lorentz structures by constructing all the SSYTs. 

Furthermore, the relationship between the spinor helicity formalism and the Young diagram can be interpreted by rules below:
\begin{equation}\label{eq: YT_translate}
    \begin{array}{c} 
    \young({{k_1}},{{k_2}}) \\ \vdots \\ \young({{k_{N-3}}},{{k_{N-2}}}) 
    \end{array} \sim
    \epsilon^{k_1\dots k_{N-2}ij}[ij]\,, \qquad \young(i,j) \sim \left\langle {ij}  \right\rangle \,,
\end{equation}
where $\epsilon$ is the Levi-Civita tensor.

For the internal group, the group tensor $T$ is obtained from the tensor product decomposition, which can be done using Young diagram outer products. Actually, the mesonic operators of ChPT are composed of building blocks of the adjoint representation of the $SU(2)$ or the $SU(3)$. For example, considering the type $u^4$, there are 4 fields of the adjoint representation and the outer products of the $SU(2)$ and the $SU(3)$ cases are
\begin{align}
\ytableausetup{boxsize=1.2em}
    SU(2): & \notag \\
    & \ytableaushort{{i_1} {j_1},} \xrightarrow{\ytableaushort{{i_2}{j_2}}} \ytableaushort{{i_1} {j_1} {i_2} {j_2}} + \ytableaushort{{i_1} {j_1} {i_2}, {j_2}} + \ytableaushort{{i_1} {j_1}, {i_2} {j_2}} \notag \\
    & \xrightarrow{\ytableaushort{{i_3} {j_3}}} \ytableaushort{{i_1} {j_1} {i_2} {j_2},{i_3} {j_3}} + \ytableaushort{{i_1} {j_1} {i_2} {i_3}, {j_2} {j_3}} + \ytableaushort{{i_1} {j_1} {i_3} {j_3}, {i_2} {j_2}} \notag \\ 
    & \xrightarrow{\ytableaushort{{i_4} {j_4}}} \ytableaushort{{i_1} {j_1} {i_2} {j_2},{i_3} {j_3} {i_4} {j_4}} + \ytableaushort{{i_1} {j_1} {i_2} {i_3}, {j_2} {j_3} {i_4} {j_4}} + \ytableaushort{{i_1} {j_1} {i_3} {j_3}, {i_2} {j_2} {i_4} {j_4}} \notag \\ 
    &= \epsilon^{i_1i_3}\epsilon^{j_1j_3}\epsilon^{i_2 i_4}\epsilon^{j_2j_4} + \epsilon^{i_1 j_2}\epsilon^{j_1j_3}\epsilon^{i_2 i_4}\epsilon^{i_3j_4} + \epsilon^{i_1i_2}\epsilon^{j_1j_2}\epsilon^{i_3 i_4}\epsilon^{j_3j_4} \,,
\end{align}

\begin{align} \label{eq: SU3_48outer_Product}
    SU(3): & \notag \\
    & \ytableaushort{{i_1} {j_1}, {k_1}} \xrightarrow{\ytableaushort{{i_2} {j_2}, {k_2}}} \ytableaushort{{i_1} {j_1} {i_2} {j_2}, {k_1} {k_2}} + \ytableaushort{{i_1} {j_1} {i_2} {j_2}, {k_1}, {k_2}} + \ytableaushort{{i_1} {j_1} {i_2}, {k_1} {j_2}{k_2}} + \ytableaushort{{i_1} {j_1} {i_2}, {k_1} {j_2}, {k_2}} +  + \ytableaushort{{i_1} {j_1} {i_2}, {k_1} {k_2}, {j_2}} + \ytableaushort{{i_1} {j_1}, {k_1} {i_2}, {j_2} {k_2}} \notag \\
    & \xrightarrow{\ytableaushort{{i_3} {j_3}, {k_3}}} \ytableaushort{{i_1} {j_1} {i_2} {j_2}, {k_1} {k_2} {i_3}, {j_3} {k_3}} + \ytableaushort{{i_1} {j_1} {i_2} {j_2}, {k_1} {i_3} {j_3}, {k_2} {k_3}} + \ytableaushort{{i_1} {j_1} {i_2} {i_3}, {1_2} {j_2}{k_2}, {j_3} {k_3}} + \ytableaushort{{i_1} {j_1} {i_2} {i_3}, {k_1} {j_2} {j_3}, {k_2} {k_3}} + \ytableaushort{{i_1} {j_1} {i_2} {i_3}, {k_1} {j_2} {k_3}, {k_2} {j_3}} \notag \\
    & + \ytableaushort{{i_1} {j_1} {i_2} {i_3}, {k_1} {k_2} {j_3}, {j_2} {k_3}} + \ytableaushort{{i_1} {j_1} {i_2} {i_3}, {k_1} {k_2} {k_3}, {j_2} {j_3}} + \ytableaushort{{i_1} {j_1} {i_3} {j_3}, {k_1} {i_2} {k_3}, {j_2} {k_2}} \notag \\
    & \xrightarrow{\ytableaushort{{i_4} {j_4}, {k_4}}} \ytableaushort{{i_1} {j_1} {i_2} {j_2}, {k_1} {k_2} {i_3} {i_4}, {j_3} {k_3} {j_4} {k_4}} + \ytableaushort{{i_1} {j_1} {i_2} {j_2}, {k_1} {i_3} {j_3} {i_4}, {k_2} {k_3} {j_4} {k_4}} + \ytableaushort{{i_1} {j_1} {i_2} {i_3}, {k_1} {j_2} {k_2} {i_4}, {j_3} {k_3} {j_4} {k_4}} + \ytableaushort{{i_1} {j_1} {i_2} {i_3}, {k_1} {j_2} {j_3} {i_4}, {k_2} {k_3} {j_4} {k_4}} + \ytableaushort{{i_1} {j_1} {i_2} {i_3}, {k_1} {j_2} {k_3} {i_4}, {k_2} {j_3} {j_4} {k_4}} \notag \\
    & + \ytableaushort{{i_1} {j_1} {i_2} {i_3}, {k_1} {k_2} {j_3} {i_4}, {j_2} {k_3} {j_4} {k_4}} + \ytableaushort{{i_1} {j_1} {i_2} {i_3}, {k_1} {k_2} {k_3} {i_4}, {j_2} {j_3} {j_4} {k_4}} + \ytableaushort{{i_1} {j_1} {i_3} {j_3}, {k_1} {i_2} {k_3} {i_4}, {j_2} {k_2} {j_4} {k_4}} \notag \\
    &= \epsilon^{i_1k_1j_3}\epsilon^{j_1k_2k_3}\epsilon^{i_2i_3j_4}\epsilon^{j_2i_4k_4} + \epsilon^{i_1k_1k_2}\epsilon^{j_1i_3k_3}\epsilon^{i_2j_3j_4}\epsilon^{j_2i_4k_4} + \epsilon^{i_1k_1j_3}\epsilon^{j_1j_2k_3}\epsilon^{i_2k_2j_4}\epsilon^{i_3i_4j_4} \notag \\
    & + \epsilon^{i_1k_1k_2}\epsilon^{j_1j_2k_3}\epsilon^{i_2j_3j_4}\epsilon^{i_3i_4k_4} + \epsilon^{i_1k_1k_2}\epsilon^{j_1j_2j_3}\epsilon^{i_2k_3j_4}\epsilon^{i_3i_4k_4} + \epsilon^{i_1k_1j_2}\epsilon^{j_1k_2k_3}\epsilon^{i_2j_3j_4}\epsilon^{i_3i_4k_4} \notag \\
    & +  \epsilon^{i_1k_1j_2}\epsilon^{j_1k_2j_3}\epsilon^{i_2k_3j_4}\epsilon^{i_3i_4k_4} + \epsilon^{i_1k_1j_2}\epsilon^{j_1i_2k_2}\epsilon^{i_3k_3j_4}\epsilon^{j_3i_4j_4}\,,
\end{align}
of which only the singlet Young tableaux are shown. 

The structures $M$ and $T$ obtained above are referred to as Lorentz y-basis. So we can get the effective operators in y-basis,
\begin{equation} \label{eq O-TG}
    \mathcal{O}^{y} = M \otimes T \,,
\end{equation}
as the product of the Lorentz and internal structures.

However, the above consideration do not account for cases where there are repeated fields among the $N$ fields. 
If repeated fields are present, the corresponding operators should be transformed into the p-basis. This transformation is achieved by applying the idempotent elements $\mathcal{Y}_{}$, which are derived from the permutation symmetries of the Lorentz and the internal structures, respectively:
\begin{equation}
    \mathcal{Y}_{}^{\mathcal{O}} = \mathcal{Y}_{}^{M} \otimes \mathcal{Y}_{}^{T} \,, \qquad \mathcal{O}_{}^{p} = \mathcal{Y}_{}^{\mathcal{O}} \mathcal{O}^{y}\, .
\end{equation}
referring to Sec. \ref{sec:2_3_Ytt_example} and Refs.~\cite{Fonseca:2019yya,Li:2020xlh,Li:2022tec} for detailed information.

Also we can initially consider the permutation symmetry for both the kinematic factor $M$ and the internal part $T$ separately. For the kinematic factor $M$ and the internal part $T$, we multiply by the idempotent elements,
\begin{equation}
    \mathcal{Y}^{M}_{} \cdot M = M^{p} \,, \qquad
    \mathcal{Y}^{T}_{} \cdot T = T^{p}\,.
\end{equation}
Therefore, the tensor product of the kinematic factor part $M^{p}$ with the internal part $T^{p}$ yields the general form of the effective operator in p basis,
\begin{equation} \label{eq O-TG_2}
    \mathcal{O}^{p} = M^{p} \otimes T^{p} \,.
\end{equation}

Furthermore, the ChPT is special and needs extra treatments since
\begin{itemize}
    \item The amplitude containing Goldstone bosons should satisfy the Adler zero condition,
    \item The external sources are off-shell and should not take their EOM into account.
\end{itemize}
Subsequently, we will discuss how to deal with such properties of the ChPT.

\subsubsection{Adler zero condition}
The amplitudes containing Goldstones satisfy the Adler zero condition \cite{Adler:1969gk}, which states that the amplitude vanishes when a Goldstone momentum $p$ becomes soft,
\begin{equation}
    \lim_{p\rightarrow 0}\mathcal{M}(p) = 0\,.
\end{equation}
The Adler zero condition constrains the Lorentz structures further, and these constraints can be expressed by a system of linear equations \cite{Sun:2022snw,Sun:2022ssa,Low:2022iim}.

Considering an operator of $N$ fields, the Lorentz basis $\{\mathcal{B}_i|i=1,2,\dots,d_N\}$ is of dimension-$d_N$. A general operator satisfies the Adler zero condition means that 
\begin{equation}
    \mathcal{M} = \sum_{i=1}^{d_N}c_i \mathcal{B}_i\,,
\end{equation}
 and 
\begin{equation}
    \lim_{p\rightarrow 0} \mathcal{M}(p) = \sum_{i=1}^{d_N}c_i \left(\lim_{p\rightarrow 0}\mathcal{B}_i\right) = \sum_{i=1}^{d_N}c_i \tilde{\mathcal{B}}_i = 0\,,
\end{equation}
where $\tilde{\mathcal{B}}_i = \lim_{p\rightarrow 0}\mathcal{B}_i$, which is generally not independent and can be decomposed in terms of the basis $\mathcal{B}_i$,
\begin{equation}
    \tilde{\mathcal{B}}_i = \sum_{j=1}^{d_N} K_{ij}\mathcal{B}_j\,.
\end{equation}
Thus the Adler zero condition becomes
\begin{equation}
    \sum_{i,j=1}^{d_N} c_i K_{ij} \mathcal{B}_j = 0\,,
\end{equation}
interchanging the summation order we can get
\begin{equation}
    \sum_{j=1}^{d_N} \left(\sum_{i=1}^{d_N} c_i K_{ij}\right)\mathcal{B}_j = 0\,,
\end{equation}
so we get a system of linear equations about the coefficients $c_i$,
\begin{equation}
    \sum_{i=1}^{d_N}c_i K_{ij} = 0\,, \quad j=1,2,\dots\,, d_N\,.
\end{equation}
The rank $r(K)$ of the coefficient matrix $K$ is the independent constraints from the Adler zero condition, and the remaining Lorentz structures are of number $d_N-r(k)$, whose explicit forms are presented by the solutions of this system of linear equations.




\subsubsection{Off-shell structure for external source} 
\label{sec:Off_Shell_Structures}

The external sources $\Sigma_\pm$ and $f_\pm$ are not dynamic degrees of freedom of the ChPT thus are of no EOM, which in the Young tensor method means the amplitudes containing them are off-shell,
\begin{align}
    D^2 \Sigma_\pm{}_i \neq 0 & \rightarrow \langle  ii\rangle [ii] \neq 0\,, \label{eq: offshell1}\\
    D_\mu f_{\pm L}^{\mu\nu}{}_{i} \neq 0 & \rightarrow \langle ii\rangle |i] \neq 0 \,,  \label{eq: offshell2} \\
    D_\mu f_{\pm R}^{\mu\nu}{}_{i} \neq 0 & \rightarrow [ ii] |i\rangle \neq 0 \label{eq: offshell3}\,.
\end{align}
In particular, the EOMs of the right-handed and the left-handed field-strength tensors are not independent, due to the Bianchi identity,
\begin{equation} \label{eq: Bian_FL_FR}
    D_\mu \tilde{f}_\pm^{\mu\nu} = 0\,,
\end{equation}
which implies
\begin{equation}
    D_\mu f_{\pm L}^{\mu\nu} - D_\mu f_{\pm R}^{\mu\nu} = 0\,.
\end{equation}

Thus to recover the EOMs of the external sources we need to relax the amplitude structures to make the conditions from Eq.~\eqref{eq: offshell1} to Eq.~\eqref{eq: offshell2} valid\footnote{The condition in Eq.~\eqref{eq: offshell3} is not included due to the Bianchi identity.}. This is done by distinguishing the field indices and the momentum indices in the amplitude in Ref.~\cite{Ren:2022tvi}. The key point there is to construct all structures after the indices are distinguished and then to reduce them to the independent ones via the relations in Eq.~\eqref{eq: relations}. More details can be found in Ref.~\cite{Ren:2022tvi} and we will use examples to illustrate this procedure further next.

\subsection{Chiral dim-8 operator examples}
\label{sec:2_3_Ytt_example}
Below, we illustrate how the Young tensor technique is applied to the construction of pure meson operators using the chiral dime-8 examples: the $SU(3) \, D f_{-} u \Sigma_{+}^2$ with 4 fields and the $SU(3) \, D u^3 \Sigma_{+}^2 $ with 5 fields.

\subsubsection{Operator with 4 fields}

First we consider the $SU(3) \, D f_{-} u \Sigma_{+}^2$ with 4 fields. Because $f_{-}$ is consisted of $F_{-L}$ and $F_{-R}$ in Eq.~\eqref{eq: u-Sigma-f}, we need to calculate this two part respectively.

For the Lorentz part, we compute the Semi-Standard Young Tableaux(SSYT). In Table.~\ref{Table:p8-4fields-SU3} we give out some basic information about the operators according to the Eq.~\eqref{eq:dictionary-WeylSpinor}. And we assigned identifiers to elements $F_{L/R}$ , $u$, $\Sigma_{+}$, and $\Sigma_{+}$, respectively, as $F_{-L/R}$, $\phi_1$, $\phi_2$, and $\phi_3$.

Attention should be drawn to the fact that $u$ is in fact $u_{\mu}$ in Eq.~\eqref{eq: u-Sigma-f}. During the classification of operators, we have adopted the convention of abbreviating $u_{\mu}$ to $u$ for the sake of simplicity. However, when it becomes necessary to perform calculations, the component $D_{\mu}$ within the $u_{\mu} \sim D_{\mu} u$ form must be explicitly stated.

\begin{table}[ht]
\centering
\begin{tabular}{@{}ccccccccc@{}}
\toprule
$p^8$ 4-fields $SU(3)$                          & $N-2$ & $n \langle \cdot \rangle$ & $\tilde{n}[\cdot]$ & helicity     & $F_{1L/R}$ & $\phi_{2}$ & $\phi_{3}$   & $\phi_{4}$   \\ \midrule
$D_{\mu} D_{\nu} F_{-L}^{\mu\nu} u \Sigma_{+}^2$ & $2$   & $2$                       & $1$                & $(-1,0,0,0)$ & $F_{-L}$     & u          & $\Sigma_{+}$ & $\Sigma_{+}$ \\
$D_{\mu} D_{\nu} F_{-R}^{\mu\nu} u \Sigma_{+}^2$ & $2$   & $1$                       & $2$                & $(+1,0,0,0)$ & $F_{-R}$     & u          & $\Sigma_{+}$ & $\Sigma_{+}$ \\ \bottomrule
\end{tabular}
\caption{Properties of an dim-8, 4-field SU(3) operator with $D_{\mu} D_{\nu} F_{-L}^{\mu\nu} \Sigma_{+}^2$ and $D_{\mu} D_{\nu} F_{-R}^{\mu\nu} \Sigma_{+}^2$. }
\label{Table:p8-4fields-SU3}
\end{table}

By using the Eq.~\eqref{eq Numbers-i} we can calculate the numbers filled in the SSYTs:
\begin{equation}
\begin{aligned}
    F_{1L} \phi_{2} \phi_{3} \phi_{4} &: \#1=3,\, \#2=\#3=\#4=1 \,, \\
    F_{1R} \phi_{2} \phi_{3} \phi_{4} &: \#1=0,\, \#2=\#3=\#4=2 \,.
\end{aligned}
\end{equation}

In this way we know $F_{1L} \phi_{2} \phi_{3} \phi_{4}$ have only one Lorentz structure:
\begin{equation}
    M_{1L} =
    \begin{ytableau} 
    1 & 1 & 1 \\
    2 & 3 & 4
    \end{ytableau} \,,
\end{equation}

Furthermore, as we have mentioned before, in Eq.~\eqref{eq: YT_translate}, there are some relationships between SSYT and spinor helicity formalism. We can read the SSYT column by column and obtain:
\begin{equation}
\begin{aligned}
    M_{1L}  
    &= [34] \langle 13 \rangle \langle 14 \rangle \\   
    &= (\lambda_{1}^{\mu} \lambda_{1}^{\nu}) (\lambda_{3\alpha} \tilde{\lambda}_{3\mu}) (\lambda_{4\nu} \tilde{\lambda}_{4}^{\alpha}) \\
    &= F_{1L}^{\mu\nu} \phi_{2} (D_{3} \phi_{3})_{\alpha\mu} (D_{4} \phi_{4})_{\nu}^{\alpha}\,.
\end{aligned}
\end{equation}

Similarly, we can determine the corresponding SSYT expressions for the operator $F_{1R} \phi_{2} \phi_{3} \phi_{4}$, which are presented in the spinor helicity formalism as follows:
\begin{equation}
\begin{aligned}
    M_{1R} &= 
    \begin{ytableau} 
    2 & 2 & 3\\
    3 & 4 & 4
    \end{ytableau}  \\
    &= (\tilde{\lambda}_{1}^{\mu} \tilde{\lambda}_{1}^{\nu}) (\lambda_{3}^{\alpha} \tilde{\lambda}_{3}^{\nu})  (\lambda_{4\alpha} \tilde{\lambda}_{4\mu})  \\
    &= F_{1R\nu}^{\mu} \phi_{2} (D_{3} \phi_{3})^{\nu\alpha} (D_{4} \phi_{4})_{\mu\alpha}\,.
\end{aligned}
\end{equation}

The $M_{1L}$ and $M_{1R}$ yields two independent Lorentz structures. However, as mentioned in Sec.~\ref{sec:Off_Shell_Structures}, we have not previously considered the EOMs and the Adler zero condition. For external sources, the absence of the EOM constraints implies that the external fields $F_{1R}$ and $\phi_{3} \phi_{4}$ are free from the EOM constraints. 

Let us first address the Adler zero condition. Taking the condition into account, the number of the resulting operators remains unchanged, given the presence of a single 
$u$. Nevertheless, it is ensured that a derivative consistently operates on $u$.

After taking into account the aforementioned EOM conditions, we recalculate the $M_{1L}$ and $M_{1R}$ components to obtain the kinematic factors $M_{L}$ and $M_{R}$ as follows:
\begin{equation}
\begin{aligned}
    M_{L}: & F_{1L}^{\mu\nu} D_{2\nu}\phi_{2} D_{3\mu} \phi_{3} \phi_{4},\\
    \quad &  F_{1L}^{\mu\nu} D_{2\nu}\phi_{2} \phi_{3} D_{4\mu} \phi_{4} ,\\
    M_{R}: & F_{1R}^{\mu\nu} D_{2\nu}\phi_{2} D_{3\mu} \phi_{3} \phi_{4}.
\end{aligned}
\end{equation}

It is important to note that due to constraint of the Bianchi identity between $F_{1 L}$ and $F_{1 R}$, as referenced in Eq.~\eqref{eq: Bian_FL_FR}, removing the EOM for $F_{1 L}$ correspondingly means that there is no need to remove the EOM for the $F_{1 R}$ component. For the scalar field $\phi$, there is no such constraint, and the EOM must be removed for all.

In summary, after considering the EOM, $F_{1 L} \phi_{2} \phi_{3} \phi_{4}$ will have two independent kinematic factors $M_{L}$, which has one more compared with the one before the EOM was removed. $F_{1 R} \phi_{2} \phi_{3} \phi_{4}$ still only has a single independent kinematic factor $M_R$.

Due to the presence of repeated fields $\phi_{3} \phi_{4}$ in the operators $F_{1 L} \phi_{2} \phi_{3} \phi_{4}$ and $F_{1 R} \phi_{2} \phi_{3} \phi_{4}$, there is an additional permutation symmetry. The Lorentz permutation symmetry generators for the $L$ and $R$ components can be expressed as:
\begin{equation}
\begin{aligned}
    L: & \left(
        \begin{array}{cc}
         0 & 1 \\
         1 & 0 \\
        \end{array}
        \right) \,, \\
    R: & \left(
        \begin{array}{c}
         -1 \\
        \end{array}
        \right) \,.
\end{aligned}
\end{equation}

Having addressed the Lorentz part $M$, we now turn our attention to the internal $T$ part. The internal $T$ is the irreducible representation of the SU(N) group. In this example, the internal $T$ can be represented as the tensor product of four dim-8 representations, with the outer product of the four 8-dimensional representations already provided in Eq.~\eqref{eq: SU3_48outer_Product}.

The relationship between the operators and the Young tableau is discussed in reference \cite{Li:2020gnx}. Here we give the Young tableau expression of $SU(3)$ operator $\Psi^A \Psi^B \Psi^C \Psi^D$:
\begin{eqnarray}
	&& (\lambda^A)_{j_1}^{l_1}\epsilon_{l_1 i_1 k_1} \Psi^A\sim \young({{i_1}}{{j_1}},{{k_1}}) \,,\\
	&& (\lambda^B)_{j_2}^{l_2}\epsilon_{l_2 i_2 k_2} \Psi^B\sim \young({{i_2}}{{j_2}},{{k_2}}) \,,\\
	&& (\lambda^C)_{j_3}^{l_3}\epsilon_{l_3 i_3 k_3} \Psi^C\sim \young({{i_3}}{{j_3}},{{k_3}}) \,,\\
	&& (\lambda^D)_{j_4}^{l_4}\epsilon_{l_4 i_4 k_4} \Psi^D\sim \young({{i_4}}{{j_4}},{{k_4}}) \,,
\end{eqnarray}
where $\lambda$ is the Gell-Mann matrix, and $\Psi$ represents the SU(3) adjoint representation field.

By contracting the previously given Eq.~\eqref{eq: SU3_48outer_Product} for the fully symmetric tensors with the conversion factor:
\begin{eqnarray}
    CF=(\lambda^A)_{j_1}^{l_1} \epsilon_{l_1 i_1 k_1}(\lambda^B)_{j_2}^{l_2}\epsilon_{l_2 i_2 k_2} (\lambda^C)_{j_3}^{l_3}\epsilon_{l_3 i_3 k_3} (\lambda^D)_{j_4}^{l_4}\epsilon_{l_4 i_4 k_4} \,,
\end{eqnarray}
we can obtain the invariant tensors. Based on these invariant tensors, we can ultimately derive eight independent matrix basis elements, as referenced in \cite{Li:2020gnx} :
\begin{eqnarray}
&T_{1}=d^{ABE}d^{CDE},\quad 
 T_{2}=d^{ABE}f^{CDE},\quad 
 T_{3}=f^{ABE}f^{CDE},\quad 
 T_{4}=\delta^{AB}\delta^{CD},\quad\nonumber \\
&T_{5}=d^{CDE}f^{ABE},\quad 
 T_{6}=\delta^{AC}\delta^{BD}\quad
 T_{7}=d^{ACE}d^{BDE},\quad
 T_{8}=d^{ACE}f^{BDE}.
\end{eqnarray}
where $f_{a b c}$ and $d_{a b c}$ are the structure constants of the $SU(3)$ :
\begin{equation} \label{eq: SU3_Structure_Constant}
\begin{aligned}
    f_{A B C}=\frac{1}{4i}Tr([\lambda_A,\lambda_B]\lambda_C) \,, \\
    d_{A B C}=\frac{1}{4}Tr(\{\lambda_A,\lambda_B\}\lambda_C) \,.
\end{aligned}
\end{equation}
Correspondingly, when there are two repeated fields, we can also determine the internal group permutation symmetry generators:
\begin{equation}
\left(
\begin{array}{cccccccc}
 1 & 0 & 0 & 0 & 0 & 0 & 0 & 0 \\
 0 & -1 & 0 & 0 & 0 & 0 & 0 & 0 \\
 0 & 0 & -1 & 0 & 0 & 0 & 0 & 0 \\
 0 & 0 & 0 & 1 & 0 & 0 & 0 & 0 \\
 0 & 0 & 0 & 0 & 1 & 0 & 0 & 0 \\
 1 & 0 & -1 & -\frac{1}{3} & 0 & \frac{1}{3} & 2 & 0 \\
 -\frac{2}{3} & 0 & -\frac{1}{3} & \frac{2}{9} & 0 & \frac{4}{9} & -\frac{1}{3} & 0 \\
 0 & 0 & 0 & 0 & 1 & 0 & 0 & -1 \\
\end{array}
\right)     \,,
\end{equation}

We then combine the Lorentz part basis $M$ with the internal part basis $T$ to obtain the m-basis:
\begin{equation} \label{eq: MLT}
M_{L}T := M_{L} \otimes T =
\left(
\begin{array}{c}
d^{ABE}d^{CDE} F_{1L}^{\mu\nu} D_{2\nu}\phi_{2} D_{3\mu} \phi_{3} \phi_{4}\\
d^{ABE}f^{CDE} F_{1L}^{\mu\nu} D_{2\nu}\phi_{2} D_{3\mu} \phi_{3} \phi_{4}\\
f^{ABE}f^{CDE} F_{1L}^{\mu\nu} D_{2\nu}\phi_{2} D_{3\mu} \phi_{3} \phi_{4}\\
\delta^{AB}\delta^{CD}  F_{1L}^{\mu\nu} D_{2\nu}\phi_{2} D_{3\mu} \phi_{3} \phi_{4}\\
d^{CDE}f^{ABE}  F_{1L}^{\mu\nu} D_{2\nu}\phi_{2} D_{3\mu} \phi_{3} \phi_{4}\\
\delta^{AC}\delta^{BD}  F_{1L}^{\mu\nu} D_{2\nu}\phi_{2} D_{3\mu} \phi_{3} \phi_{4}\\
d^{ACE}d^{BDE}  F_{1L}^{\mu\nu} D_{2\nu}\phi_{2} D_{3\mu} \phi_{3} \phi_{4}\\
d^{ACE}f^{BDE}  F_{1L}^{\mu\nu} D_{2\nu}\phi_{2} D_{3\mu} \phi_{3} \phi_{4}\\
d^{ABE}d^{CDE} F_{1L}^{\mu\nu} D_{2\nu}\phi_{2} \phi_{3} D_{4\mu}\phi_{4}  \\
d^{ABE}f^{CDE} F_{1L}^{\mu\nu} D_{2\nu}\phi_{2} \phi_{3} D_{4\mu}\phi_{4}  \\
f^{ABE}f^{CDE} F_{1L}^{\mu\nu} D_{2\nu}\phi_{2} \phi_{3} D_{4\mu}\phi_{4}  \\
\delta^{AB}\delta^{CD} F_{1L}^{\mu\nu} D_{2\nu}\phi_{2} \phi_{3} D_{4\mu}\phi_{4}  \\
d^{CDE}f^{ABE} F_{1L}^{\mu\nu} D_{2\nu}\phi_{2} \phi_{3} D_{4\mu}\phi_{4}  \\
\delta^{AC}\delta^{BD} F_{1L}^{\mu\nu} D_{2\nu}\phi_{2} \phi_{3} D_{4\mu} \phi_{4} \\
d^{ACE}d^{BDE} F_{1L}^{\mu\nu} D_{2\nu}\phi_{2} \phi_{3} D_{4\mu}\phi_{4}  \\
d^{ACE}f^{BDE} F_{1L}^{\mu\nu} D_{2\nu}\phi_{2} \phi_{3} D_{4\mu}\phi_{4}  \\ 
\end{array}
\right)     \,,
\end{equation}

\begin{equation} \label{eq: MRT}
M_{R}T := M_{R} \otimes T =
\left(
\begin{array}{c}
d^{ABE}d^{CDE} F_{1R}^{\mu\nu} D_{2\nu}\phi_{2} D_{3\mu} \phi_{3} \phi_{4}\\
d^{ABE}f^{CDE} F_{1R}^{\mu\nu} D_{2\nu}\phi_{2} D_{3\mu} \phi_{3} \phi_{4}\\
f^{ABE}f^{CDE} F_{1R}^{\mu\nu} D_{2\nu}\phi_{2} D_{3\mu} \phi_{3} \phi_{4}\\
\delta^{AB}\delta^{CD}  F_{1R}^{\mu\nu} D_{2\nu}\phi_{2} D_{3\mu} \phi_{3} \phi_{4}\\
d^{CDE}f^{ABE}  F_{1R}^{\mu\nu} D_{2\nu}\phi_{2} D_{3\mu} \phi_{3} \phi_{4}\\
\delta^{AC}\delta^{BD}  F_{1R}^{\mu\nu} D_{2\nu}\phi_{2} D_{3\mu} \phi_{3} \phi_{4}\\
d^{ACE}d^{BDE}  F_{1R}^{\mu\nu} D_{2\nu}\phi_{2} D_{3\mu} \phi_{3} \phi_{4}\\
d^{ACE}f^{BDE}  F_{1R}^{\mu\nu} D_{2\nu}\phi_{2} D_{3\mu} \phi_{3} \phi_{4}\\
\end{array}
\right)     \,,
\end{equation}
at this point, we have not yet considered the permutation symmetry. 

The formula of the idempotent element is
\begin{equation}
    \mathcal{Y} = \frac{1}{n!} \sum_{\sigma \in S_{n}} F(\sigma) \sigma \,,
\end{equation}
where $n$ is the number of repeated fields, $\sigma$ is group elements of the permutation group $S_n$, $F(\sigma)$ is the expansion coefficient.

To calculate the permutation symmetry, we first determine the idempotent elements for both the Lorentz part:
\begin{equation}
\begin{aligned}
    \mathcal{Y}^{M_{L}} \left[ 
    \ytableausetup{smalltableaux}
    \begin{ytableau} 
    3 & 4
    \end{ytableau}
    \right] &= 
    \frac{1}{2} \left( 
    \begin{pmatrix}
    1 & 0 \\0 & 1
    \end{pmatrix}
    +
    \begin{pmatrix}
    0 & 1 \\1 & 0
    \end{pmatrix} \right) = 
    \begin{pmatrix}
    \frac{1}{2} & \frac{1}{2} \\\frac{1}{2} & \frac{1}{2}
    \end{pmatrix} \,,  \\ 
    \mathcal{Y}^{M_{L}} \left[ 
    \ytableausetup{smalltableaux}
    \begin{ytableau} 
    3 \\ 4
    \end{ytableau}
    \right] &= 
    \frac{1}{2} \left( 
    \begin{pmatrix}
    1 & 0 \\0 & 1
    \end{pmatrix}
    -
    \begin{pmatrix}
    0 & 1 \\1 & 0
    \end{pmatrix} \right) = 
    \begin{pmatrix}
    \frac{1}{2} & -\frac{1}{2} \\-\frac{1}{2} & \frac{1}{2}
    \end{pmatrix} \,,  \\ 
    \mathcal{Y}^{M_{R}} \left[ 
    \ytableausetup{smalltableaux}
    \begin{ytableau} 
    3 & 4
    \end{ytableau}
    \right] &= 
    \frac{1}{2} \left( 
    1 +(-1) \right) = 0 \,, \\
    \mathcal{Y}^{M_{R}} \left[ 
    \ytableausetup{smalltableaux}
    \begin{ytableau} 
    3 \\ 4
    \end{ytableau}
    \right] &= 
    \frac{1}{2} \left( 
    1-(-1) \right) = 1 \,, 
\end{aligned}
\end{equation}
and the internal part: 
\begin{equation}
\begin{aligned}
    \mathcal{Y}^{T} \left[ 
    \ytableausetup{smalltableaux}
    \begin{ytableau} 
    3 & 4
    \end{ytableau}
    \right] &= 
    \frac{1}{2} \left( 
    \left(
    \begin{array}{cccccccc}
     1 & 0 & 0 & 0 & 0 & 0 & 0 & 0 \\
     0 & 1 & 0 & 0 & 0 & 0 & 0 & 0 \\
     0 & 0 & 1 & 0 & 0 & 0 & 0 & 0 \\
     0 & 0 & 0 & 1 & 0 & 0 & 0 & 0 \\
     0 & 0 & 0 & 0 & 1 & 0 & 0 & 0 \\
     0 & 0 & 0 & 0 & 0 & 1 & 0 & 0 \\
     0 & 0 & 0 & 0 & 0 & 0 & 1 & 0 \\
     0 & 0 & 0 & 0 & 0 & 0 & 0 & 1 \\
    \end{array}
    \right) +
    \left(
    \begin{array}{cccccccc}
     1 & 0 & 0 & 0 & 0 & 0 & 0 & 0 \\
     0 & -1 & 0 & 0 & 0 & 0 & 0 & 0 \\
     0 & 0 & -1 & 0 & 0 & 0 & 0 & 0 \\
     0 & 0 & 0 & 1 & 0 & 0 & 0 & 0 \\
     0 & 0 & 0 & 0 & 1 & 0 & 0 & 0 \\
     1 & 0 & -1 & -\frac{1}{3} & 0 & \frac{1}{3} & 2 & 0 \\
     -\frac{2}{3} & 0 & -\frac{1}{3} & \frac{2}{9} & 0 & \frac{4}{9} & -\frac{1}{3} & 0 \\
     0 & 0 & 0 & 0 & 1 & 0 & 0 & -1 \\
    \end{array}
    \right)
    \right)   \\ 
    &= \left(
    \begin{array}{cccccccc}
     1 & 0 & 0 & 0 & 0 & 0 & 0 & 0 \\
     0 & 0 & 0 & 0 & 0 & 0 & 0 & 0 \\
     0 & 0 & 0 & 0 & 0 & 0 & 0 & 0 \\
     0 & 0 & 0 & 1 & 0 & 0 & 0 & 0 \\
     0 & 0 & 0 & 0 & 1 & 0 & 0 & 0 \\
     \frac{1}{2} & 0 & -\frac{1}{2} & -\frac{1}{6} & 0 & \frac{2}{3} & 1 & 0 \\
     -\frac{1}{3} & 0 & -\frac{1}{6} & \frac{1}{9} & 0 & \frac{2}{9} & \frac{1}{3} & 0 \\
     0 & 0 & 0 & 0 & \frac{1}{2} & 0 & 0 & 0 \\
    \end{array}
    \right) \,,
\end{aligned}
\end{equation}
\begin{equation}
\begin{aligned}
    \mathcal{Y}^{T} \left[ 
    \ytableausetup{smalltableaux}
    \begin{ytableau} 
    3 \\ 4
    \end{ytableau}
    \right] &= 
    \frac{1}{2} \left( 
    \left(
    \begin{array}{cccccccc}
     1 & 0 & 0 & 0 & 0 & 0 & 0 & 0 \\
     0 & 1 & 0 & 0 & 0 & 0 & 0 & 0 \\
     0 & 0 & 1 & 0 & 0 & 0 & 0 & 0 \\
     0 & 0 & 0 & 1 & 0 & 0 & 0 & 0 \\
     0 & 0 & 0 & 0 & 1 & 0 & 0 & 0 \\
     0 & 0 & 0 & 0 & 0 & 1 & 0 & 0 \\
     0 & 0 & 0 & 0 & 0 & 0 & 1 & 0 \\
     0 & 0 & 0 & 0 & 0 & 0 & 0 & 1 \\
    \end{array}
    \right) -   
    \left(
    \begin{array}{cccccccc}
     1 & 0 & 0 & 0 & 0 & 0 & 0 & 0 \\
     0 & -1 & 0 & 0 & 0 & 0 & 0 & 0 \\
     0 & 0 & -1 & 0 & 0 & 0 & 0 & 0 \\
     0 & 0 & 0 & 1 & 0 & 0 & 0 & 0 \\
     0 & 0 & 0 & 0 & 1 & 0 & 0 & 0 \\
     1 & 0 & -1 & -\frac{1}{3} & 0 & \frac{1}{3} & 2 & 0 \\
     -\frac{2}{3} & 0 & -\frac{1}{3} & \frac{2}{9} & 0 & \frac{4}{9} & -\frac{1}{3} & 0 \\
     0 & 0 & 0 & 0 & 1 & 0 & 0 & -1 \\
    \end{array}
    \right)
    \right)   \\ 
    &= \left(
    \begin{array}{cccccccc}
     0 & 0 & 0 & 0 & 0 & 0 & 0 & 0 \\
     0 & 1 & 0 & 0 & 0 & 0 & 0 & 0 \\
     0 & 0 & 1 & 0 & 0 & 0 & 0 & 0 \\
     0 & 0 & 0 & 0 & 0 & 0 & 0 & 0 \\
     0 & 0 & 0 & 0 & 0 & 0 & 0 & 0 \\
     -\frac{1}{2} & 0 & \frac{1}{2} & \frac{1}{6} & 0 & \frac{1}{3} & -1 & 0 \\
     \frac{1}{3} & 0 & \frac{1}{6} & -\frac{1}{9} & 0 & -\frac{2}{9} & \frac{2}{3} & 0 \\
     0 & 0 & 0 & 0 & -\frac{1}{2} & 0 & 0 & 1 \\
    \end{array}
    \right) \,.
\end{aligned}
\end{equation}
By taking the direct product of $\mathcal{Y}^{M_{L}}$ with $\mathcal{Y}^{T}$, we obtain the overall permutation symmetry idempotent elements matrix
\begin{equation}
\mathcal{Y}^{\mathcal{O}_{L}} = 
\mathcal{Y}^{M_{L}} \left[ 
    \ytableausetup{smalltableaux}
    \begin{ytableau} 
    3 & 4
    \end{ytableau}
    \right]
    \otimes
\mathcal{Y}^{T} \left[ 
    \ytableausetup{smalltableaux}
    \begin{ytableau} 
    3 & 4
    \end{ytableau}
    \right] \\ +
    \mathcal{Y}^{M_{L}} \left[ 
    \ytableausetup{smalltableaux}
    \begin{ytableau} 
    3 \\ 4
    \end{ytableau}
    \right]
    \otimes
\mathcal{Y}^{T} \left[ 
    \ytableausetup{smalltableaux}
    \begin{ytableau} 
    3 \\ 4
    \end{ytableau}
    \right] \\ \,,
\end{equation}   
\begin{equation}
\mathcal{Y}^{\mathcal{O}_{L}} = 
\left(
\begin{array}{cccccccccccccccc} \label{eq: YOL_Idempotent_Elements}
 \frac{1}{2} & 0 & 0 & 0 & 0 & 0 & 0 & 0 & \frac{1}{2} & 0 & 0 & 0 & 0 & 0 & 0 & 0 \\
 0 & \frac{1}{2} & 0 & 0 & 0 & 0 & 0 & 0 & 0 & -\frac{1}{2} & 0 & 0 & 0 & 0 & 0 & 0 \\
 0 & 0 & \frac{1}{2} & 0 & 0 & 0 & 0 & 0 & 0 & 0 & -\frac{1}{2} & 0 & 0 & 0 & 0 & 0 \\
 0 & 0 & 0 & \frac{1}{2} & 0 & 0 & 0 & 0 & 0 & 0 & 0 & \frac{1}{2} & 0 & 0 & 0 & 0 \\
 0 & 0 & 0 & 0 & \frac{1}{2} & 0 & 0 & 0 & 0 & 0 & 0 & 0 & \frac{1}{2} & 0 & 0 & 0 \\
 0 & 0 & 0 & 0 & 0 & \frac{1}{2} & 0 & 0 & \frac{1}{2} & 0 & -\frac{1}{2} & -\frac{1}{6} & 0 & \frac{1}{6} & 1 & 0 \\
 0 & 0 & 0 & 0 & 0 & 0 & \frac{1}{2} & 0 & -\frac{1}{3} & 0 & -\frac{1}{6} & \frac{1}{9} & 0 & \frac{2}{9} & -\frac{1}{6} & 0 \\
 0 & 0 & 0 & 0 & 0 & 0 & 0 & \frac{1}{2} & 0 & 0 & 0 & 0 & \frac{1}{2} & 0 & 0 & -\frac{1}{2} \\
 \frac{1}{2} & 0 & 0 & 0 & 0 & 0 & 0 & 0 & \frac{1}{2} & 0 & 0 & 0 & 0 & 0 & 0 & 0 \\
 0 & -\frac{1}{2} & 0 & 0 & 0 & 0 & 0 & 0 & 0 & \frac{1}{2} & 0 & 0 & 0 & 0 & 0 & 0 \\
 0 & 0 & -\frac{1}{2} & 0 & 0 & 0 & 0 & 0 & 0 & 0 & \frac{1}{2} & 0 & 0 & 0 & 0 & 0 \\
 0 & 0 & 0 & \frac{1}{2} & 0 & 0 & 0 & 0 & 0 & 0 & 0 & \frac{1}{2} & 0 & 0 & 0 & 0 \\
 0 & 0 & 0 & 0 & \frac{1}{2} & 0 & 0 & 0 & 0 & 0 & 0 & 0 & \frac{1}{2} & 0 & 0 & 0 \\
 \frac{1}{2} & 0 & -\frac{1}{2} & -\frac{1}{6} & 0 & \frac{1}{6} & 1 & 0 & 0 & 0 & 0 & 0 & 0 & \frac{1}{2} & 0 & 0 \\
 -\frac{1}{3} & 0 & -\frac{1}{6} & \frac{1}{9} & 0 & \frac{2}{9} & -\frac{1}{6} & 0 & 0 & 0 & 0 & 0 & 0 & 0 & \frac{1}{2} & 0 \\
 0 & 0 & 0 & 0 & \frac{1}{2} & 0 & 0 & -\frac{1}{2} & 0 & 0 & 0 & 0 & 0 & 0 & 0 & \frac{1}{2} \\
\end{array}
\right)\,,
\end{equation} 
although every row of this idempotent elements matrix contains non-zero component, the rows are not completely linearly independent from one another. It is possible to eliminate several rows through elementary row operations. Ultimately, the rank of the matrix is calculated to be 8, hence there are 8 independent operators corresponding to the $L$ part.

In the same way, we present the result of $\mathcal{Y}^{\mathcal{O}_{R}}$ below:
\begin{equation}
\mathcal{Y}^{\mathcal{O}_{R}} = 
\mathcal{Y}^{M_{R}} \left[ 
    \ytableausetup{smalltableaux}
    \begin{ytableau} 
    3 & 4
    \end{ytableau}
    \right]
    \otimes
\mathcal{Y}^{T} \left[ 
    \ytableausetup{smalltableaux}
    \begin{ytableau} 
    3 & 4
    \end{ytableau}
    \right] \\ +
    \mathcal{Y}^{M_{R}} \left[ 
    \ytableausetup{smalltableaux}
    \begin{ytableau} 
    3 \\ 4
    \end{ytableau}
    \right]
    \otimes
\mathcal{Y}^{T} \left[ 
    \ytableausetup{smalltableaux}
    \begin{ytableau} 
    3 \\ 4
    \end{ytableau}
    \right] =    \mathcal{Y}^{M_{R}} \left[ 
    \ytableausetup{smalltableaux}
    \begin{ytableau} 
    3 \\ 4
    \end{ytableau}
    \right]
    \otimes
\mathcal{Y}^{T} \left[ 
    \ytableausetup{smalltableaux}
    \begin{ytableau} 
    3 \\ 4
    \end{ytableau}
    \right] \\ \,,
\end{equation}   
\begin{equation} \label{eq: YOR_Idempotent_Elements}
\mathcal{Y}^{\mathcal{O}_{R}} = 
\left(
\begin{array}{cccccccc}
 0 & 0 & 0 & 0 & 0 & 0 & 0 & 0 \\
 0 & 1 & 0 & 0 & 0 & 0 & 0 & 0 \\
 0 & 0 & 1 & 0 & 0 & 0 & 0 & 0 \\
 0 & 0 & 0 & 0 & 0 & 0 & 0 & 0 \\
 0 & 0 & 0 & 0 & 0 & 0 & 0 & 0 \\
 -\frac{1}{2} & 0 & \frac{1}{2} & \frac{1}{6} & 0 & \frac{1}{3} & -1 & 0 \\
 \frac{1}{3} & 0 & \frac{1}{6} & -\frac{1}{9} & 0 & -\frac{2}{9} & \frac{2}{3} & 0 \\
 0 & 0 & 0 & 0 & -\frac{1}{2} & 0 & 0 & 1 \\
\end{array}
\right)\,.
\end{equation}  
By the same logic, we can deduce the rank of the corresponding $R$ part to be 4. And considering both the L and R components, we can ascertain that the total number of independent operators is 12.

Finally, by taking the direct product of the permutation symmetry idempotent elements $\mathcal{Y}^{\mathcal{O}}$ in Eq.~\eqref{eq: YOL_Idempotent_Elements}, Eq.~\eqref{eq: YOR_Idempotent_Elements} with the effective operator $ M_{L/R} \otimes T$ in Eq.~\eqref{eq: MLT}, Eq.~\eqref{eq: MRT}, we can get the effective operators $\mathcal{O}_{p}^{L/R}$ with permutation symmetry:
\begin{equation}
    \mathcal{O}_{p}^{L/R} = \mathcal{Y}^{\mathcal{O}_{L/R}} \cdot (M_{L/R} \otimes T) \,.
\end{equation}

By selecting appropriate bases, we can enumerate all twelve operators :

$F_{1 L} \phi_{2} \phi_{3} \phi_{4}:$
\begin{equation} \label{eq: F1Lphi}
\begin{array}{l}
\Sigma_{+}^{C} F_{-L}^{\mu \nu A} \left( D_{\nu} u^{A} \right) \left( D_{\mu} \Sigma_{+}^{C} \right) \,,
\\
\Sigma_{+}^{B} F_{-L}^{\mu \nu A} \left( D_{\nu} u^{B} \right) \left( D_{\mu} \Sigma_{+}^{A} \right) \,,
\\
\Sigma_{+}^{A} F_{-L}^{\mu \nu A} \left( D_{\nu} u^{B} \right) \left( D_{\mu} \Sigma_{+}^{B} \right) \,,
\\
d^{ABE} d^{CDE} \Sigma_{+}^{D} F_{-L}^{\mu \nu A} \left( D_{\nu} u^{B} \right) \left( D_{\mu} \Sigma_{+}^C \right) \,,
\\
d^{ABE} f^{CDE} \Sigma_{+}^{D} F_{-L}^{\mu \nu A} \left( D_{\nu} u^{B} \right) \left( D_{\mu} \Sigma_{+}^{C} \right) \,,
\\
d^{CDE} f^{ABE} \Sigma_{+}^{D} F_{-L}^{\mu \nu A} \left( D_{\nu} u^{B} \right) \left( D_{\mu} \Sigma_{+}^{C} \right) \,,
\\
d^{ACE} f^{BDE} \Sigma_{+}^{D} F_{-L}^{\mu \nu A} \left( D_{\nu} u^{B} \right) \left( D_{\mu} \Sigma_{+}^{C} \right) \,,
\\
f^{ABE} f^{CDE} \Sigma_{+}^{D} F_{-L}^{\mu \nu A} \left( D_{\nu} u^{B} \right) \left( D_{\mu} \Sigma_{+}^{C} \right) \,,
\end{array}
\end{equation}

$F_{1 R} \phi_{2} \phi_{3} \phi_{4} :$
\begin{equation} \label{eq: F1Rphi}
\begin{array}{l}
d^{ABE} d^{CDE} \Sigma_{+}^{C} F_{-R}^{\mu \nu D} \left( D_{\nu} u^{A} \right) \left( D_{\mu} \Sigma_{+}^B \right) \,,
\\
d^{ABE} f^{CDE} \Sigma_{+}^{C} F_{-R}^{\mu \nu D} \left( D_{\nu} u^{A} \right) \left( D_{\mu} \Sigma_{+}^{B} \right) \,,
\\
d^{CDE} f^{ABE} \Sigma_{+}^{C} F_{-R}^{\mu \nu D} \left( D_{\nu} u^{A} \right) \left( D_{\mu} \Sigma_{+}^{B} \right) \,,
\\
f^{ABE} f^{CDE} \Sigma_{+}^{C} F_{-R}^{\mu \nu D} \left( D_{\nu} u^{A} \right) \left( D_{\mu} \Sigma_{+}^{B} \right) \,,
\end{array}
\end{equation}
this tells the presence of a total of 12 independent bases elements, agreed with results from the Hilbert Series in Sec.\ref{sec:3_6_HS_OC}.

\subsubsection{Operator with 5 fields}

We now consider example of the five-field operator the  $SU(3)\, D u^3 \Sigma_{+}^2$. Some basic information about this operator is presented in Table.~\ref{table: Inf_D_u3_Sigma2}. Similar to the previous example, we will first calculate the kinematic factor part $M$ using the SSYT method, followed by the consideration of the internal part $T$. Due to the presence of three repeated fields $u$ and two repeated fields $\Sigma_{+}$ in the operator $D u^3 \Sigma_{+}^2$, the permutation symmetry must also be taken into account to obtain an independent and complete operator structure.
\begin{table}[ht]
\centering
\begin{tabular}{@{}llllllllll@{}}
\toprule
$p^8$ 5-Fields $SU(3)$                             & $N-2$ & $n \langle \cdot \rangle$ & $\tilde{n}[\cdot]$ & helicity      & $\phi_{1}$ & $\phi_{2}$ & $\phi_{4}$ & $\phi_{4}$   & $\phi_{5}$   \\ \midrule
$D_{\mu} D_{\nu} D^{\mu} D^{\nu} u^3 \Sigma_{+}^2$ & $3$   & $2$                       & $2$                & $(0,0,0,0,0)$ & $u$        & $u$        & $u$        & $\Sigma_{+}$ & $\Sigma_{+}$ \\ \bottomrule
\end{tabular}
\caption{Properties of an 8-Dimension, 5-Field SU(3) operator with $D_{\mu} D_{\nu} D^{\mu} D^{\nu} u^3 \Sigma_{+}^2$ components for N, n, $\tilde{n}$, helicity and definition.}
\label{table: Inf_D_u3_Sigma2}
\end{table}

Firstly, we consider the kinematic factor part $M$. In this case, with three Goldstones $u_{\mu}$ and one $D_{\mu}$, the Adler zero condition significantly restricts the Lorentz structure of the operator. Without considering the Adler zero and the EOM, the SSYT method yields 16 Lorentz structures:
\begin{equation}
\begin{aligned}
&\phi_{1} \phi_{2} \phi_{3} \phi_{4} \phi_{5} D_{4}^{\mu} D_{4}^{\nu} D_{5}^{\mu} D_{5}^{\nu},\, 
\phi_{1} \phi_{2} \phi_{3} \phi_{4} \phi_{5} D_{3}^{\mu} D_{3}^{\nu} D_{5}^{\mu} D_{5}^{\nu},\, 
\phi_{1} \phi_{2} \phi_{3} \phi_{4} \phi_{5} D_{3}^{\mu} D_{3}^{\nu} D_{4}^{\mu} D_{4}^{\nu},\, 
\\
&\phi_{1} \phi_{2} \phi_{3} \phi_{4} \phi_{5} D_{3}^{\mu} D_{3}^{\nu} D_{4}^{\mu} D_{5}^{\nu},\, 
\phi_{1} \phi_{2} \phi_{3} \phi_{4} \phi_{5} D_{3}^{\mu} D_{4}^{\nu} D_{5}^{\mu} D_{5}^{\nu},\,
\phi_{1} \phi_{2} \phi_{3} \phi_{4} \phi_{5} D_{3}^{\mu} D_{4}^{\nu} D_{4}^{\mu} D_{5}^{\nu},\, 
\\
&\phi_{1} \phi_{2} \phi_{3} \phi_{4} \phi_{5} D_{2}^{\mu} D_{2}^{\nu} D_{5}^{\mu} D_{5}^{\nu},\,
\phi_{1} \phi_{2} \phi_{3} \phi_{4} \phi_{5} D_{2}^{\mu} D_{2}^{\nu} D_{4}^{\mu} D_{4}^{\nu},\,
\phi_{1} \phi_{2} \phi_{3} \phi_{4} \phi_{5} D_{2}^{\mu} D_{2}^{\nu} D_{4}^{\mu} D_{5}^{\nu},\, 
\\
&\phi_{1} \phi_{2} \phi_{3} \phi_{4} \phi_{5} D_{2}^{\mu} D_{3}^{\nu} D_{5}^{\mu} D_{4}^{\nu},\, 
\phi_{1} \phi_{2} \phi_{3} \phi_{4} \phi_{5} D_{2}^{\mu} D_{4}^{\nu} D_{3}^{\mu} D_{5}^{\nu},\, 
\phi_{1} \phi_{2} \phi_{3} \phi_{4} \phi_{5} D_{2}^{\mu} D_{3}^{\nu} D_{4}^{\mu} D_{5}^{\nu},\,
\\
&\phi_{1} \phi_{2} \phi_{3} \phi_{4} \phi_{5} D_{2}^{\mu} D_{4}^{\nu} D_{5}^{\mu} D_{5}^{\nu},\,
\phi_{1} \phi_{2} \phi_{3} \phi_{4} \phi_{5} D_{2}^{\mu} D_{3}^{\nu} D_{5}^{\mu} D_{5}^{\nu},\, 
\phi_{1} \phi_{2} \phi_{3} \phi_{4} \phi_{5} D_{2}^{\mu} D_{3}^{\nu} D_{4}^{\mu} D_{4}^{\nu},\,
\\
&i \phi_{1} \phi_{2} \phi_{3} \phi_{4} \phi_{5} D_{4}^{\lambda} D_{2}^{\mu} D_{3}^{\nu} D_{5}^{\rho} \epsilon^{\lambda \mu \nu \rho}.
\end{aligned}
\end{equation}
For the sake of clarity in the expressions, all indices are written as superscripts.

Upon incorporating the Adler zero condition, it introduces nine additional constraints, resulting in only seven independent bases:
\begin{equation}
\begin{aligned}
&\phi_{1} \phi_{2} \phi_{3} \phi_{4} \phi_{5} D_{1}^{\mu} D_{3}^{\nu} D_{2}^{\mu} D_{1}^{\nu},\,
\phi_{1} \phi_{2} \phi_{3} \phi_{4} \phi_{5} D_{1}^{\mu} D_{3}^{\nu} D_{2}^{\mu} D_{2}^{\nu},\,
\phi_{1} \phi_{2} \phi_{3} \phi_{4} \phi_{5} D_{1}^{\mu} D_{2}^{\nu} D_{3}^{\mu} D_{3}^{\nu},\,
\\
&\phi_{1} \phi_{2} \phi_{3} \phi_{4} \phi_{5} D_{1}^{\mu} D_{3}^{\nu} D_{2}^{\mu} D_{4}^{\nu},\,
\phi_{1} \phi_{2} \phi_{3} \phi_{4} \phi_{5} D_{1}^{\mu} D_{2}^{\nu} D_{3}^{\mu} D_{4}^{\nu},\,
\phi_{1} \phi_{2} \phi_{3} \phi_{4} \phi_{5} D_{1}^{\mu} D_{2}^{\nu} D_{4}^{\mu} D_{3}^{\nu},\,
\\
&\phi_{1} \phi_{2} \phi_{3} \phi_{4} \phi_{5} D_{3}^{\lambda} D_{1}^{\mu} D_{2}^{\nu} D_{4}^{\rho} \epsilon^{\lambda \mu \nu \rho},
\end{aligned}
\end{equation}
where the fields $\phi_{1,2,3}$ correspond to the Goldstones $u$. It is evident that upon imposing the Adler zero condition each $u$ must be accompanied by at least one derivative $D^{\mu}$ to form the representation $D^{\mu}u$.


Moreover, after removing the EOM constraints from the external field $\Sigma_{+}$, the number of bases remains unchanged. This is because an extra derivative acting on $\Sigma_{+}$ does not constitute an EOM, and thus Eq.~\eqref{eq: offshell1} neither increases nor decreases the number of constraints.

Next, we consider the internal group $T$. The three Goldstones $u$ correspond to three $SU(3)$ 8-dimensional representations, while the two $\Sigma$ correspond to two $SU(3)$ 8-dimensional representations. By performing a tensor product decomposition using the outer product of the corresponding Young diagram, we obtain 32 singlet representations, corresponding to 32 bases:
\begin{equation}
\begin{aligned}
&\delta^{DE} d_{abc}^{ABC},\, d_{abc}^{ACF} d_{abc}^{BFG} d_{abc}^{DEG},\, d_{abc}^{ACF} d_{abc}^{BEG} d_{abc}^{DFG},\, f_{abc}^{ACG} d_{abc}^{BFG} d_{abc}^{DEF},\,
\\
&d_{abc}^{ACF} d_{abc}^{BFG} f_{abc}^{DEG},\, d_{abc}^{ACF} f_{abc}^{BEG} d_{abc}^{DFG},\, f_{abc}^{ACG} d_{abc}^{BEF} d_{abc}^{DFG},\, d_{abc}^{ACF} d_{abc}^{BEG} f_{abc}^{DFG},\,
\\
&f_{abc}^{ACF} d_{abc}^{BFG} f_{abc}^{DEG},\, f_{abc}^{ACF} f_{abc}^{BEG} d_{abc}^{DFG},\, d_{abc}^{ACF} f_{abc}^{BFG} f_{abc}^{DEG},\, f_{abc}^{ACF} f_{abc}^{BEG} f_{abc}^{DEG},\,
\\
&f_{abc}^{ACG} d_{abc}^{BEF} f_{abc}^{DFG},\, f_{abc}^{ACF} f_{abc}^{BFG} f_{abc}^{DEG},\, f_{abc}^{ACF} f_{abc}^{BEG} f_{abc}^{DFG},\, \delta^{AC} d_{abc}^{BDE},\,
\\
&\delta^{BE} f_{abc}^{ACD},\, d_{abc}^{ABF} d_{abc}^{CFG} d_{abc}^{DEG},\, d_{abc}^{ABF} d_{abc}^{CFG} f_{abc}^{DEG},\, d_{abc}^{ABF} f_{abc}^{CFG} f_{abc}^{DEG},\,
\\
&f_{abc}^{ABF} f_{abc}^{CFG} f_{abc}^{DEG},\, d_{abc}^{ABF} d_{abc}^{CEG} d_{abc}^{DFG},\, f_{abc}^{ABG} d_{abc}^{CFG} d_{abc}^{DFG},\, f_{abc}^{ABG} d_{abc}^{CEF} d_{abc}^{DFG},\,
\\
&d_{abc}^{ABF} f_{abc}^{CEG} f_{abc}^{DFG},\, f_{abc}^{ABG} f_{abc}^{CFG} d_{abc}^{DEF},\, f_{abc}^{ABF} f_{abc}^{CEG} f_{abc}^{DFG},\, \delta^{AC} f_{abc}^{BDE},\, 
\\
&d_{abc}^{ABF} f_{abc}^{CEG} d_{abc}^{DFG},\, f_{abc}^{ABF} f_{abc}^{CEG} d_{abc}^{DFG},\, d_{abc}^{ADF} d_{abc}^{BFG} d_{abc}^{CEG},\, f_{abc}^{ADG} d_{abc}^{BFG} d_{abc}^{CEF},\,
\\
&f_{abc}^{ADG} d_{abc}^{BFG} d_{abc}^{DEF}\,.
\end{aligned}
\end{equation}

Since there are three repeated fields $u$ and two repeated fields $\Sigma_{+}$ in the operator $D u^3 \Sigma_{+}^2$, the permutation symmetry idempotent elements corresponding to each are $\mathcal{Y}^u$ and $\mathcal{Y}^{\Sigma_{+}}$, respectively. Therefore, the total permutation symmetry idempotent elements are:
\begin{equation} \label{eq: yu_ySigma}
    \mathcal{Y}^\mathcal{O} = (\mathcal{Y}^{Mu} \cdot \mathcal{Y}^{M\Sigma_{+}} ) \otimes (\mathcal{Y}^{Tu} \cdot \mathcal{Y}^{T\Sigma_{+}})\,.
\end{equation}

Given the number of bases, the idempotent elements for the kinematic factor part $M$ form a $7\times7$ matrix, while the internal group $T$ part yields a $32\times32$ idempotent matrix.
The multiplication of these two parts results in the total permutation symmetry idempotent elements in Eq.~\eqref{eq: yu_ySigma}. 

It is evident that the matrix of total idempotent elements is $(7\times7) \otimes (32\times32)$.
Subsequent linear combinations yield a matrix of rank 19.
Thus, we obtain 19 independent and complete operators that take into account the permutation symmetry:
\begin{equation}
\begin{aligned}
&
d^{ABC} \Sigma_{+}^{D} \left(D_{\nu} u^{C}\right) \left(D^{\mu} u^{B}\right) \left(D_{\mu} D^{\nu} u^{A}\right)\,, 
\\ &
d^{ACF} d^{BFG} d^{DEG} \Sigma_{+}^{D} \Sigma_{+}^{E} \left(D_{\mu} u^{A}\right) \left(D_{\nu} u^{C}\right) \left(D^{\mu} D^{\nu} u^{B}\right)\,, 
\\ &
d^{ACF} d^{BEG} d^{DFG} \Sigma_{+}^{D} \Sigma_{+}^{E} \left(D_{\nu} u^{C}\right) \left(D^{\mu} u^{B}\right) \left(D_{\mu} D^{\nu} u^{A}\right)\,, 
\\ &
d^{ACF} d^{BEG} d^{DFG} \Sigma_{+}^{D} \Sigma_{+}^{E} \left(D_{\mu} u^{A}\right) \left(D_{\nu} u^{C}\right) \left(D^{\mu} D^{\nu} u^{B}\right)\,, 
\\ &
d^{ACF} d^{BEG} d^{DFG} \Sigma_{+}^{E} \left(D_{\mu} u^{A}\right) \left(D_{\nu} u^{C}\right) \left(D^{\nu} \Sigma_{+}^{D}\right) \left(D^{\mu} u^{B}\right)\,, 
\\ &
d^{ACF} d^{BEG} d^{DFG} \Sigma_{+}^{E} \left(D_{\mu} u^{A}\right) \left(D_{\nu} u^{B}\right) \left(D^{\nu} \Sigma_{+}^{D}\right) \left(D^{\mu} u^{C}\right)\,, 
\\ &
d^{BFG} d^{DEF} f^{ACG} \Sigma_{+}^{D} \Sigma_{+}^{E} \left(D_{\nu} u^{C}\right) \left(D^{\mu} u^{B}\right) \left(D_{\mu} D^{\nu} u^{A}\right)\,, 
\\ &
d^{ACF} d^{BFG} f^{DEG} \Sigma_{+}^{E} \left(D_{\mu} u^{A}\right) \left(D_{\nu} u^{C}\right) \left(D^{\nu} \Sigma_{+}^{D}\right) \left(D^{\mu} u^{B}\right)\,, 
\\ &
d^{ACF} d^{BFG} f^{DEG} \Sigma_{+}^{E} \left(D_{\mu} u^{A}\right) \left(D_{\nu} u^{B}\right) \left(D^{\nu} \Sigma_{+}^{D}\right) \left(D^{\mu} u^{C}\right)\,, 
\\ &
d^{ACF} d^{DFG} f^{BEG} \Sigma_{+}^{E} \left(D_{\mu} u^{A}\right) \left(D_{\nu} u^{C}\right) \left(D^{\nu} \Sigma_{+}^{D}\right) \left(D^{\mu} u^{B}\right)\,, 
\\ &
d^{ACF} d^{DFG} f^{BEG} \Sigma_{+}^{E} \left(D_{\mu} u^{A}\right) \left(D_{\nu} u^{B}\right) \left(D^{\nu} \Sigma_{+}^{D}\right) \left(D^{\mu} u^{C}\right)\,, \\ &
d^{BEF} d^{DFG} f^{ACG} \Sigma_{+}^{D} \Sigma_{+}^{E} \left(D_{\nu} u^{C}\right) \left(D^{\mu} u^{B}\right) \left(D_{\mu} D^{\nu} u^{A}\right)\,, 
\\ &
d^{BEF} d^{DFG} f^{ACG} \Sigma_{+}^{E} \left(D_{\mu} u^{A}\right) \left(D_{\nu} u^{C}\right) \left(D^{\nu} \Sigma_{+}^{D}\right) \left(D^{\mu} u^{B}\right)\,, 
\\ &
d^{BEF} d^{DFG} f^{ACG} \Sigma_{+}^{E} \left(D_{\rho} \Sigma_{+}^{D}\right) \left(D_{\mu} u^{A}\right) \left(D_{\nu} u^{B}\right) \left(D_{\lambda} u^{C}\right) \epsilon^{\lambda \mu \nu \rho}\,, 
\\ &
d^{BFG} f^{ACF} f^{DEG} \Sigma_{+}^{E} \left(D_{\mu} u^{A}\right) \left(D_{\nu} u^{C}\right) \left(D^{\nu} \Sigma_{+}^{D}\right) \left(D^{\mu} u^{B}\right)\,, 
\\ &
d^{BFG} f^{ACF} f^{DEG} \Sigma_{+}^{E} \left(D_{\rho} \Sigma_{+}^{D}\right) \left(D_{\mu} u^{A}\right) \left(D_{\nu} u^{B}\right) \left(D_{\lambda} u^{C}\right) \epsilon^{\lambda \mu \nu \rho}\,, 
\\ &
d^{DFG} f^{ACF} f^{BEG} \Sigma_{+}^{D} \Sigma_{+}^{E} \left(D_{\nu} u^{C}\right) \left(D^{\mu} u^{B}\right) \left(D_{\mu} D^{\nu} u^{A}\right)\,, 
\\ &
d^{DFG} f^{ACF} f^{BEG} \Sigma_{+}^{E} \left(D_{\rho} \Sigma_{+}^{D}\right) \left(D_{\mu} u^{A}\right) \left(D_{\nu} u^{B}\right) \left(D_{\lambda} u^{C}\right) \epsilon^{\lambda \mu \nu \rho}.
\end{aligned}
\label{eq:DSigmau_basis}
\end{equation}

If additional fields are introduced, the complexity of the operator also increases. Luckily, for more complicated operators, we can use $ABC4EFT$ to calculate the corresponding kinematic factor $M$ and internal part $T$, as well as the overall operator form.

%


%
\section{Pure Mesonic Operators}
\label{sec:3_OB}
\subsection{CP eigen-operators and Hilbert series}
 Discrete symmetries like the parity and the charge conjugation are important when constructing an effective operator. Just as the ChPT Lagrangian can be divided into four disconnected parts based on their $CP$ eigenvalues, the effective operators can also be categorized into four classes according to their $CP$ relations,
\begin{equation}
    C+P+\,,\quad C+P-\,,\quad C-P+\,,\quad C-P-\,.\quad
\end{equation}
For a building block $\mathcal{O}$, $\mathcal{O} = u_\mu\,,f_\pm\,,\Sigma_\pm$, it transforms as follows under the $C$ and $P$ transformation:
\begin{equation}
\begin{aligned}
    P : \quad & \mathcal{O} \overset{P}{\rightarrow} \eta_{\mathcal{O}}\mathcal{O},\\
   C : \quad & \mathcal{O} \overset{C}{\rightarrow} (-1)^{\mathcal{O}}\mathcal{O}^{T},
\end{aligned}
\end{equation}
here $\eta_{\mathcal{O}}$ and $(-1)^{\mathcal{O}}$ are the intrinsic charge of $\mathcal{O}$ in Table \ref{Table:BuildingBlock-Transformations}. The effective operators we have constructed are composed of these building blocks, which consequently possess different properties under $CP$ transformation. Also we need to point that although $F_{\pm L\mu\nu}$ and $F_{\pm R\mu\nu}$  are used during the operator construction in Young tensor technique, we now use the $CP$ eigenstate notation for $f$ in the following calculation, 
\begin{equation}
    \begin{aligned}
        &f_{\pm\mu\nu} = F_{\pm L\mu\nu}+F_{\pm R\mu\nu}\,, \\
        &\Tilde{f}_{\pm\mu\nu} = F_{\pm R\mu\nu}-F_{\pm L\mu\nu}\,, \\
    \end{aligned}
    \label{eq:fTildef}
\end{equation}
and the definition of $\Tilde{f}_{\pm\mu\nu}$ is in Eq.~\eqref{eq: u-Sigma-f}. Thus, the final result will be composed of $f_{\pm\mu\nu}$ or $\Tilde{f}_{\pm\mu\nu}$.

The results generated by the Young tensor technique can all be converted into the form of trace basis. For example, an effective operator which composed of four building blocks can take the form of $\langle\mathcal{A}\mathcal{B}\mathcal{C}\mathcal{D}\rangle$ or $\langle\mathcal{A}\mathcal{B}\rangle\langle\mathcal{C}\mathcal{D}\rangle$. Based on its $C$ transformation properties, we can divide it into two parts: one has eigenvalue $C$- and the other has eigenvalue $C$+,
\begin{equation}
    \begin{aligned}
        C+ :\quad & \langle\mathcal{A}\mathcal{B}\mathcal{C}\mathcal{D}\rangle + \langle\mathcal{A}\mathcal{D}\mathcal{C}\mathcal{B}\rangle\,,\\
        &\langle\mathcal{A}\mathcal{B}\rangle\langle\mathcal{C}\mathcal{D}\rangle \,,\\
        C- :\quad & \langle\mathcal{A}\mathcal{B}\mathcal{C}\mathcal{D}\rangle - \langle\mathcal{A}\mathcal{D}\mathcal{C}\mathcal{B}\rangle\,,
    \end{aligned}
\end{equation}
where, we have taken into account the trace cyclic property.

Sometimes, we can change an operator's $P$ parity without altering its $C$ parity by simply adding $\epsilon_{\mu\nu\rho\lambda}$.

The complete, independent, and $CP$-satisfying operator we obtained is consistent with the Hilbert series, which is another algebra method that presents the number of independent operators of each type. It takes the form of an infinite series~\cite{Henning:2015alf, Lehman:2015via, Lehman:2015coa, Henning:2015daa, Henning:2017fpj, Marinissen:2020jmb}
\begin{equation}
    HS = 1 + n_1 Q_1 + n_2 Q_2 + \dots\,,
\end{equation}
where $Q_1\,, Q_2\,,\dots$ are different types, and $n_1\,, n_2\,,\dots$ are the numbers of the independent operators of each type respectively.

The Hilbert series can be obtained via the orthogonal relation of the characters of the compact group $G$.
In detail, we suppose there are $N$ fields in the representation $\mathbf{r}$ of the group $G$, and the number of the the irreducible representation $\mathbf{r}'$ composed of these fields is determined by the orthogonal relation
\begin{equation}
    n_{\mathbf{r}'} = \int dz \chi*_{\mathbf{r}'}(z)\left(\chi_\mathbf{r}(z)\right)^N\,,
\end{equation}
where $\chi_\mathbf{r}$ and $\chi_{\mathbf{r}'}$ are the characters of the representations $\mathbf{r} \,, \mathbf{r}'$ respectively and $z$ are the maximal ring variables of the compact group $G$, which is $G = SO(3,1) \times SU(N_f)_L \times SU(N_f)_R$ in the ChPT\footnote{The Lorentz group $SO(3,1)$ algebra is isomorphic to $SU(2) \times SU(2)$.}.

Besides the compact group, discrete symmetries such as $C$ and $P$ can be embedded in the Hilbert series as well~\cite{Graf:2020yxt, Sun:2022aag}. When $C$ and $P$ symmetries are considered, the original group $G$ is extended to 4 disjoint sectors,
\begin{equation}
    G \rightarrow G^{++} \sqcup G^{+-} \sqcup G^{-+} \sqcup G^{--}\,.
\end{equation}
There is one Hilbert series for each sector, 
\begin{equation}
    HS \rightarrow \left\{\begin{array}{l}
HS^{++} \\
HS^{-+} \\
HS^{+-} \\
HS^{--} 
    \end{array}\right. \,,
\end{equation}
and the $C$ and $P$ invariant operators are given by the average of them,
\begin{equation}
    HS^{C+P+} = \frac{1}{4}(HS^{++}+HS^{-+}+HS^{--}+HS^{+-})\,.
\end{equation}

As a complementary check, we adopt the Hilbert series result of the mesonic operators of the ChPT in Ref.~\cite{Graf:2020yxt} as a crosscheck. At the same time, in order to further distinguish $CP$, we have further classified the Hilbert sequence to obtain sequence results for different $CP$ values.

\subsection{Trace basis via Cayley-Hamilton theorem}

Regarding to the internal structure, the invariant tensor basis generated by the Young tensor technique can be converted into the form of trace basis equivalently. When only three fields are in consideration, we can use the definitions of the structural constants from earlier Eq.~\eqref{eq: SU3_Structure_Constant} along with the properties of commutators and anticommutators to convert the results into the form of trace basis.

\begin{equation} 
\begin{aligned}
    f_{A B C}=\frac{1}{4i}Tr([\lambda_A,\lambda_B]\lambda_C) \,, \\
    d_{A B C}=\frac{1}{4}Tr(\{\lambda_A,\lambda_B\}\lambda_C) \,.
\end{aligned}
\end{equation}

When the number of fields is larger than three, the results will then take the form of multiple products of $d$ or $f$. Below, we present the transformation between the invariant tensor basis and the trace basis.
In the case of four fields, we can get,
\begin{equation} \label{eq: ff_fd_dd}
\begin{aligned}
f^{ABE} f^{CDE} &= \frac{1}{2} \left[Tr(\lambda^A \lambda^B \lambda^C \lambda^D )-Tr(\lambda^B \lambda^A \lambda^C \lambda^D )-Tr(\lambda^C \lambda^A \lambda^B \lambda^D )+Tr(\lambda^C \lambda^B \lambda^A \lambda^D )\right],
\\
f^{ABE} d^{CDE} &= \frac{1}{2} \left[Tr(\lambda^A \lambda^B \lambda^C \lambda^D )-Tr(\lambda^B \lambda^A \lambda^C \lambda^D )+Tr(\lambda^C \lambda^A \lambda^B \lambda^D )-Tr(\lambda^C \lambda^B \lambda^A \lambda^D )\right],
\\
d^{ABE} d^{CDE} &= \frac{1}{6} \left[-\delta^{AB} \delta^{CD}+3 Tr(\lambda^A \lambda^B \lambda^C \lambda^D )+3 Tr(\lambda^B \lambda^A \lambda^C \lambda^D )\right. \\
                &\left. \qquad \qquad \qquad \qquad +3 Tr(\lambda^C \lambda^A \lambda^B \lambda^D )+3 Tr(\lambda^C \lambda^B \lambda^A \lambda^D )\right],
\end{aligned}
\end{equation}
and for the case of five fields,
\begin{equation} \label{eq: fff_ddd}
\begin{aligned}
d^{ACF} d^{BFG} d^{DEG} &= \frac{1}{2} \left[-2 \delta^{DE} Tr(\lambda^A \lambda^B \lambda^C )-2 \delta^{DE} Tr(\lambda^B \lambda^A \lambda^C )-2 \delta^{AC} Tr(\lambda^B \lambda^D \lambda^E ) \right.\\
& -2 \delta^{AC} Tr(\lambda^D \lambda^B \lambda^E )+3 Tr(\lambda^A \lambda^C \lambda^B \lambda^D \lambda^E )+3 Tr(\lambda^B \lambda^A \lambda^C \lambda^D \lambda^E ) \\
& +3 Tr(\lambda^B \lambda^C \lambda^A \lambda^D \lambda^E ) +3 Tr(\lambda^C \lambda^A \lambda^B \lambda^D \lambda^E )+3 Tr(\lambda^D \lambda^A \lambda^C \lambda^B \lambda^E ) \\
& \left. +3 Tr(\lambda^D \lambda^B \lambda^A \lambda^C \lambda^E )) +3 Tr(\lambda^D \lambda^B \lambda^C \lambda^A \lambda^E )+3 Tr(\lambda^D \lambda^C \lambda^A \lambda^B \lambda^E )\right],
\\
d^{ACF} d^{BFG} f^{DEG} 
& = \frac{1}{2} \left[-2 \delta^{AC} (Tr(\lambda^B \lambda^D \lambda^E )+2 \delta^{AC} (Tr(\lambda^D \lambda^B \lambda^E )+3 (Tr(\lambda^A \lambda^C \lambda^B \lambda^D \lambda^E )  \right.\\ 
& +3 Tr(\lambda^B \lambda^A \lambda^C \lambda^D \lambda^E )+3 (Tr(\lambda^B \lambda^C \lambda^A \lambda^D \lambda^E )+3 (Tr(\lambda^C \lambda^A \lambda^B \lambda^D \lambda^E )\\ 
& -3 Tr(\lambda^D \lambda^A \lambda^C \lambda^B \lambda^E ))-3 (Tr(\lambda^D \lambda^B \lambda^A \lambda^C \lambda^E )-3 (Tr(\lambda^D \lambda^B \lambda^C \lambda^A \lambda^E )\\ 
&\left.-3 Tr(\lambda^D \lambda^C \lambda^A \lambda^B \lambda^E ) \right],
\\
d^{ACF} f^{BFG} f^{DEG} 
&= \frac{1}{4} \left[-(Tr(\lambda^A \lambda^C \lambda^B \lambda^D \lambda^E )+Tr(\lambda^B \lambda^A \lambda^C \lambda^D \lambda^E )+Tr(\lambda^B \lambda^C \lambda^A \lambda^D \lambda^E ) \right.\\ 
&-Tr(\lambda^C \lambda^A \lambda^B \lambda^D \lambda^E )+Tr(\lambda^D \lambda^A \lambda^C \lambda^B \lambda^E )-Tr(\lambda^D \lambda^B \lambda^A \lambda^C \lambda^E )\\ 
&\left.-Tr(\lambda^D \lambda^B \lambda^C \lambda^A \lambda^E )+Tr(\lambda^D \lambda^C \lambda^A \lambda^B \lambda^E ) \right],
\\
f^{ACF} f^{BFG} f^{DEG} 
&= \frac{1}{4} \left[-(Tr(\lambda^A \lambda^C \lambda^B \lambda^D \lambda^E )+Tr(\lambda^B \lambda^A \lambda^C \lambda^D \lambda^E )-Tr(\lambda^B \lambda^C \lambda^A \lambda^D \lambda^E ) \right.\\
&+Tr(\lambda^C \lambda^A \lambda^B \lambda^D \lambda^E )+Tr(\lambda^D \lambda^A \lambda^C \lambda^B \lambda^E )-Tr(\lambda^D \lambda^B \lambda^A \lambda^C \lambda^E )\\ 
&\left.+Tr(\lambda^D \lambda^B \lambda^C \lambda^A \lambda^E )-Tr(\lambda^D \lambda^C \lambda^A \lambda^B \lambda^E )\right].
\end{aligned}
\end{equation}

Unlike the invariant tensor basis, 
after converting the results into trace basis, we note that there are additional constraints on these trace bases, which can be eliminated by the Cayley-Hamilton theorem.

The Cayley-Hamilton theorem states that: for any square matrix $\mathcal{A}$ with eigenvalue $\lambda$ and corresponding characteristic polynomial $\phi_{\mathcal{A}}(\lambda)$, if $\lambda$ is replaced with $\mathcal{A}$, then $\phi_{\mathcal{A}}(\mathcal{A})$ must equal to zero matrix $\mathcal{O}$:
  \begin{equation}
       det(\lambda I_{n\times n}-\mathcal{A})|_{\lambda=\mathcal{A}} = \mathcal{O} ,
  \end{equation}
where $I_{n\times n}$ is the ${n\times n}$ identity matrix.

Here we directly present the Cayley-Hamilton theorem corresponding to $2\times2$ and $3\times3$ matrices $\mathcal{A}$ respectively:
\begin{equation}
\label{eq:ch-2}
  \mathcal{A}^2 = \langle{\mathcal{A}}\rangle \mathcal{A} + \frac{1}{2}(\langle{\mathcal{A}^2}\rangle-\langle{\mathcal{A}}\rangle^2)\mathbb{I}_{2\times 2}\,,
\end{equation}
\begin{equation}
\label{eq:ch-3}
  \mathcal{A}^3 = \mathcal{A}^2\langle{\mathcal{A}}\rangle-\frac{1}{2}\mathcal{A}(\langle{\mathcal{A}}\rangle^2-\langle{\mathcal{A}^2}\rangle)+\frac{1}{6}(\langle{\mathcal{A}}\rangle^3-3\langle{\mathcal{A}^2}\rangle\langle{\mathcal{A}}\rangle+2\langle{\mathcal{A}^3}\rangle)\mathbb{I}_{3\times 3}\,.
\end{equation}
It should be pointed out that when constructing meson operators for the $SU(2)$ and the $SU(3)$ case, the matrices $\mathcal{A}$ we consider would appear as the adjoint representation of the group, thus they are traceless:
\begin{equation}
  \langle\mathcal{A}\rangle = 0,
\end{equation}
this greatly simplifies Eq.~\eqref{eq:ch-2} and Eq.~\eqref{eq:ch-3}. At this point, we have:
\begin{equation}
\label{eq:ch-2.1}
  \mathcal{A}^2 = \frac{1}{2}(\langle{\mathcal{A}^2}\rangle-\langle{\mathcal{A}}\rangle^2)\mathbb{I}_{2\times 2}\,,
\end{equation}
\begin{equation}
\label{eq:ch-3.1}
  \mathcal{A}^3 = \frac{1}{2}\mathcal{A}\langle\mathcal{A}^2\rangle + \frac{1}{3}\langle\mathcal{A}^3\rangle\mathbb{I}_{3\times 3} .
\end{equation}
More complicated cases can be found in Appendix.~\ref{app:A1_Cayley} for detailed information.

\subsubsection{CP eigen-operators for 4 fields}
In the next two subsubsections, we will convert the results obtained from the previous examples into trace basis form, then combining them to obtain the independent operators, and further categorizing them into four classes based on their $CP$ transformation properties.  Additionally, the correctness of the number of results can be confirmed by comparison with the Hilbert series.

Let's take a look at the situation of the $SU(3) $ $\, D f_{-} u \Sigma_{+}^2$ case. First for the internal structure we convert the invariant tensor basis into the trace basis.  
Some techniques, including the IBP transformation and the properties of symmetry and antisymmetry index are used,
for example:
\begin{equation}
    \langle D^{\mu} \Sigma_{+} \Sigma_{+} f_{-\mu\nu} u^{\nu} \rangle + \langle D^{\mu} f_{-\mu\nu} \Sigma_{+} \Sigma_{+} u^{\nu} \rangle 
    + \langle D^\mu u^\nu f_{-\mu\nu}\Sigma_+\Sigma_+\rangle= 0\,,
\end{equation}
where due to the antisymmetry index of $f$, the third term which the derivative $D$ act on $u$ is equal to $0$, 
\begin{equation}
\langle D^\mu u^\nu f_{-\mu\nu}\Sigma_+\Sigma_+\rangle = 0 \,.
\end{equation}

With the above treatment, according to  Eq.~\eqref{eq: ff_fd_dd}, we can easily change Eq.~\eqref{eq: F1Lphi} and Eq.~\eqref{eq: F1Rphi} to the form below:
\begin{equation}
    \begin{aligned}
        &\langle D_\mu\Sigma_+\Sigma_+\rangle\langle u_\nu F_{-L}^{\mu\nu}\rangle\,,\quad 
        \langle u_\nu\Sigma_+\rangle\langle F_{-L}^{\mu\nu}D_\mu\Sigma_+\rangle\,,\quad 
        \langle u_\nu D_\mu\Sigma_+\rangle\langle\Sigma_+ F_{-L}^{\mu\nu}\rangle ,\\
        &\frac{1}{2} \langle\{F_{-L}^{\mu\nu},u_\nu\}\{D_\mu\Sigma_+,\Sigma_+\}\rangle-\frac{1}{6}\langle F_{-L}^{\mu\nu}u_\nu\rangle\langle D_\nu\Sigma_+\Sigma_+\rangle ,\\
        &\frac{1}{2} \langle\{F_{-L}^{\mu\nu},u_\nu\}[D_\mu\Sigma_+,\Sigma_+]\rangle\,,\quad 
        \frac{1}{2} \langle[F_{-L}^{\mu\nu},u_\nu]\{D_\mu\Sigma_+,\Sigma_+\}\rangle ,\\
        &\frac{1}{2} \langle[F_{-L}^{\mu\nu},u_\nu][D_\mu\Sigma_+,\Sigma_+]\rangle\,,\quad
        \frac{1}{2} \langle[F_{-L}^{\mu\nu},D_\mu\Sigma_+][u_\nu,\Sigma_+]\rangle,
    \end{aligned}
\end{equation}
and:
\begin{equation}
    \begin{aligned}
        &\frac{1}{2} \langle\{u_\nu,D_\mu\Sigma_+\}\{F_{-R}^{\mu\nu},\Sigma_+\}\rangle-\frac{1}{6}\langle u_\nu D_\nu\Sigma_+\rangle\langle\Sigma_+ F_{-R}^{\mu\nu}\rangle\, ,\\
        &\frac{1}{2} \langle\{u_\nu,D_\mu\Sigma_+\}[F_{-R}^{\mu\nu},\Sigma_+]\rangle\,,\quad 
        \frac{1}{2} \langle[u_\nu,D_\mu\Sigma_+]\{F_{-R}^{\mu\nu},\Sigma_+\}\rangle ,\\
        &\frac{1}{2} \langle[u_\nu,D_\mu\Sigma_+][F_{-R}^{\mu\nu},\Sigma_+]\rangle,
    \end{aligned}
\end{equation}
in which we have used a simplified notation: $D_\nu u \sim u_\nu$. 

Next thing to do is to expand the aforementioned equations and then perform a linear combination to obtain the independent terms  with the notation $f$ and $\Tilde{f}$ according to Eq.\ref{eq:fTildef}. We will use the form that clearly states $\epsilon_{\mu\nu\rho\lambda}$.

 Two types of results will be obtained: one is a trace of length four, and the other is the product of two traces, each of them are of length two. Now, we directly present the independent terms: 

\begin{equation}
\begin{aligned}
    & \langle u_\nu\Sigma_+\rangle\langle f_-^{\mu\nu}D_\mu\Sigma_+\rangle\,,
    && \langle f_-^{\mu\nu}\Sigma_+\rangle\langle D_\mu\Sigma_+u_\nu\rangle\,,
    && \langle D_\mu\Sigma_+\Sigma_+ u_\nu f_-^{\mu\nu}\rangle\,,   \\
    & \langle D_\mu\Sigma_+ f_-^{\mu\nu} u_\nu\Sigma_+\rangle\,
    && \langle D_\mu\Sigma_+ u_\nu\Sigma_+f_-^{\mu\nu}\rangle\,,
    && \langle D_\mu\Sigma_+f_-^{\mu\nu}\Sigma_+ u_\nu\rangle\,,\\
    & \langle D_\mu\Sigma_+\Sigma_+f_-^{\mu\nu} u_\nu\rangle\,,
    && \langle D_\mu\Sigma_+ u_\nu f_-^{\mu\nu}\Sigma_+\rangle\,, 
    && 
\end{aligned}
\end{equation}
and:
\begin{equation}
\begin{aligned}
    &\langle D^\mu\Sigma_+f_-^{\nu\rho}\Sigma_+u^\lambda\rangle\epsilon_{\mu\nu\rho\lambda}\,, 
    \langle D^\mu\Sigma_+ u^\nu f_-^{\rho\lambda}\Sigma_+\rangle\epsilon_{\mu\nu\rho\lambda}\,, \\
    &\langle D^\mu\Sigma_+u^\nu f_-^{\rho\lambda}\Sigma_+\rangle\epsilon_{\mu\nu\rho\lambda}\,,
    \langle D^\mu\Sigma_+\Sigma_+f_-^{\nu\rho}u^\lambda \rangle\epsilon_{\mu\nu\rho\lambda}\,.
\end{aligned}
\end{equation}

In the following, we will move on to categorize these terms based on their $CP$ transformation properties. For example, some of them satisfy the following specific $CP$ relations:
\begin{equation}
\begin{aligned}
    &\langle u_\nu\Sigma_+\rangle\langle f_-^{\mu\nu}D_\mu\Sigma_+\rangle\overset{P}{\rightarrow}\langle u_\nu\Sigma_+\rangle\langle f_-^{\mu\nu}D_\mu\Sigma_+\rangle,\\
    &\langle u_\nu\Sigma_+\rangle\langle f_-^{\mu\nu}D_\mu\Sigma_+\rangle\overset{C}{\rightarrow}\langle u_\nu\Sigma_+\rangle\langle f_-^{\mu\nu}D_\mu\Sigma_+\rangle,
\end{aligned}
\end{equation}
in which, obviously this basis transform exactly into itself, while some of them don't transform into itself: 
\begin{equation}
\begin{aligned}
    &\langle D_\mu\Sigma_+\Sigma_+ u_\nu f_-^{\mu\nu}\rangle\overset{P}{\rightarrow}\langle D_\mu\Sigma_+\Sigma_+ u_\nu f_-^{\mu\nu}\rangle,\\
    &\langle D_\mu\Sigma_+\Sigma_+ u_\nu f_-^{\mu\nu}\rangle\overset{C}{\rightarrow}\langle D_\mu\Sigma_+ f_-^{\mu\nu}u_\nu\Sigma_+\rangle,
\end{aligned}
\end{equation}
therefore, such a combination:
\begin{equation}
    \langle D_\mu\Sigma_+\Sigma_+ u_\nu f_-^{\mu\nu}\rangle+\langle D_\mu\Sigma_+ f_-^{\mu\nu}u_\nu\Sigma_+\rangle,
\end{equation}
is $C$+, and we can abbreviate it as,
\begin{equation}
    \langle D_\mu\Sigma_+\Sigma_+ u_\nu f_-^{\mu\nu}\rangle+h.c..
\end{equation}
A similar discussion can be applied to other operators.

Finally, we obtain the complete, independent set of pure meson operators for the type $Df\Sigma_+^2u$ that satisfy the $CP$ relations in Table.~\ref{table: Df_Sigmap_2u}.

\begin{table} [h] 
    \centering
    \begin{tabular}{@{}cc|cc@{}}
    \hline
       &$C+P+$&$C+P-$\\ \hline
         &$\langle u_\nu\Sigma_+\rangle\langle f_-^{\mu\nu}D_\mu\Sigma_+\rangle$
         &$\langle D^\mu\Sigma_+f_-^{\nu\rho}\Sigma_+u^\lambda\rangle\epsilon_{\mu\nu\rho\lambda}$+h.c.\\
         &$\langle f_-^{\mu\nu}\Sigma_+\rangle\langle D_\mu\Sigma_+u_\nu\rangle$&$\langle D^\mu\Sigma_+u^\nu f_-^{\rho\lambda}\Sigma_+\rangle\epsilon_{\mu\nu\rho\lambda}$+h.c.\\ 
         &$\langle D_\mu\Sigma_+\Sigma_+ u_\nu f_-^{\mu\nu}\rangle$+h.c.&\\ 
         &$\langle D_\mu\Sigma_+ u_\nu\Sigma_+f_-^{\mu\nu}\rangle$+h.c.&\\
         &$\langle D_\mu\Sigma_+\Sigma_+f_-^{\mu\nu} u_\nu\rangle$+h.c.&\\ \hline
         &$C-P+$&$C-P-$ \\ \hline 
         &$\langle D_\mu\Sigma_-\Sigma_+ u_\nu f_-^{\mu\nu}\rangle$-h.c. &$\langle D^\mu\Sigma_+f_-^{\nu\rho}\Sigma_+u^\lambda\rangle\epsilon_{\mu\nu\rho\lambda}$-h.c.\\
         &$\langle D_\mu\Sigma_- u_\nu\Sigma_+f_-^{\mu\nu}\rangle$-h.c.&$\langle D^\mu\Sigma_+u^\nu f_-^{\rho\lambda}\Sigma_+\rangle\epsilon_{\mu\nu\rho\lambda}$-h.c.\\
         &$\langle D_\mu\Sigma_-\Sigma_+f_-^{\mu\nu} u_\nu\rangle$-h.c.&\\
         \hline
    \end{tabular}
    \caption{The complete, independent meason operators of the type $Df\Sigma_+^2u$.}
    \label{table: Df_Sigmap_2u}
\end{table}

\subsubsection{CP eigen-operators for 5 fields}
Now, let's take a look at another example with the $SU(3) \, D u^3 \Sigma_{+}^2 $ of 5 fields in Eq.~(\ref{eq:DSigmau_basis}). When it comes to dealing with five fields, we should mention that the Cayley-Hamilton relation are somewhat much more complicated, and it would involve 12 sets of the Cayley-Hamilton relation equations. Fortunately, we can utilize the symmetry of the repeated fields to simplify the Cayley-Hamilton relation equations to 7 sets with regard to $\mathcal{ABCDD}$ in this case. Here ${A}$, ${B}$, and ${C}$ denote distinct fields, while ${DD}$ denotes a pair of identical fields.

For terms that do not include $\epsilon_{\mu\nu\rho\lambda}$ , we directly present the corresponding independent bases after linear combination:
\begin{equation}
    \begin{aligned}
         &\langle u^\mu u_\mu u_\nu\Sigma_+ D^\nu\Sigma_+\rangle\,,\quad\quad \quad
         \langle u^\mu u_\mu D^\nu\Sigma_+\Sigma_+ u_\nu\rangle\,,\quad\quad\quad
         \langle u^\mu u_\mu\Sigma_+ u_\nu D^\nu\Sigma_+\rangle\,,\quad\quad\quad \\
         &\langle u^\mu u_\mu D^\nu\Sigma_+ u_\nu\Sigma_+\rangle\,,\quad\quad\quad
         \langle u^\mu u_\mu u_\nu D^\nu\Sigma_+\Sigma_+\rangle\,,\quad\quad\quad
         \langle u^\mu u_\mu\Sigma_+ D^\nu\Sigma_+u_\nu\rangle\,,\quad\quad\quad \\
         &\langle u^\mu u_\nu u_\mu\Sigma_+ D^\nu\Sigma_+\rangle\,,\quad\quad \quad
         \langle u^\mu u_\nu u_\mu D^\nu\Sigma_+\Sigma_+\rangle\,,\quad\quad \quad
         \langle u^\mu\Sigma_+ u_\mu u_\nu D^\nu\Sigma_+\rangle\,,\quad\quad\quad\\
         &\langle u^\mu\Sigma_+ u_\mu D^\nu\Sigma_+ u_\nu\rangle\,,\quad\quad\quad
         \langle u^\mu D^\nu\Sigma_+ u_\mu\Sigma_+ u_\nu\rangle\,,\quad\quad\quad
         \langle u^\mu D^\nu\Sigma_+ u_\mu u_\nu\Sigma_+\rangle\,,\quad\quad\quad\\
         &\langle u_\mu u^\mu u_\nu\rangle\langle\Sigma_+D^\nu\Sigma_+\rangle\,,\quad\quad\,\,
         \langle u_\mu u^\mu\Sigma_+\rangle\langle D^\nu\Sigma_+u_\nu\rangle\,,\quad\quad\,\,
         \langle u_\mu u^\mu\rangle\langle u^\nu\Sigma_+D^\nu\Sigma_+\rangle\,,\quad\quad\,\,\\
         &\langle u_\mu u^\mu\rangle\langle u^\nu D^\nu\Sigma_+\Sigma_+\rangle\,,\quad\quad\, 
    \end{aligned}
\end{equation}

Regarding another type of anomalous operators containing $\epsilon_{\mu\nu\rho\lambda}$, the situation is even simpler, for there are three repeated fields $u$, also due to the properties of the fully antisymmetric tensor $\epsilon_{\mu\nu\rho\lambda}$, the derivative should not act on $u$ during the IBP transformation. This property will impose more strict restrictions on the form of the trace basis here.

The independent anomalous basis including $\epsilon_{\mu\nu\rho\lambda}$ are:
\begin{equation}
    \begin{aligned}
         &\langle D^\rho\Sigma_+ \Sigma_+u^\lambda u^\mu u^\nu\rangle\epsilon_{\rho\lambda\mu\nu}\,,\quad
         \langle D^\rho\Sigma_+ u^\lambda\Sigma_+ u^\mu u^\nu\rangle\epsilon_{\rho\lambda\mu\nu}\,,\quad
         \langle D^\rho\Sigma_+\Sigma_+\rangle\langle u^\lambda u^\mu u^\nu\rangle\epsilon_{\rho\lambda\mu\nu}. 
    \end{aligned}
\end{equation}

After getting these independent basis, we can further reduce them based on the $CP$ transformation relations to obtain Table.~\ref{tab:my_label_2}. 
\begin{table} [ht] 
    \centering
    \begin{tabular}{cc|cc} \hline
       &$C+P+$&$C+P-$\\ \hline
         &$\langle D^\rho\Sigma_+ \Sigma_+u^\lambda u^\mu u^\nu\rangle\epsilon_{\rho\lambda\mu\nu}$&$\langle u_\mu u^\mu u_\nu\rangle\langle\Sigma_+D^\nu\Sigma_+\rangle$\\
         &$ \langle D^\rho\Sigma_+ u^\lambda\Sigma_+ u^\mu u^\nu\rangle\epsilon_{\rho\lambda\mu\nu}$&$\langle u_\mu u^\mu \Sigma_+\rangle\langle\Sigma_+D^\nu u_\nu\rangle$\\ 
         &&$\langle u_\mu u^\mu\rangle\langle D^\nu\Sigma_+\Sigma_+u^\nu\rangle$+h.c.\\
         &&$\langle D^\nu\Sigma_+u^\mu u^\nu\Sigma_+ u_\mu\rangle$+h.c.\\
         &&$\langle D^\nu\Sigma_+\Sigma_+u^\mu u_\mu u^\nu\rangle$+h.c.\\
         &&$\langle D^\nu\Sigma_+u^\mu u_\mu\Sigma_+ u^\nu\rangle$+h.c.\\
         &&$\langle D^\nu\Sigma_+u^\mu u_\mu u^\nu\Sigma_+\rangle$+h.c.\\ 
         &&$\langle D^\nu\Sigma_+u^\mu u^\nu u_\mu\Sigma_+\rangle$+h.c.\\ 
         &&$\langle D^\nu\Sigma_+u^\mu\Sigma_+ u_\mu u^\nu\rangle$+h.c.\\ \hline
         &$C-P+$&$C-P-$\\ \hline
         &$\langle D^\rho\Sigma_+\Sigma_+\rangle\langle u^\lambda u^\mu u^\nu\rangle\epsilon_{\rho\lambda\mu\nu}$-h.c.&$\langle u_\mu u^\mu\rangle\langle D^\nu\Sigma_+\Sigma_+u^\nu\rangle$-h.c.\\
         &&$\langle D^\nu\Sigma_+u^\mu u^\nu\Sigma_+ u_\mu\rangle$-h.c.\\
         &&$\langle D^\nu\Sigma_+\Sigma_+u^\mu u_\mu u^\nu\rangle$-h.c.\\
         &&$\langle D^\nu\Sigma_+u^\mu u_\mu\Sigma_+ u^\nu\rangle$-h.c.\\
         &&$\langle D^\nu\Sigma_+u^\mu u_\mu u^\nu\Sigma_+\rangle$-h.c.\\
         &&$\langle D^\nu\Sigma_+u^\mu u^\nu u_\mu\Sigma_+\rangle$-h.c.\\
         &&$\langle D^\nu\Sigma_+u^\mu\Sigma_+ u_\mu u^\nu\rangle$-h.c.\\ \hline
    \end{tabular}
    \caption{The complete, independent meason operators of the type $D\Sigma_+^2u^3$.}
    \label{tab:my_label_2}
\end{table}


%



%

%
\subsection{Operator counting}
%

The complete CP-eigen bases of purely mesonic operators up to $\mathcal{O}(p^8)$ that we constructed are consistent with the Hilbert series. For the convenience of reference, we have listed the operator counting for different $CP$ sectors, categorized by their types within the $SU(2)$ and the $SU(3)$ groups at orders $\mathcal{O}(p^4)$, $\mathcal{O}(p^6)$, and $\mathcal{O}(p^8)$ in Tables \ref{tab: su2p4}, \ref{tab: su3p4}, \ref{tab: su2p6}, \ref{tab: su3p6}, \ref{tab: su2p8}, and \ref{tab: su3p8}, respectively.

\label{sec:3_6_HS_OC}
\begin{table}[ht]
\centering
\caption{The number of different $CP$ types of operators in $p^4$ order in the $SU(2)$ case. }
\begin{tabular}{@{}cccccccccccccc@{}}
\toprule
     &$f^2$  &$fu^2$ &$u^4$ &$\Sigma^2$ &$\Sigma u^2$  &All      \\ \midrule
C+P+ &2      &1      &2     &4          &1             &10                               \\
C+P- &2      &1      &0     &2          &1             &6                             \\
C-P+ &1      &1      &0     &0          &0             &2                                     \\
C-P- &1      &1      &0     &0          &0             &2                                      \\ \hline 
All  &6      &4      &2     &6          &2             &20                               \\ \bottomrule
\end{tabular}
\label{tab: su2p4}
\end{table}
\begin{table}[ht]
\centering
\caption{The number of different $CP$ types of operators in $p^4$ order in the $SU(3)$ case. }
\begin{tabular}{@{}cccccccccccccc@{}}
\toprule
     &$f^2$  &$fu^2$ &$u^4$ &$\Sigma^2$ &$\Sigma u^2$  &All      \\ \midrule
C+P+ &2      &1      &3     &4          &2             &12                               \\
C+P- &2      &1      &0     &2          &2             &7                             \\
C-P+ &1      &1      &0     &0          &0             &2                                     \\
C-P- &1      &1      &0     &0          &0             &2                                      \\ \hline 
All  &6      &4      &3     &6          &4             &23                               \\ \bottomrule                                         
\end{tabular}
\label{tab: su3p4}
\end{table}

\begin{table}[ht]
\setlength\tabcolsep{6pt} 
\fontsize{9pt}{11pt}\selectfont
\centering
\caption{The number of different $CP$ types of operators in $p^6$ order in the $SU(2)$ case. }
\begin{tabular}{@{}cccccccccccccc@{}}
\toprule
     &$D^2f^2$  &$f^3$ &$f^2\Sigma$ &$D^2\Sigma^2$ &$\Sigma^3$ &$Df^2u$  &$Df\Sigma u$ &$D\Sigma^2u$ &$D^2fu^2$  &$f^2u^2$       \\ \midrule
C+P+ &2         &2     &6           & 4          & 5         & 2       & 2           &2            &1          &10                                       \\
C+P- &0         &2     &6           & 2          & 5         & 2       & 2           &2            &0          &6                                       \\
C-P+ &0         &2     &2           & 0          & 0         & 2       & 2           &1            &0          &5                                          \\
C-P- &1         &2     &2           & 0          & 0         & 2       & 2           &2            &1          &5                                           \\ \hline 
All  &3         &8     &16          & 6          & 10        & 8       & 8           &7            &2          &26                                          \\ \bottomrule
     &$f\Sigma u^2$&$\Sigma^2u^2$&$D^2\Sigma u^2$  &$Dfu^3$ &$D\Sigma u^3$ &$D^2u^4$    &$fu^4$     &$\Sigma u^4$   &$Du^5$      &$u^6$    &All \\ \midrule
C+P+ &4            &6            &1                &3       &2             &2           &2          &2              &0           &3        &61                \\
C+P- &4            &3            &1                &1       &2             &0           &2          &2              &0           &0        &42               \\
C-P+ &4            &0            &0                &1       &0             &0           &2          &0              &0           &0        &21              \\
C-P- &4            &0            &0                &3       &0             &0           &2          &0              &1           &0        &27
             \\ \hline
All  &16           &9            &2                & 8      &4             &2           &8          &4              &1           &3        &151           \\ \bottomrule
\end{tabular}\
\label{tab: su2p6}
\end{table}
\begin{table}[ht]
\setlength\tabcolsep{6pt} 
\fontsize{9pt}{11pt}\selectfont
\centering
\caption{The number of different $CP$ types of operators in $p^6$ order in the $SU(3)$ case. }
\begin{tabular}{@{}cccccccccccccc@{}}
\toprule
     &$D^2f^2$ &$f^3$ &$f^2\Sigma$ &$D^2\Sigma^2$ &$\Sigma^3$ &$Df^2u$  &$Df\Sigma u$ &$D\Sigma^2u$ &$D^2fu^2$  &$f^2u^2$    \\ \midrule
C+P+ & 2       & 2    &10          & 4            & 7         & 4       & 3           &3             &1         &20                                       \\
C+P- & 0       & 2    &10          & 2            & 7         & 4       & 3           &2             &0         &14                                       \\
C-P+ & 0       & 2    &4           & 0            & 0         & 4       & 3           &1             &0         &10                                          \\
C-P- & 1       & 2    &4           & 0            & 0         & 4       & 3           &2             &1         &12                                       \\ \hline 
All  & 3       & 8    &36          & 6            & 14        & 16      & 12          &8             &2         &56                                          \\ \bottomrule
     &$f\Sigma u^2$  &$\Sigma^2u^2$ &$D^2\Sigma u^2$ &$Dfu^3$ &$D\Sigma u^3$ &$D^2u^4$    &$fu^4$     &$\Sigma u^4$   &$Du^5$      &$u^6$     &All \\ \midrule
C+P+ &10             &10            &2                &7      &4             &3           &8          &8              &1           &8         &117             \\
C+P- &10             &6             &2                &4      &4             &0           &8          &8              &2           &1         &89              \\
C-P+ &10             &0             &0                &4      &1             &0           &8          &1              &0           &0         &48                \\
C-P- &10             &1             &0                &7      &1             &0           &8          &1              &2           &2         &61                \\ \hline
All  &40             &17            &4                &22     &10            &3           &32         &18             &5           &11        &315               \\ \bottomrule
\end{tabular}
\label{tab: su3p6}
\end{table}
\newpage

\begin{table}[ht]
\setlength\tabcolsep{5.5pt} 
\fontsize{9pt}{11pt}\selectfont
\centering
\caption{The number of different $CP$ types of operators in $p^8$ order in the $SU(2)$ case. }
\begin{tabular}{@{}cccccccccccc@{}}
\toprule
     &$D^4f^2$ &$f^4$  &$D^2f^3$ &$D^2f^2\Sigma$ &$f^3\Sigma$ &$f^2\Sigma^2$  &$D^4\Sigma^2$ &$D^2f\Sigma^2$ &$D^2\Sigma^3$      \\ \midrule
C+P+ & 2       & 18    &6        & 17            & 6         & 22             & 4            &4              &10                          \\
C+P- & 0       & 11    &6        & 17            & 6         & 22             & 2            &3              &10                           \\
C-P+ & 0       & 8     &6        & 11            & 6         & 10             & 0            &3              &1                    \\
C-P- & 1       & 8     &6        & 11            & 6         & 10             & 0            &4              &1                        \\ \hline 
All  & 3       & 45    &24       & 56            & 24        & 64             & 6            &14             &22                  \\ \bottomrule
       &$\Sigma^4$ &$Df^3 u$   &$D^3f^2u$  &$Df^2\Sigma u$   &$D^3f\Sigma u$  &$D\Sigma^3 u$  &$D^3\Sigma^2 u$  &$Df\Sigma^2 u$ &$D^4fu^2$            \\ \midrule
C+P+   &12         &28         &4          &40               &4               &8              &4                &18             &1                   \\
C+P-   &8          &26         &3          &40               &4               &8              &4                &16             &0 \\
C-P+   &0          &26         &2          &40               &4               &4              &2                &16             &0\\
C-P-   &0          &28         &4          &40               &4               &4              &2                &18             &1     \\ \hline
All    &20         &108        &13         &160              &16              &24             &12               &68             &2\\ \bottomrule
     &$f^3u^2$ &$D^2f\Sigma u^2$&$f^2\Sigma u^2$   &$D^2\Sigma^2u^2$  &$f\Sigma^2u^2$   &$\Sigma^3 u^2$ &$D^2f^2u^2$ &$D^4\Sigma u^2$ &$D^3fu^3$   \\ \midrule
C+P+ &24       &22              &34                &20                &14               &8              &42          &1               &6                 \\
C+P- &24       &22              &34                &13                &14               &8              &28          &1               &2                 \\
C-P+ &24       &22              &24                &2                 &14               &1              &25          &0               &2                \\
C-P- &24       &22              &24                &2                 &14               &1              &29          &0               &6                \\ \hline
All  &96       &88              &116               &37                &56               &18             &124         &2               &16                \\ \bottomrule
      &$Df^2 u^3$&$D^3\Sigma u^3$ &$Df\Sigma u^3$ &$D\Sigma^2u^3$&$D^4 u^4$   &$f^2u^4$  &$D^2fu^4$   &$f\Sigma u^4$  &$\Sigma^2 u^4$     \\ \midrule
C+P+  &33        &4               &30             &8             &3           &28        &14          &10             &12                   \\
C+P-  &33        &4               &30             &8             &0           &18        &11          &10             &6                        \\
C-P+  &30        &1               &30             &6             &0           &16        &11          &10             &0                      \\
C-P-  &32        &1               &30             &8             &0           &16        &14          &10             &0                      \\ \hline
All   &128       &10              &120            &30            &3           &78        &50          &40             &18                      \\ \bottomrule
       &$D^2\Sigma u^4$&$Dfu^5$ &$D\Sigma u^5$   &$D^3u^5$   &$D^2u^6$ &$f u^6$   &$\Sigma u^6$  &$Du^7$     &$u^8$    &All\\ \midrule
C+P+   &7              &12      &7               &0          &9        &4         &3             &0          &4        &567                     \\
C+P-   &7              &8       &7               &0          &2        &4         &3             &0          &0        &483                      \\
C-P+   &5              &8       &2               &0          &0        &4         &0             &0          &0        &376                     \\
C-P-   &5              &12      &2               &2          &0        &4         &0             &2          &0        &408                        \\ \hline
All    &24             &40      &18              &2          &11       &12        &6             &2          &4       &1834                       \\ \bottomrule
\end{tabular}
\label{tab: su2p8}
\end{table}



\begin{table}[ht]
\setlength\tabcolsep{5.5pt} 
\fontsize{9pt}{11pt}\selectfont
\centering
\caption{The number of different $CP$ types of operators in $p^8$ order in the $SU(3)$ case. }
\begin{tabular}{@{}ccccccccccccc@{}}
\toprule
     &$D^4f^2$ &$f^4$  &$D^2f^3$ &$D^2f^2\Sigma$ &$f^3\Sigma$ &$f^2\Sigma^2$  &$D^4\Sigma^2$ &$D^2f\Sigma^2$ &$D^2\Sigma^3$        \\ \midrule
C+P+ & 2       &32     &8        &27             &14          &41             &4             &5              &13\\
C+P- & 0       &21     &8        &27             &14          &41             &2             &3              &13\\
C-P+ & 0       &14     &8        &18             &14          &20             &0             &3              &1\\
C-P- & 1       &15     &8        &18             &14          &20             &0             &5              &1\\ \hline 
All  & 3       &82     &32       &90             &56          &122            &6             &16             &28\\ \bottomrule
     &$\Sigma^4$&$Df^3 u$   &$D^3f^2u$  &$Df^2\Sigma u$   &$D^3f\Sigma u$  &$D\Sigma^3 u$  &$D^3\Sigma^2 u$  &$Df\Sigma^2 u$ &$D^4fu^2$   \\ \midrule
C+P+ &17        &77         &8          &100              &6               &15             &6                &40             &1\\
C+P- &12        &73         &7          &100              &6               &15             &6                &37             &0\\
C-P+ &0         &73         &6          &100              &6               &8              &2                &37             &0\\
C-P- &0         &77         &7          &100              &6               &8              &2                &40             &1\\ \hline 
All  &29        &300        &28         &400              &24              &46             &16               &154            &2\\ \bottomrule
     &$D^2f^2u^2$ &$f^3u^2$&$D^2f\Sigma u^2$&$f^2\Sigma u^2$   &$D^2\Sigma^2u^2$  &$f\Sigma^2u^2$   &$\Sigma^3 u^2$  &$D^4\Sigma u^2$  &$D^3fu^3$        \\ \midrule
C+P+ &91          &107     &55              &138               &36                &51               &23              &2                &14  \\
C+P- &71          &107     &55              &138               &26                &51               &23              &2                &10\\
C-P+ &61          &107      &55              &110               &6                 &51               &3               &0                &10\\
C-P- &69          &107     &55              &110               &8                 &51               &3               &0                &14\\ \hline 
All  &292         &428     &220             &496               &76                &204              &52              &4                &48\\ \bottomrule
     &$Df^2 u^3$ &$D^3\Sigma u^3$   &$Df\Sigma u^3$&$D\Sigma^2u^3$&$D^4 u^4$   &$f^2u^4$  &$D^2fu^4$   &$f\Sigma u^4$  &$\Sigma^2 u^4$        \\ \midrule
C+P+ &181        &8                 &133           &36            &5           &165       &58          &88             &43\\
C+P- &179        &8                 &133           &34            &0           &137       &60          &88             &30\\
C-P+ &171        &4                 &133           &22            &0           &115       &60          &88             &7\\
C-P- &173        &4                 &133           &25            &0           &129       &58          &88             &12\\ \hline 
All  &704        &24                &532           &117           &5           &546       &236         &352            &92\\ \bottomrule
    &$D^2\Sigma u^4$  &$Dfu^5$ &$D\Sigma u^5$   &$D^3u^5$&$D^2u^6$ &$f u^6$   &$\Sigma u^6$  &$Du^7$     &$u^8$    &All\\ \midrule
C+P+&35               &99      &37              &3       &31       &45        &26            &13         &20       &1959\\
C+P-&35               &88      &37              &7       &11       &45        &26            &19         &4        &1809\\
C-P+&16               &88      &30              &1       &9        &45        &9             &9          &0        &1520\\
C-P-&16               &99      &30              &5       &9        &45        &9             &13         &6        &1594\\ \hline
All &102              &374     &134             &16      &60       &180       &70            &54         &30       &6882\\ \bottomrule
\end{tabular}
\label{tab: su3p8}
\end{table}



%
\section{Conclusion}
\label{sec:4_Conclusions}




We constructed the complete and independent operator sets including external sources up to $\mathcal{O}(p^8)$ in the purely mesonic sector of the ChPT for both the $SU(2)$ and $SU(3)$ cases. They are classified by different $CP$ eigenvalues using the trace basis. For the $SU(2)$ case, we obtained totally 640 $C$+$P$+ operators, 532 $C$+$P$- operators, 399 $C$-$P$+ operators, and 437 $C$-$P$- operators, in which the numbers are 567, 483, 376, and 408 for the $\mathcal{O}(p^8)$ ones,
while for the $SU(3)$ case, the total numbers of different $CP$ eigenvalues are 2090, 1906, 1570, and 1657, and the $\mathcal{O}(p^8)$ operator numbers are 1959, 1809, 1520, and 1594 respectively.

We adopted the Young tensor technique as an efficient method to construct effective operators with certain extensions and advantages compared with other literatures:
\begin{itemize}
    \item Compared with traditional operator construction in  Ref.~\cite{Fearing:1994ga,Bijnens:1999hw,Bijnens:1999sh,Ebertshauser:2001nj,Bijnens:2023hyv}, the Lorentz structures here are obtained via on-shell amplitudes from the SSYTs of the primary Young diagrams, which eliminates the all the redundancies automatically. And we obtain the $\mathcal{O}(p^8)$ operator set with other CP eigenvalues for the first time.
    \item Compared with the on-shell amplitude construction of purely mesonic operators~\cite{Low:2022iim}, we extended the Young tensor technique to manage the amplitude with both the on-shell Goldstone boson and the off-shell external sources~\cite{Ren:2022tvi}, and thus obtained the on-shell amplitude basis including the external sources.
    \item Applying the Young tensor technique to internal symmetries, one constructs the invariant tensor basis. However, for the $SU(N)$ adjoint representations, the $CP$ eigenstates prefer the trace basis. We have shown the equivalence between two bases and the transformation between bases is explicitly written.
    
\end{itemize}

Finally, the high-dimension operators obtained in this paper would have various applications. Firstly, the $\mathcal{O}(p^8)$ operators determine the exact low-energy coefficients of the ChPT at next-to-next-to-next-to-leading order~\cite{Bijnens:2017wba}. Secondly, the operators here including all $CP$ eigenvalues can be used to probe $CP$-violation in pure mesonic interactions such as the decay of the $\eta\,, \eta'$~\cite{Akdag:2022sbn} and the calculation of non-vanishing EDM~\cite{Engel:2013lsa,deVries:2011an,Bsaisou:2014oka}. The Young tensor technique is so general and can be used for other sectors such as the meson-baryon sector of the ChPT~\cite{SCQ}, and so on. 


\acknowledgments

We would like to express our gratitude to Chuan-Qiang Song for the valuable assistance provided and for the insightful discussions throughout the course of this work. This work is supported by the National Science Foundation of China under Grants No. 12347105,
No. 12375099 and No. 12047503, and the National Key Research and Development Program of China
Grant No. 2020YFC2201501, No. 2021YFA0718304.

\bibliographystyle{JHEP}
\bibliography{biblio.bib}

\providecommand{\href}[2]{#2}\begingroup\raggedright\begin{thebibliography}{10}

\bibitem{Weinberg:1968de}
S.~Weinberg, \emph{{Nonlinear realizations of chiral symmetry}},
  \href{https://doi.org/10.1103/PhysRev.166.1568}{\emph{Phys. Rev.} {\bfseries
  166} (1968) 1568}.

\bibitem{Weinberg:1978kz}
S.~Weinberg, \emph{{Phenomenological Lagrangians}},
  \href{https://doi.org/10.1016/0378-4371(79)90223-1}{\emph{Physica A}
  {\bfseries 96} (1979) 327}.

\bibitem{Gasser:1983yg}
J.~Gasser and H.~Leutwyler, \emph{{Chiral Perturbation Theory to One Loop}},
  \href{https://doi.org/10.1016/0003-4916(84)90242-2}{\emph{Annals Phys.}
  {\bfseries 158} (1984) 142}.

\bibitem{Gasser:1984gg}
J.~Gasser and H.~Leutwyler, \emph{{Chiral Perturbation Theory: Expansions in
  the Mass of the Strange Quark}},
  \href{https://doi.org/10.1016/0550-3213(85)90492-4}{\emph{Nucl. Phys. B}
  {\bfseries 250} (1985) 465}.

\bibitem{Gasser:1987rb}
J.~Gasser, M.E.~Sainio and A.~Svarc, \emph{{Nucleons with chiral loops}},
  \href{https://doi.org/10.1016/0550-3213(88)90108-3}{\emph{Nucl. Phys. B}
  {\bfseries 307} (1988) 779}.

\bibitem{Coleman:1969sm}
S.R.~Coleman, J.~Wess and B.~Zumino, \emph{{Structure of phenomenological
  Lagrangians. 1.}},
  \href{https://doi.org/10.1103/PhysRev.177.2239}{\emph{Phys. Rev.} {\bfseries
  177} (1969) 2239}.

\bibitem{Callan:1969sn}
C.G.~Callan, Jr., S.R.~Coleman, J.~Wess and B.~Zumino, \emph{{Structure of
  phenomenological Lagrangians. 2.}},
  \href{https://doi.org/10.1103/PhysRev.177.2247}{\emph{Phys. Rev.} {\bfseries
  177} (1969) 2247}.

\bibitem{scherer2012primer}
S.~Scherer, \emph{A primer for chiral perturbation theory}, Springer (2012).

\bibitem{Wess:1971yu}
J.~Wess and B.~Zumino, \emph{{Consequences of anomalous Ward identities}},
  \href{https://doi.org/10.1016/0370-2693(71)90582-X}{\emph{Phys. Lett. B}
  {\bfseries 37} (1971) 95}.

\bibitem{Witten:1983tw}
E.~Witten, \emph{{Global Aspects of Current Algebra}},
  \href{https://doi.org/10.1016/0550-3213(83)90063-9}{\emph{Nucl. Phys. B}
  {\bfseries 223} (1983) 422}.

\bibitem{Fearing:1994ga}
H.W.~Fearing and S.~Scherer, \emph{{Extension of the chiral perturbation theory
  meson Lagrangian to order p(6)}},
  \href{https://doi.org/10.1103/PhysRevD.53.315}{\emph{Phys. Rev. D} {\bfseries
  53} (1996) 315} [\href{https://arxiv.org/abs/hep-ph/9408346}{{\ttfamily
  hep-ph/9408346}}].

\bibitem{Bijnens:1999hw}
J.~Bijnens, G.~Colangelo and G.~Ecker, \emph{{Renormalization of chiral
  perturbation theory to order p**6}},
  \href{https://doi.org/10.1006/aphy.1999.5982}{\emph{Annals Phys.} {\bfseries
  280} (2000) 100} [\href{https://arxiv.org/abs/hep-ph/9907333}{{\ttfamily
  hep-ph/9907333}}].

\bibitem{Bijnens:1999sh}
J.~Bijnens, G.~Colangelo and G.~Ecker, \emph{{The Mesonic chiral Lagrangian of
  order p**6}},
  \href{https://doi.org/10.1088/1126-6708/1999/02/020}{\emph{JHEP} {\bfseries
  02} (1999) 020} [\href{https://arxiv.org/abs/hep-ph/9902437}{{\ttfamily
  hep-ph/9902437}}].

\bibitem{Ebertshauser:2001nj}
T.~Ebertshauser, H.W.~Fearing and S.~Scherer, \emph{{The Anomalous chiral
  perturbation theory meson Lagrangian to order p**6 revisited}},
  \href{https://doi.org/10.1103/PhysRevD.65.054033}{\emph{Phys. Rev. D}
  {\bfseries 65} (2002) 054033}
  [\href{https://arxiv.org/abs/hep-ph/0110261}{{\ttfamily hep-ph/0110261}}].

\bibitem{Bijnens:2001bb}
J.~Bijnens, L.~Girlanda and P.~Talavera, \emph{{The Anomalous chiral Lagrangian
  of order p**6}}, \href{https://doi.org/10.1007/s100520100887}{\emph{Eur.
  Phys. J. C} {\bfseries 23} (2002) 539}
  [\href{https://arxiv.org/abs/hep-ph/0110400}{{\ttfamily hep-ph/0110400}}].

\bibitem{Bijnens:2017wba}
J.~Bijnens and N.~Hermansson~Truedsson, \emph{{The Pion Mass and Decay Constant
  at Three Loops in Two-Flavour Chiral Perturbation Theory}},
  \href{https://doi.org/10.1007/JHEP11(2017)181}{\emph{JHEP} {\bfseries 11}
  (2017) 181} [\href{https://arxiv.org/abs/1710.01901}{{\ttfamily
  1710.01901}}].

\bibitem{Bijnens:2018lez}
J.~Bijnens, N.~Hermansson-Truedsson and S.~Wang, \emph{{The order p$^{8}$
  mesonic chiral Lagrangian}},
  \href{https://doi.org/10.1007/JHEP01(2019)102}{\emph{JHEP} {\bfseries 01}
  (2019) 102} [\href{https://arxiv.org/abs/1810.06834}{{\ttfamily
  1810.06834}}].

\bibitem{Hermansson-Truedsson:2020rtj}
N.~Hermansson-Truedsson, \emph{{Chiral Perturbation Theory at NNNLO}},
  \href{https://doi.org/10.3390/sym12081262}{\emph{Symmetry} {\bfseries 12}
  (2020) 1262} [\href{https://arxiv.org/abs/2006.01430}{{\ttfamily
  2006.01430}}].

\bibitem{Bijnens:2023hyv}
J.~Bijnens, N.~Hermansson-Truedsson and J.~Ruiz-Vidal, \emph{{The anomalous
  chiral Lagrangian at order p$^{8}$}},
  \href{https://doi.org/10.1007/JHEP01(2024)009}{\emph{JHEP} {\bfseries 01}
  (2024) 009} [\href{https://arxiv.org/abs/2310.20547}{{\ttfamily
  2310.20547}}].

\bibitem{Akdag:2022sbn}
H.~Akdag, B.~Kubis and A.~Wirzba, \emph{{C and CP violation in effective field
  theories}}, \href{https://doi.org/10.1007/JHEP06(2023)154}{\emph{JHEP}
  {\bfseries 06} (2023) 154}
  [\href{https://arxiv.org/abs/2212.07794}{{\ttfamily 2212.07794}}].

\bibitem{Engel:2013lsa}
J.~Engel, M.J.~Ramsey-Musolf and U.~van Kolck, \emph{{Electric Dipole Moments
  of Nucleons, Nuclei, and Atoms: The Standard Model and Beyond}},
  \href{https://doi.org/10.1016/j.ppnp.2013.03.003}{\emph{Prog. Part. Nucl.
  Phys.} {\bfseries 71} (2013) 21}
  [\href{https://arxiv.org/abs/1303.2371}{{\ttfamily 1303.2371}}].

\bibitem{Bsaisou:2014oka}
J.~Bsaisou, U.-G.~Mei\ss{}ner, A.~Nogga and A.~Wirzba, \emph{{P- and
  T-Violating Lagrangians in Chiral Effective Field Theory and Nuclear Electric
  Dipole Moments}},
  \href{https://doi.org/10.1016/j.aop.2015.04.031}{\emph{Annals Phys.}
  {\bfseries 359} (2015) 317}
  [\href{https://arxiv.org/abs/1412.5471}{{\ttfamily 1412.5471}}].

\bibitem{deVries:2011an}
J.~de~Vries, R.~Higa, C.P.~Liu, E.~Mereghetti, I.~Stetcu, R.G.E.~Timmermans
  et~al., \emph{{Electric Dipole Moments of Light Nuclei From Chiral Effective
  Field Theory}}, \href{https://doi.org/10.1103/PhysRevC.84.065501}{\emph{Phys.
  Rev. C} {\bfseries 84} (2011) 065501}
  [\href{https://arxiv.org/abs/1109.3604}{{\ttfamily 1109.3604}}].

\bibitem{Li:2020gnx}
H.-L.~Li, Z.~Ren, J.~Shu, M.-L.~Xiao, J.-H.~Yu and Y.-H.~Zheng, \emph{{Complete
  set of dimension-eight operators in the standard model effective field
  theory}}, \href{https://doi.org/10.1103/PhysRevD.104.015026}{\emph{Phys. Rev.
  D} {\bfseries 104} (2021) 015026}
  [\href{https://arxiv.org/abs/2005.00008}{{\ttfamily 2005.00008}}].

\bibitem{Li:2020xlh}
H.-L.~Li, Z.~Ren, M.-L.~Xiao, J.-H.~Yu and Y.-H.~Zheng, \emph{{Complete set of
  dimension-nine operators in the standard model effective field theory}},
  \href{https://doi.org/10.1103/PhysRevD.104.015025}{\emph{Phys. Rev. D}
  {\bfseries 104} (2021) 015025}
  [\href{https://arxiv.org/abs/2007.07899}{{\ttfamily 2007.07899}}].

\bibitem{Li:2022tec}
H.-L.~Li, Z.~Ren, M.-L.~Xiao, J.-H.~Yu and Y.-H.~Zheng, \emph{{Operators for
  generic effective field theory at any dimension: on-shell amplitude basis
  construction}}, \href{https://doi.org/10.1007/JHEP04(2022)140}{\emph{JHEP}
  {\bfseries 04} (2022) 140}
  [\href{https://arxiv.org/abs/2201.04639}{{\ttfamily 2201.04639}}].

\bibitem{Ren:2022tvi}
Z.~Ren and J.-H.~Yu, \emph{{A complete set of the dimension-8
  Green\textquoteright{}s basis operators in the Standard Model effective field
  theory}}, \href{https://doi.org/10.1007/JHEP02(2024)134}{\emph{JHEP}
  {\bfseries 02} (2024) 134}
  [\href{https://arxiv.org/abs/2211.01420}{{\ttfamily 2211.01420}}].

\bibitem{Li:2020tsi}
H.-L.~Li, Z.~Ren, M.-L.~Xiao, J.-H.~Yu and Y.-H.~Zheng, \emph{{Low energy
  effective field theory operator basis at d \ensuremath{\leq} 9}},
  \href{https://doi.org/10.1007/JHEP06(2021)138}{\emph{JHEP} {\bfseries 06}
  (2021) 138} [\href{https://arxiv.org/abs/2012.09188}{{\ttfamily
  2012.09188}}].

\bibitem{Sun:2022aag}
H.~Sun, Y.-N.~Wang and J.-H.~Yu, \emph{{Hilbert Series and Operator Counting on
  the Higgs Effective Field Theory}},
  \href{https://arxiv.org/abs/2211.11598}{{\ttfamily 2211.11598}}.

\bibitem{Sun:2022snw}
H.~Sun, M.-L.~Xiao and J.-H.~Yu, \emph{{Complete NNLO operator bases in Higgs
  effective field theory}},
  \href{https://doi.org/10.1007/JHEP04(2023)086}{\emph{JHEP} {\bfseries 04}
  (2023) 086} [\href{https://arxiv.org/abs/2210.14939}{{\ttfamily
  2210.14939}}].

\bibitem{Sun:2022ssa}
H.~Sun, M.-L.~Xiao and J.-H.~Yu, \emph{{Complete NLO operators in the Higgs
  effective field theory}},
  \href{https://doi.org/10.1007/JHEP05(2023)043}{\emph{JHEP} {\bfseries 05}
  (2023) 043} [\href{https://arxiv.org/abs/2206.07722}{{\ttfamily
  2206.07722}}].

\bibitem{Li:2023wdz}
H.-L.~Li, Z.~Ren, M.-L.~Xiao, J.-H.~Yu and Y.-H.~Zheng, \emph{{On-shell
  operator construction in the effective field theory of gravity}},
  \href{https://doi.org/10.1007/JHEP10(2023)019}{\emph{JHEP} {\bfseries 10}
  (2023) 019} [\href{https://arxiv.org/abs/2305.10481}{{\ttfamily
  2305.10481}}].

\bibitem{Li:2021tsq}
H.-L.~Li, Z.~Ren, M.-L.~Xiao, J.-H.~Yu and Y.-H.~Zheng, \emph{{Operator bases
  in effective field theories with sterile neutrinos: d \ensuremath{\leq} 9}},
  \href{https://doi.org/10.1007/JHEP11(2021)003}{\emph{JHEP} {\bfseries 11}
  (2021) 003} [\href{https://arxiv.org/abs/2105.09329}{{\ttfamily
  2105.09329}}].

\bibitem{Song:2023jqm}
H.~Song, H.~Sun and J.-H.~Yu, \emph{{Complete EFT Operator Bases for Dark
  Matter and Weakly-Interacting Light Particle}},
  \href{https://arxiv.org/abs/2306.05999}{{\ttfamily 2306.05999}}.

\bibitem{Song:2023lxf}
H.~Song, H.~Sun and J.-H.~Yu, \emph{{Effective field theories of axion, ALP and
  dark photon}}, \href{https://doi.org/10.1007/JHEP01(2024)161}{\emph{JHEP}
  {\bfseries 01} (2024) 161}
  [\href{https://arxiv.org/abs/2305.16770}{{\ttfamily 2305.16770}}].

\bibitem{Low:2022iim}
I.~Low, J.~Shu, M.-L.~Xiao and Y.-H.~Zheng, \emph{{Amplitude/operator basis in
  chiral perturbation theory}},
  \href{https://doi.org/10.1007/JHEP01(2023)031}{\emph{JHEP} {\bfseries 01}
  (2023) 031} [\href{https://arxiv.org/abs/2209.00198}{{\ttfamily
  2209.00198}}].

\bibitem{Cheung:2014dqa}
C.~Cheung, K.~Kampf, J.~Novotny and J.~Trnka, \emph{{Effective Field Theories
  from Soft Limits of Scattering Amplitudes}},
  \href{https://doi.org/10.1103/PhysRevLett.114.221602}{\emph{Phys. Rev. Lett.}
  {\bfseries 114} (2015) 221602}
  [\href{https://arxiv.org/abs/1412.4095}{{\ttfamily 1412.4095}}].

\bibitem{Cheung:2015ota}
C.~Cheung, K.~Kampf, J.~Novotny, C.-H.~Shen and J.~Trnka, \emph{{On-Shell
  Recursion Relations for Effective Field Theories}},
  \href{https://doi.org/10.1103/PhysRevLett.116.041601}{\emph{Phys. Rev. Lett.}
  {\bfseries 116} (2016) 041601}
  [\href{https://arxiv.org/abs/1509.03309}{{\ttfamily 1509.03309}}].

\bibitem{Low:2019ynd}
I.~Low and Z.~Yin, \emph{{Soft Bootstrap and Effective Field Theories}},
  \href{https://doi.org/10.1007/JHEP11(2019)078}{\emph{JHEP} {\bfseries 11}
  (2019) 078} [\href{https://arxiv.org/abs/1904.12859}{{\ttfamily
  1904.12859}}].

\bibitem{Dai:2020cpk}
L.~Dai, I.~Low, T.~Mehen and A.~Mohapatra, \emph{{Operator Counting and Soft
  Blocks in Chiral Perturbation Theory}},
  \href{https://doi.org/10.1103/PhysRevD.102.116011}{\emph{Phys. Rev. D}
  {\bfseries 102} (2020) 116011}
  [\href{https://arxiv.org/abs/2009.01819}{{\ttfamily 2009.01819}}].

\bibitem{Kampf:2021jvf}
K.~Kampf, \emph{{The ChPT: top-down and bottom-up}},
  \href{https://doi.org/10.1007/JHEP12(2021)140}{\emph{JHEP} {\bfseries 12}
  (2021) 140} [\href{https://arxiv.org/abs/2109.11574}{{\ttfamily
  2109.11574}}].

\bibitem{Lehman:2015via}
L.~Lehman and A.~Martin, \emph{{Hilbert Series for Constructing Lagrangians:
  expanding the phenomenologist's toolbox}},
  \href{https://doi.org/10.1103/PhysRevD.91.105014}{\emph{Phys. Rev. D}
  {\bfseries 91} (2015) 105014}
  [\href{https://arxiv.org/abs/1503.07537}{{\ttfamily 1503.07537}}].

\bibitem{Henning:2015alf}
B.~Henning, X.~Lu, T.~Melia and H.~Murayama, \emph{{2, 84, 30, 993, 560, 15456,
  11962, 261485, ...: Higher dimension operators in the SM EFT}},
  \href{https://doi.org/10.1007/JHEP08(2017)016}{\emph{JHEP} {\bfseries 08}
  (2017) 016} [\href{https://arxiv.org/abs/1512.03433}{{\ttfamily
  1512.03433}}].

\bibitem{Henning:2015daa}
B.~Henning, X.~Lu, T.~Melia and H.~Murayama, \emph{{Hilbert series and operator
  bases with derivatives in effective field theories}},
  \href{https://doi.org/10.1007/s00220-015-2518-2}{\emph{Commun. Math. Phys.}
  {\bfseries 347} (2016) 363}
  [\href{https://arxiv.org/abs/1507.07240}{{\ttfamily 1507.07240}}].

\bibitem{Marinissen:2020jmb}
C.B.~Marinissen, R.~Rahn and W.J.~Waalewijn, \emph{{..., 83106786, 114382724,
  1509048322, 2343463290, 27410087742, ... efficient Hilbert series for
  effective theories}},
  \href{https://doi.org/10.1016/j.physletb.2020.135632}{\emph{Phys. Lett. B}
  {\bfseries 808} (2020) 135632}
  [\href{https://arxiv.org/abs/2004.09521}{{\ttfamily 2004.09521}}].

\bibitem{Graf:2020yxt}
L.~Graf, B.~Henning, X.~Lu, T.~Melia and H.~Murayama, \emph{{2, 12, 117, 1959,
  45171, 1170086, \textellipsis{}: a Hilbert series for the QCD chiral
  Lagrangian}}, \href{https://doi.org/10.1007/JHEP01(2021)142}{\emph{JHEP}
  {\bfseries 01} (2021) 142}
  [\href{https://arxiv.org/abs/2009.01239}{{\ttfamily 2009.01239}}].

\bibitem{Bijnens:2022zqo}
J.~Bijnens, S.B.~Gudnason, J.~Yu and T.~Zhang, \emph{{Hilbert series and
  higher-order Lagrangians for the O(N) model}},
  \href{https://doi.org/10.1007/JHEP05(2023)061}{\emph{JHEP} {\bfseries 05}
  (2023) 061} [\href{https://arxiv.org/abs/2212.07901}{{\ttfamily
  2212.07901}}].

\bibitem{Henning:2019enq}
B.~Henning and T.~Melia, \emph{{Constructing effective field theories via their
  harmonics}}, \href{https://doi.org/10.1103/PhysRevD.100.016015}{\emph{Phys.
  Rev. D} {\bfseries 100} (2019) 016015}
  [\href{https://arxiv.org/abs/1902.06754}{{\ttfamily 1902.06754}}].

\bibitem{Fonseca:2019yya}
R.M.~Fonseca, \emph{{Enumerating the operators of an effective field theory}},
  \href{https://doi.org/10.1103/PhysRevD.101.035040}{\emph{Phys. Rev. D}
  {\bfseries 101} (2020) 035040}
  [\href{https://arxiv.org/abs/1907.12584}{{\ttfamily 1907.12584}}].

\bibitem{Adler:1969gk}
S.L.~Adler, \emph{{Axial vector vertex in spinor electrodynamics}},
  \href{https://doi.org/10.1103/PhysRev.177.2426}{\emph{Phys. Rev.} {\bfseries
  177} (1969) 2426}.

\bibitem{Lehman:2015coa}
L.~Lehman and A.~Martin, \emph{{Low-derivative operators of the Standard Model
  effective field theory via Hilbert series methods}},
  \href{https://doi.org/10.1007/JHEP02(2016)081}{\emph{JHEP} {\bfseries 02}
  (2016) 081} [\href{https://arxiv.org/abs/1510.00372}{{\ttfamily
  1510.00372}}].

\bibitem{Henning:2017fpj}
B.~Henning, X.~Lu, T.~Melia and H.~Murayama, \emph{{Operator bases,
  $S$-matrices, and their partition functions}},
  \href{https://doi.org/10.1007/JHEP10(2017)199}{\emph{JHEP} {\bfseries 10}
  (2017) 199} [\href{https://arxiv.org/abs/1706.08520}{{\ttfamily
  1706.08520}}].

\bibitem{SCQ}
C.-Q.~Song, H.~Sun and J.-H.~Yu, ``Complete cp-eigen bases of meson-baryon
  interactions in chiral lagrangian up to $\mathcal{O}(p^5)$.''.

\end{thebibliography}\endgroup

\appendix

\section{Cayley-Hamilton Theorem}
\label{app:A1_Cayley}

Below, we demonstrate how to use the Cayley-Hamilton theorem when reducing redundancies in the trace basis. Since all involved are traces of multiple fields, in order to obtain the Cayley-Hamilton relations satisfied by multiple fields, we need to set \footnote{depending on the number of fields considered.}:
\begin{equation} \mathcal{A}=\mathcal{A}+\mathcal{B}+\mathcal{C}+\mathcal{D}+\mathcal{E}+\dots \,,
\end{equation}
and then substitute it into the expansion. Then, we extract the terms proportional to the product of fields from both sides, which gives us the Cayley-Hamilton relation that needed. We directly present the Cayley-Hamilton relations corresponding to four and five fields in the examples.
For four fields, there is only one independent constraint:
\begin{equation}
\begin{aligned}
& -\langle\mathcal{A}\mathcal{D}\rangle\langle\mathcal{B}\mathcal{C}\rangle - \langle\mathcal{A}\mathcal{C}\rangle\langle\mathcal{B}\mathcal{D}\rangle - \langle\mathcal{A}\mathcal{B}\rangle\langle\mathcal{C}\mathcal{D}\rangle  \\
& + \langle{\mathcal{A}\mathcal{B}\mathcal{C}\mathcal{D}}\rangle + \langle{\mathcal{A}\mathcal{C}\mathcal{B}\mathcal{D}}\rangle + \langle{\mathcal{A}\mathcal{B}\mathcal{D}\mathcal{C}}\rangle + \langle{\mathcal{A}\mathcal{C}\mathcal{D}\mathcal{B}}\rangle + \langle{\mathcal{A}\mathcal{D}\mathcal{B}\mathcal{C}}\rangle + \langle{\mathcal{A}\mathcal{D}\mathcal{C}\mathcal{B}}\rangle \\
& = 0\,, 
\end{aligned}
\end{equation}
so among the 9 trace bases, only 8 are independent. Among these, one could choose the following as the independent bases:
\begin{equation}
\begin{aligned}
    &\langle\mathcal{A}\mathcal{B}\rangle\langle\mathcal{C}\mathcal{D}\rangle,\langle\mathcal{A}\mathcal{D}\rangle\langle\mathcal{B}\mathcal{C}\rangle,\\
    &\langle{\mathcal{A}\mathcal{B}\mathcal{C}\mathcal{D}}\rangle,\langle{\mathcal{A}\mathcal{C}\mathcal{B}\mathcal{D}}\rangle,\langle{\mathcal{A}\mathcal{B}\mathcal{D}\mathcal{C}}\rangle,\\
    &\langle{\mathcal{A}\mathcal{C}\mathcal{D}\mathcal{B}}\rangle,\langle{\mathcal{A}\mathcal{D}\mathcal{B}\mathcal{C}}\rangle,\langle{\mathcal{A}\mathcal{D}\mathcal{C}\mathcal{B}}\rangle.
\end{aligned}
\end{equation}

When there are four fields involved, the Cayley-Hamilton relation is given by a single equation. However, the situation becomes more complicated when there are more than four fields. In such cases, the Cayley-Hamilton relation will yield multiple equations. For example, in the case with five fields, the Cayley-Hamilton relation for five fields will consist of 12 sets of Cayley-Hamilton relations. After removing the redundancy using the same method, we obtain independent trace bases for the situation with five different fields. Below, we directly present the reduced independent bases:
\begin{equation}
    \begin{aligned}
  & \langle \mathcal{A}\mathcal{B}\mathcal{C}\mathcal{D}\mathcal{E} \rangle \,,\quad \langle \mathcal{A}\mathcal{B}\mathcal{C}\mathcal{E}\mathcal{D} \rangle \,,\quad \langle \mathcal{A}\mathcal{B}\mathcal{D}\mathcal{C}\mathcal{E} \rangle \,,\quad \langle \mathcal{A}\mathcal{B}\mathcal{D}\mathcal{E}\mathcal{C} \rangle \,,\quad \langle \mathcal{A}\mathcal{B}\mathcal{E}\mathcal{C}\mathcal{D} \rangle \,,\quad \langle \mathcal{A}\mathcal{B}\mathcal{E}\mathcal{D}\mathcal{C} \rangle \,,\notag \\
  & \langle \mathcal{A}\mathcal{C}\mathcal{B}\mathcal{D}\mathcal{E} \rangle \,,\quad \langle \mathcal{A}\mathcal{C}\mathcal{B}\mathcal{E}\mathcal{D} \rangle \,,\quad \langle \mathcal{A}\mathcal{C}\mathcal{D}\mathcal{B}\mathcal{E} \rangle \,,\quad \langle \mathcal{A}\mathcal{C}\mathcal{D}\mathcal{E}\mathcal{B} \rangle \,,\quad \langle \mathcal{A}\mathcal{C}\mathcal{E}\mathcal{B}\mathcal{D} \rangle \,,\quad \langle \mathcal{A}\mathcal{C}\mathcal{E}\mathcal{D}\mathcal{B} \rangle \,,\notag \\
  & \langle \mathcal{A}\mathcal{D}\mathcal{B}\mathcal{C}\mathcal{E} \rangle \,,\quad \langle \mathcal{A}\mathcal{D}\mathcal{B}\mathcal{E}\mathcal{C} \rangle \,,\quad \langle \mathcal{A}\mathcal{D}\mathcal{C}\mathcal{B}\mathcal{E} \rangle \,,\quad \langle \mathcal{A}\mathcal{D}\mathcal{C}\mathcal{E}\mathcal{B} \rangle \,,\quad \langle \mathcal{A}\mathcal{D}\mathcal{E}\mathcal{B}\mathcal{C} \rangle \,,\quad \langle \mathcal{A}\mathcal{D}\mathcal{E}\mathcal{C}\mathcal{B} \rangle \,,\notag \\
  & \langle \mathcal{A}\mathcal{E}\mathcal{B}\mathcal{C}\mathcal{D} \rangle \,,\quad \langle \mathcal{A}\mathcal{E}\mathcal{B}\mathcal{D}\mathcal{C} \rangle \,,\quad \langle \mathcal{A}\mathcal{E}\mathcal{C}\mathcal{B}\mathcal{D} \rangle \,,\quad \langle \mathcal{A}\mathcal{E}\mathcal{C}\mathcal{D}\mathcal{B} \rangle \,,\quad \langle \mathcal{A}\mathcal{E}\mathcal{D}\mathcal{B}\mathcal{C} \rangle \,,\quad \langle \mathcal{A}\mathcal{E}\mathcal{D}\mathcal{C}\mathcal{B} \rangle \,,\notag \\
  & \langle \mathcal{A}\mathcal{B}\mathcal{C} \rangle\langle \mathcal{D}\mathcal{E} \rangle \,,\quad \langle \mathcal{A}\mathcal{B}\mathcal{D} \rangle\langle \mathcal{C}\mathcal{E} \rangle \,,\quad \langle \mathcal{A}\mathcal{C}\mathcal{D} \rangle\langle \mathcal{B}\mathcal{E} \rangle \,,\quad \langle \mathcal{A}\mathcal{C}\mathcal{B} \rangle\langle \mathcal{D}\mathcal{E} \rangle \,, \notag \\
  & \langle \mathcal{A}\mathcal{D}\mathcal{B} \rangle\langle \mathcal{C}\mathcal{E} \rangle \,,\quad \langle \mathcal{A}\mathcal{E}\mathcal{B} \rangle\langle \mathcal{C}\mathcal{D} \rangle \,,\quad \langle \mathcal{A}\mathcal{B} \rangle\langle \mathcal{C}\mathcal{D}\mathcal{E} \rangle \,,\quad \langle \mathcal{A}\mathcal{B} \rangle\langle \mathcal{C}\mathcal{E}\mathcal{D} \rangle \,.
\end{aligned}
\end{equation}

For a set of the Cayley-Hamilton equation involving repeated fields that appears in the main content, we present their forms here.

The Cayley-Hamilton relation for $\mathcal{ABCDD}$ is:
\begin{equation}
    \begin{aligned}
        1) \quad & \langle \mathcal{A}\mathcal{B}\mathcal{D}\mathcal{C}\mathcal{D}\rangle +\langle \mathcal{A}\mathcal{B}\mathcal{D}\mathcal{D}\mathcal{C}\rangle +\langle \mathcal{A}\mathcal{C}\mathcal{B}\mathcal{D}\mathcal{D}\rangle +\langle \mathcal{A}\mathcal{C}\mathcal{D}\mathcal{B}\mathcal{D}\rangle +\langle \mathcal{A}\mathcal{D}\mathcal{B}\mathcal{D}\mathcal{C}\rangle +\langle \mathcal{A}\mathcal{D}\mathcal{C}\mathcal{B}\mathcal{D}\rangle  \notag \\
           = & \langle \mathcal{A}\mathcal{B}\mathcal{D}\rangle\langle \mathcal{C}\mathcal{D}\rangle +\langle \mathcal{A}\mathcal{C}\mathcal{D}\rangle\langle \mathcal{B}\mathcal{D}\rangle +\langle \mathcal{A}\mathcal{C}\rangle\langle \mathcal{B}\mathcal{D}\mathcal{D}\rangle +\langle \mathcal{A}\mathcal{D}\mathcal{C}\rangle\langle \mathcal{B}\mathcal{D}\rangle +\langle \mathcal{A}\mathcal{D}\rangle\langle \mathcal{B}\mathcal{D}\mathcal{C}\rangle \,, \\
    2) \quad & \langle \mathcal{A}\mathcal{C}\mathcal{D}\mathcal{B}\mathcal{D}\rangle +\langle \mathcal{A}\mathcal{C}\mathcal{D}\mathcal{D}\mathcal{B}\rangle +\langle \mathcal{A}\mathcal{D}\mathcal{B}\mathcal{C}\mathcal{D}\rangle +\langle \mathcal{A}\mathcal{D}\mathcal{B}\mathcal{D}\mathcal{C}\rangle +\langle \mathcal{A}\mathcal{D}\mathcal{C}\mathcal{D}\mathcal{B}\rangle +\langle \mathcal{A}\mathcal{D}\mathcal{D}\mathcal{B}\mathcal{C}\rangle  \notag \\
           = & \langle \mathcal{A}\mathcal{C}\mathcal{D}\rangle\langle \mathcal{B}\mathcal{D}\rangle +\langle \mathcal{A}\mathcal{C}\rangle\langle \mathcal{B}\mathcal{D}\mathcal{D}\rangle +\langle \mathcal{A}\mathcal{D}\mathcal{B}\rangle\langle \mathcal{C}\mathcal{D}\rangle +\langle \mathcal{A}\mathcal{D}\mathcal{C}\rangle\langle \mathcal{B}\mathcal{D}\rangle +\langle \mathcal{A}\mathcal{D}\rangle\langle \mathcal{B}\mathcal{C}\mathcal{D}\rangle \,, \\
    3) \quad & \langle \mathcal{A}\mathcal{B}\mathcal{C}\mathcal{D}\mathcal{D}\rangle +\langle \mathcal{A}\mathcal{B}\mathcal{D}\mathcal{D}\mathcal{C}\rangle +\langle \mathcal{A}\mathcal{C}\mathcal{B}\mathcal{D}\mathcal{D}\rangle +\langle \mathcal{A}\mathcal{C}\mathcal{D}\mathcal{D}\mathcal{B}\rangle +\langle \mathcal{A}\mathcal{D}\mathcal{D}\mathcal{B}\mathcal{C}\rangle +\langle \mathcal{A}\mathcal{D}\mathcal{D}\mathcal{C}\mathcal{B}\rangle  \notag \\
           = & \langle \mathcal{A}\mathcal{B}\mathcal{C}\rangle\langle \mathcal{D}\mathcal{D}\rangle +\langle \mathcal{A}\mathcal{B}\rangle\langle \mathcal{C}\mathcal{D}\mathcal{D}\rangle +\langle \mathcal{A}\mathcal{C}\mathcal{B}\rangle\langle \mathcal{D}\mathcal{D}\rangle +\langle \mathcal{A}\mathcal{C}\rangle\langle \mathcal{B}\mathcal{D}\mathcal{D}\rangle +\langle \mathcal{A}\mathcal{D}\mathcal{D}\rangle\langle \mathcal{B}\mathcal{C}\rangle \,, \\
    4) \quad & \langle \mathcal{A}\mathcal{B}\mathcal{C}\mathcal{D}\mathcal{D}\rangle +\langle \mathcal{A}\mathcal{B}\mathcal{D}\mathcal{C}\mathcal{D}\rangle +\langle \mathcal{A}\mathcal{C}\mathcal{D}\mathcal{B}\mathcal{D}\rangle +\langle \mathcal{A}\mathcal{C}\mathcal{D}\mathcal{D}\mathcal{B}\rangle +\langle \mathcal{A}\mathcal{D}\mathcal{B}\mathcal{C}\mathcal{D}\rangle +\langle \mathcal{A}\mathcal{D}\mathcal{C}\mathcal{D}\mathcal{B}\rangle  \notag \\
           = & \langle \mathcal{A}\mathcal{B}\mathcal{D}\rangle\langle \mathcal{C}\mathcal{D}\rangle +\langle \mathcal{A}\mathcal{B}\rangle\langle \mathcal{C}\mathcal{D}\mathcal{D}\rangle +\langle \mathcal{A}\mathcal{C}\mathcal{D}\rangle\langle \mathcal{B}\mathcal{D}\rangle +\langle \mathcal{A}\mathcal{D}\mathcal{B}\rangle\langle \mathcal{C}\mathcal{D}\rangle +\langle \mathcal{A}\mathcal{D}\rangle\langle \mathcal{B}\mathcal{C}\mathcal{D}\rangle \,, \\
    5) \quad & \langle \mathcal{A}\mathcal{B}\mathcal{D}\mathcal{C}\mathcal{D}\rangle +\langle \mathcal{A}\mathcal{B}\mathcal{D}\mathcal{D}\mathcal{C}\rangle +\langle \mathcal{A}\mathcal{D}\mathcal{B}\mathcal{D}\mathcal{C}\rangle +\langle \mathcal{A}\mathcal{D}\mathcal{C}\mathcal{B}\mathcal{D}\rangle +\langle \mathcal{A}\mathcal{D}\mathcal{C}\mathcal{D}\mathcal{B}\rangle +\langle \mathcal{A}\mathcal{D}\mathcal{D}\mathcal{C}\mathcal{B}\rangle  \notag \\
           = & \langle \mathcal{A}\mathcal{B}\mathcal{D}\rangle\langle \mathcal{C}\mathcal{D}\rangle +\langle \mathcal{A}\mathcal{B}\rangle\langle \mathcal{C}\mathcal{D}\mathcal{D}\rangle +\langle \mathcal{A}\mathcal{D}\mathcal{B}\rangle\langle \mathcal{C}\mathcal{D}\rangle +\langle \mathcal{A}\mathcal{D}\mathcal{C}\rangle\langle \mathcal{B}\mathcal{D}\rangle +\langle \mathcal{A}\mathcal{D}\rangle\langle \mathcal{B}\mathcal{D}\mathcal{C}\rangle \,, \\
    6) \quad & \langle \mathcal{A}\mathcal{B}\mathcal{C}\mathcal{D}\mathcal{D}\rangle +\langle \mathcal{A}\mathcal{D}\mathcal{B}\mathcal{C}\mathcal{D}\rangle +\langle \mathcal{A}\mathcal{D}\mathcal{D}\mathcal{B}\mathcal{C}\rangle  \notag \\
           = & 2\langle \mathcal{A}\mathcal{D}\mathcal{D}\rangle\langle \mathcal{B}\mathcal{C}\rangle +2\langle \mathcal{A}\mathcal{D}\rangle\langle \mathcal{B}\mathcal{C}\mathcal{D}\rangle +\langle \mathcal{A}\mathcal{B}\mathcal{C}\rangle\langle \mathcal{D}\mathcal{D}\rangle \,, \\
    7) \quad & \langle \mathcal{A}\mathcal{C}\mathcal{B}\mathcal{D}\mathcal{D}\rangle +\langle \mathcal{A}\mathcal{D}\mathcal{C}\mathcal{B}\mathcal{D}\rangle +\langle \mathcal{A}\mathcal{D}\mathcal{D}\mathcal{C}\mathcal{B}\rangle  \notag \\
           = & 2\langle \mathcal{A}\mathcal{D}\mathcal{D}\rangle\langle \mathcal{B}\mathcal{C}\rangle +2\langle \mathcal{A}\mathcal{D}\rangle\langle \mathcal{B}\mathcal{D}\mathcal{C}\rangle +\langle \mathcal{A}\mathcal{C}\mathcal{B}\rangle\langle \mathcal{D}\mathcal{D}\rangle \,.
    \end{aligned}
\end{equation}

The Cayley-Hamilton relation for $\mathcal{ABCCC}$ is:
\begin{equation}
    \begin{aligned}
    1) \quad & \langle \mathcal{A}\mathcal{B}\mathcal{C}\mathcal{C}\mathcal{C}\rangle +\langle \mathcal{A}\mathcal{C}\mathcal{B}\mathcal{C}\mathcal{C}\rangle +\langle \mathcal{A}\mathcal{C}\mathcal{C}\mathcal{B}\mathcal{C}\rangle  \notag \\ 
    = &  2\langle \mathcal{A}\mathcal{C}\mathcal{C}\rangle\langle \mathcal{B}\mathcal{C}\rangle +2\langle \mathcal{A}\mathcal{C}\rangle\langle \mathcal{B}\mathcal{C}\mathcal{C}\rangle +\langle \mathcal{A}\mathcal{B}\mathcal{C}\rangle\langle \mathcal{C}\mathcal{C}\rangle  \,, \\
   2) \quad & \langle \mathcal{A}\mathcal{C}\mathcal{B}\mathcal{C}\mathcal{C}\rangle +\langle \mathcal{A}\mathcal{C}\mathcal{C}\mathcal{B}\mathcal{C}\rangle +\langle \mathcal{A}\mathcal{C}\mathcal{C}\mathcal{C}\mathcal{B}\rangle  \notag \\ 
    = &  2\langle \mathcal{A}\mathcal{C}\mathcal{C}\rangle\langle \mathcal{B}\mathcal{C}\rangle +2\langle \mathcal{A}\mathcal{C}\rangle\langle \mathcal{B}\mathcal{C}\mathcal{C}\rangle +\langle \mathcal{A}\mathcal{C}\mathcal{B}\rangle\langle \mathcal{C}\mathcal{C}\rangle \,, \\
   3) \quad & 2\langle \mathcal{A}\mathcal{B}\mathcal{C}\mathcal{C}\mathcal{C}\rangle +2\langle \mathcal{A}\mathcal{C}\mathcal{C}\mathcal{C}\mathcal{B}\rangle +\langle \mathcal{A}\mathcal{C}\mathcal{B}\mathcal{C}\mathcal{C}\rangle +\langle \mathcal{A}\mathcal{C}\mathcal{C}\mathcal{B}\mathcal{C}\rangle  \notag \\ 
    = &  \langle \mathcal{A}\mathcal{B}\mathcal{C}\rangle\langle \mathcal{C}\mathcal{C}\rangle +\langle \mathcal{A}\mathcal{B}\rangle\langle \mathcal{C}\mathcal{C}\mathcal{C}\rangle +\langle \mathcal{A}\mathcal{C}\mathcal{B}\rangle\langle \mathcal{C}\mathcal{C}\rangle +\langle \mathcal{A}\mathcal{C}\mathcal{C}\rangle\langle \mathcal{B}\mathcal{C}\rangle +\langle \mathcal{A}\mathcal{C}\rangle\langle \mathcal{B}\mathcal{C}\mathcal{C}\rangle \,. \\
    \end{aligned}
\end{equation}

\section{Chiral Dim-4 Operator List}
\label{app:p4}
\subsection{Chiral dim-4 operator list of SU(2)}

\setcounter{magicrownumbers}{0}
\begin{center}
\begin{longtable}{|c|c|c|c|c|c|}
    \caption{Chiral dim-4 operators of SU(2).} \\
    \hline
        &&$C+P+$&$C+P-$&$C-P+$&$C-P-$\\
        \hline
        Type
        &\myrownumber
        &${\mathcal{O}^{++}}_i^{(4)}$
        &${\mathcal{O}^{+-}}_i^{(4)}$
        &${\mathcal{O}^{-+}}_i^{(4)}$
        &${\mathcal{O}^{--}}_i^{(4)}$\\
        \hline
        $f^2$
        &\myrownumber
        &$\langle f_{+\mu\nu} f_{+}^{\mu\nu}\rangle$
        &$\langle f_+^{\mu\nu} f_+^{\rho\lambda}\rangle\epsilon_{\mu\nu\rho\lambda}$&$\langle f_+^{\mu\nu} f_-^{\rho\lambda}\rangle\epsilon_{\mu\nu\rho\lambda}$
        &$\langle f_{+\mu\nu} f_-^{\mu\nu}\rangle$\\
        &\myrownumber
        &$\langle f_{-\mu\nu} f_{-}^{\mu\nu}\rangle$
        &$\langle f_-^{\mu\nu} f_-^{\rho\lambda}\rangle\epsilon_{\mu\nu\rho\lambda}$
        &
        &\\
        \hline 
        $fu^2$
        &\myrownumber
        &$\langle u_{\mu}u_{\nu} f_{+}^{\mu\nu}\rangle$
        &$\langle f_+^{\mu\nu}u^\rho u^\lambda\rangle\epsilon_{\mu\nu\rho\lambda}$
        &$\langle f_-^{\mu\nu}u^\rho u^\lambda\rangle\epsilon_{\mu\nu\rho\lambda}$
        &$\langle u_{\mu}u_{\nu}f_{-}^{\mu \nu}\rangle$\\
        \hline
        $u^4$
        &\myrownumber
        &$\langle u_{\mu}u_{\nu} \rangle  \langle u^{\mu}u^{\nu} \rangle$
        &
        &
        &\\
        &\myrownumber
        &$\langle u_{\mu}u^{\mu} \rangle  \langle u_{\nu}u^{\nu} \rangle$
        &
        &
        &\\
\hline   
    \end{longtable} 
\end{center}

\subsection{Chiral dim-4 operator list of SU(3)}
\setcounter{magicrownumbers}{0}
\begin{center}
    \begin{longtable}{|c|c|c|c|c|c|}
    \caption{Chiral dim-4 operators of SU(3).} \\
    \hline
        &&$C+P+$&$C+P-$&$C-P+$&$C-P-$\\
        \hline
        Type
        &\myrownumber
        &${\mathcal{O}^{++}}_i^{(4)}$
        &${\mathcal{O}^{+-}}_i^{(4)}$
        &${\mathcal{O}^{-+}}_i^{(4)}$
        &${\mathcal{O}^{--}}_i^{(4)}$\\
        \hline
        $f^2$
        &\myrownumber
        &$\langle f_{+\mu\nu} f_{+}^{\mu\nu}\rangle$ 
        &$\langle f_+^{\mu\nu} f_+^{\rho\lambda}\rangle\epsilon_{\mu\nu\rho\lambda}$
        &$\langle f_+^{\mu\nu}  f_+^{\rho\lambda}\rangle\epsilon_{\mu\nu\rho\lambda}$
        &$\langle f_{+\mu\nu} f_-^{\mu\nu}\rangle$\\
        
        &\myrownumber
        &$\langle f_{-\mu\nu} f_{-}^{\mu\nu}\rangle$
        &$\langle f_-^{\mu\nu}  f_-^{\rho\lambda}\rangle\epsilon_{\mu\nu\rho\lambda}$
        &
        &\\
        \hline
        $fu^2$
        &\myrownumber
        &$\langle u_{\mu}u_{\nu} f_{+}^{\mu \nu}\rangle$
        &$\langle u^{\mu}u^{\nu}f_-^{\rho\lambda}\rangle\epsilon_{\mu\nu\rho\lambda}$
        &$\langle u^{\mu}u^{\nu}f_-^{\rho\lambda}\rangle\epsilon_{\mu\nu\rho\lambda}$
        &$\langle u_{\mu}u_{\nu} f_{-}^{\mu \nu}\rangle$\\
        \hline
        $u^4$
        &\myrownumber
        &$\langle u^\mu u_\mu u^\nu u_\nu \rangle$
        &
        &
        &\\
        &\myrownumber
        &$\langle u^\mu u^\nu u_\nu u_\mu \rangle$
        &
        &
        &\\
\hline
    \end{longtable} 
\end{center}

\newpage

\begin{landscape}
\section{Chiral Dim-6 Operator List}
\label{app:p6}
\subsection{Chiral dim-6 operator list of SU(2)}

\setcounter{magicrownumbers}{0}
\begin{center}
    \begin{longtable}{|c|c|c|c|c|c|}
    \caption{Chiral dim-6 operators of SU(2).} \\
    \hline
        &&$C+P+$&$C+P-$&$C-P+$&$C-P-$\\
        \hline
         Type&$i$&${\mathcal{O}^{++}}_i^{(6)}$&${\mathcal{O}^{+-}}_i^{(6)}$&${\mathcal{O}^{-+}}_i^{(6)}$&${\mathcal{O}^{--}}_i^{(6)}$\\
         \hline
        {$D^2f^2$}
        &\myrownumber&$\langle D^\mu D_\lambda f_-^{\nu\lambda} f_{-\mu\nu}\rangle$&&&$\langle D^\mu D_\lambda f_-^{\nu\lambda} f_{+\mu\nu}\rangle$\\
        &\myrownumber&$\langle D^\mu D_\lambda f_+^{\nu\lambda} f_{+\mu\nu}\rangle$&&&\\
         \hline
        {$f^3$}
        &\myrownumber
        &$\langle[f_{+\mu\nu},f_{+\lambda}^\nu] f_+^{\lambda\mu}\rangle$
        &$\langle [f_+^{\mu\alpha},f_+^{\nu\rho}]f_{+\alpha}^\lambda\rangle\epsilon_{\mu\nu\rho\lambda}$
        &$\langle [f_-^{\mu\alpha},f_-^{\nu\rho}]f_{-\alpha}^\lambda \rangle\epsilon_{\mu\nu\rho\lambda}$
        &$\langle[f_{-\mu\nu},f_{-\lambda}^\nu] f_-^{\lambda\mu}\rangle$\\
        &\myrownumber
        &$\langle[f_{-\mu\nu},f_{-\lambda}^\nu] f_+^{\lambda\mu}\rangle$
        &$\langle [f_-^{\mu\alpha},f_-^{\nu\rho}]f_{+\alpha}^\lambda\rangle\epsilon_{\mu\nu\rho\lambda}$
        &$\langle [f_+^{\mu\alpha},f_+^{\nu\rho}]f_{-\alpha}^\lambda\rangle\epsilon_{\mu\nu\rho\lambda}$
        &$\langle[f_{-\mu\nu},f_{+\lambda}] f_+^{\lambda\mu}\rangle$\\
         \hline
         {$f^2\Sigma$}
        &\myrownumber
        &$\langle\Sigma_-\rangle\langle f_-^{\mu\nu}f_-^{\rho\lambda}\rangle\epsilon_{\mu\nu\rho\lambda}$
        &$\langle\Sigma_-\rangle\langle f_-^{\mu\nu} f_{-\mu\nu}\rangle$
        &$\langle\Sigma_-\rangle\langle f_{-\mu\nu} f_+^{\mu\nu}\rangle$
        &$\langle\Sigma_-\rangle\langle f_-^{\mu\nu}f_+^{\rho\lambda}\rangle\epsilon_{\mu\nu\rho\lambda}$\\
\hline
       
    \end{longtable}
\end{center}

\newpage

\subsection{Chiral dim-6 operator list of SU(3)}

\setcounter{magicrownumbers}{0}
\begin{center}
    \begin{longtable}{|c|c|c|c|c|c|}
    \caption{Chiral dim-6 operators of SU(3).} \\
    \hline
        &&$C+P+$&$C+P-$&$C-P+$&$C-P-$\\
        \hline
         Type&$i$&${\mathcal{O}^{++}}_i^{(6)}$&${\mathcal{O}^{+-}}_i^{(6)}$&${\mathcal{O}^{-+}}_i^{(6)}$&${\mathcal{O}^{--}}_i^{(6)}$\\
         \hline
        {$D^2f^2$}
        &\myrownumber
        &$\langle D^\mu D_\lambda f_-^{\nu\lambda} f_{-\mu\nu}\rangle$& &
        &$\langle D^\mu D_\lambda f_-^{\nu\lambda} f_{+\mu\nu}\rangle$\\
        &\myrownumber
        &$\langle D^\mu D_\lambda f_+^{\nu\lambda} f_{+\mu\nu}\rangle$&&&\\
         \hline
        {$f^3$}
        &\myrownumber
        &$\langle[f_{-\mu\nu},f_{-\lambda}^\nu] f_+^{\lambda\mu}\rangle$
        &$\langle f_+^{\mu\nu}f_-^{\rho\alpha}f_{-\alpha}^\lambda\rangle\epsilon_{\mu\nu\rho\lambda}$
        &$\langle f_-^{\mu\nu}f_+^{\rho\alpha}f_{+\alpha}^\lambda\rangle\epsilon_{\mu\nu\rho\lambda}$
        &$\langle f_{\mu\nu}[f_{-\lambda}^\nu,f_-^{\lambda\mu}]\rangle$\\
        &\myrownumber
        &$\langle[f_{+\mu\nu},f_{+\lambda}^\nu] f_+^{\lambda\mu}\rangle$
        &$\langle f_+^{\mu\nu}f_+^{\rho\alpha}f_{+\alpha}^\lambda\rangle\epsilon_{\mu\nu\rho\lambda}$
        &$\langle f_-^{\mu\nu}f_-^{\rho\alpha}f_{-\alpha}^\lambda\rangle\epsilon_{\mu\nu\rho\lambda}$
        &$\langle f_{\-\mu\nu}[f_{+\lambda}^\nu,f_+^{\lambda\mu}]\rangle$\\
         \hline 
        {$f^2\Sigma$}
        &\myrownumber
        &$\langle\Sigma_-\rangle\langle f_-^{\mu\nu}f_-^{\rho\lambda}\rangle\epsilon_{\mu\nu\rho\lambda}$
        &$\langle\Sigma_-\rangle\langle f_-^{\mu\nu} f_{-\mu\nu}\rangle$
        &$\langle\Sigma_-\rangle\langle f_{-\mu\nu} f_+^{\mu\nu}\rangle$
        &$\langle\{f_{-\mu\nu},f_+^{\rho\lambda}\}\Sigma_-\rangle\epsilon_{\mu\nu\rho\lambda}$\\
\hline
        
    \end{longtable}
\end{center}

\newpage
\end{landscape}

\begin{landscape}
\section{Chiral Dim-8 Operator List}
\label{app:p8}
\subsection{Chiral dim-8 operator list of SU(2)}

\setcounter{magicrownumbers}{0}
\setlength\tabcolsep{6pt} 
\fontsize{8pt}{10pt}\selectfont
\begin{center}
    \begin{longtable}{|c|c|c|c|c|c|}
    \caption{Chiral dim-8 operators of SU(2).} \\
    \hline
        &&$C+P+$&$C+P-$&$C-P+$&$C-P-$\\
        \hline
         Type&$i$&${\mathcal{O}^{++}}_i^{(8)}$&${\mathcal{O}^{+-}}_i^{(8)}$&${\mathcal{O}^{-+}}_i^{(8)}$&${\mathcal{O}^{--}}_i^{(8)}$\\
         \hline
        {$D^4f^2$}
        &\myrownumber
        &$\langle f_-^{\mu\nu}D^\rho D^\lambda D_\rho D_\lambda f_{-\mu\nu}\rangle$
        &
        &
        &$\langle f_-^{\mu\nu}D^\rho D^\lambda D_\rho D_\lambda f_{+\mu\nu}\rangle$
        \\
        &\myrownumber
        &$\langle f_+^{\mu\nu}D^\rho D^\lambda D_\rho D_\lambda f_{+\mu\nu}\rangle$
        &
        &
        &
        \\
        \hline
        {$f^4$}
        &\myrownumber
        &$\langle f_{-\mu\nu} f_-^{\nu\rho}\rangle\langle f_{-\lambda\rho} f_-^{\lambda\mu}\rangle$
        &$\langle f_+^{\mu\nu}f_+^{\rho\lambda}\rangle\langle f_{+\alpha\beta}f_+^{\alpha\beta}\rangle\epsilon_{\mu\nu\rho\lambda}$
        &$\langle f_-^{\mu\alpha}f_-^{\nu\beta}\rangle\langle f_{-\alpha}^\rho f_{+\beta}^\lambda\rangle\epsilon_{\mu\nu\rho\lambda}$
        &$\langle f_{-\mu\nu}f_-^{\nu\rho}\rangle\langle f_{-\rho\lambda}f_+^{\mu\lambda}\rangle$
        \\
        &\myrownumber
        &$\langle f_{-\lambda\rho} f_{-\mu\nu}\rangle\langle f_-^{\lambda\rho} f_-^{\mu\nu}\rangle$
        &$\langle f_+^{\mu\nu} f_{+\alpha\beta}\rangle\langle f_+^{\rho\lambda} f_+^{\alpha\beta}\rangle\epsilon_{\mu\nu\rho\lambda}$
        &$\langle f_-^{\mu\alpha}f_{-\alpha}^\nu\rangle\langle f_-^{\rho\beta}f_{+\beta}^\lambda\rangle\epsilon_{\mu\nu\rho\lambda}$
        &$\langle f_{-\mu\nu}f_{-\rho\lambda}\rangle\langle f_-^{\mu\nu}f_+^{\rho\lambda}\rangle$
        \\
\hline
    \end{longtable}
\end{center}

\subsection{Chiral dim-8 operator list of SU(3)}

\setcounter{magicrownumbers}{0}
\setlength\tabcolsep{6pt} 
\fontsize{8pt}{10pt}\selectfont
\begin{center}
\begin{longtable}{|c|c|c|c|c|c|}
    \caption{Chiral dim-8 operators of SU(3).} \\
    \hline
        &&$C+P+$&$C+P-$&$C-P+$&$C-P-$\\
\hline
         Type&$i$&${\mathcal{O}^{++}}_i^{(8)}$&${\mathcal{O}^{+-}}_i^{(8)}$&${\mathcal{O}^{-+}}_i^{(8)}$&${\mathcal{O}^{--}}_i^{(8)}$\\
\hline
        {$D^4 f^2$}
        &\myrownumber&$\langle f_-^{\rho\lambda}D^\mu D^\nu D_\nu D_\mu f_{-\rho\lambda}\rangle$
        &&&$\langle f_-^{\rho\lambda}D^\mu D^\nu D_\nu D_\mu f_{+\rho\lambda}\rangle$\\
        &\myrownumber&$\langle f_+^{\rho\lambda}D^\mu D^\nu D_\nu D_\mu f_{+\rho\lambda}\rangle$
        &&&\\
        \hline
        {$D^4 \Sigma^2$}
        &\myrownumber&$\langle\Sigma_+D^4\Sigma_+\rangle$
        &$\langle\Sigma_- D^4 \Sigma_+\rangle$&&\\
        &\myrownumber&$\langle\Sigma_-D^4\Sigma_-\rangle$
        &$\langle\Sigma_-\rangle D^4\langle \Sigma_+\rangle$&&\\
        &\myrownumber&$\langle\Sigma_-\rangle D^4\langle \Sigma_-\rangle$&&&\\
\hline
    \end{longtable}

\end{center}

\end{landscape}

\end{document}